\documentclass[final,5p,twocolumn]{elsarticle}
\makeatletter
\def\ps@pprintTitle{%
 \let\@oddhead\@empty
 \let\@evenhead\@empty
 \def\@oddfoot{}%
 \let\@evenfoot\@oddfoot}
\makeatother
\usepackage{graphicx}
\usepackage{amssymb}
\usepackage{amsthm}
\theoremstyle{definition}
\newtheorem{defn}{Definition}

\usepackage{lineno}

\usepackage{algorithmic}
\usepackage{mathtools}
\usepackage{multirow}
\usepackage{array}
\usepackage{tabularx}
\usepackage{booktabs}
\usepackage{threeparttable}
\usepackage{makecell}
\usepackage{lscape}
\usepackage{longtable}
\usepackage{enumitem}
\usepackage{subcaption}
\usepackage{graphicx}
\usepackage{xurl}
\usepackage{soul} 

\usepackage{txfonts}

    \usepackage{verbatim}
    \usepackage{color}

    \newcounter{DaveCommentCounter}
    \newcommand{\nrs}[1]{\textcolor{black}{#1}}

    \newcounter{AshfaqCommentCounter}
\usepackage[dvipsnames]{xcolor}
\usepackage[Algorithm,ruled]{algorithm}

\journal{Journal of Systems and Software}

\begin{document}

\begin{frontmatter}


\title{SWFC-ART: A Cost-effective Approach for Fixed-Size-Candidate-Set Adaptive Random Testing through Small World Graphs}

\author[ujs]{Muhammad Ashfaq}
\ead{5102180325@stmail.ujs.edu.cn}

\author[ujs,macao]{Rubing Huang\corref{mycorrespondingauthor}}
\ead{rbhuang@ujs.edu.cn, rbhuang@must.edu.mo}

\author[unn]{Dave Towey}
\ead{dave.towey@nottingham.edu.cn}

\author[ttu]{Michael Omari}
\ead{michael.omari@ttu.edu.gh}

\author[russia]{Dmitry Yashunin}
\ead{Dmitry.Yashunin@harman.com}

\author[ghana]{Patrick Kwaku Kudjo}
\ead{kudjo@upsamail.edu.gh}

\author[macao]{Tao Zhang}
\ead{tazhang@must.edu.mo}

\cortext[mycorrespondingauthor]{Corresponding author}

\address[ujs]{School of Computer Science and Communication Engineering, Jiangsu University, Zhenjiang, Jiangsu, 212013, China}
\address[macao]{Faculty of Information Technology, Macau University of Science and Technology, Macau 999078, China}
\address[unn]{School of Computer Science, University of Nottingham Ningbo, China}
\address[ttu]{Department of Computer Science, Takoradi Technical University, Takoradi, Western Region, Ghana}
\address[russia]{Harman X, Nizhny Novgorod, 24 Salganskaya street, 603105, Russia}
\address[ghana]{University of Professional Studies, Accra, Ghana}

\begin{abstract}
Adaptive random testing (ART) improves the failure-detection effectiveness of random testing by leveraging properties of the clustering of failure-causing inputs of most faulty programs: ART uses a sampling mechanism that evenly spreads test cases within a software's input domain.
The widely-used Fixed-Sized-Candidate-Set ART (FSCS-ART) sampling strategy faces a quadratic time cost, which worsens as the dimensionality of the software input domain increases.
In this paper, we propose an approach based on small world graphs that can enhance the computational efficiency of FSCS-ART:
SWFC-ART.
To efficiently perform nearest neighbor queries for candidate test cases, SWFC-ART incrementally constructs a hierarchical navigable small world graph for previously executed, non-failure-causing test cases.
Moreover,  SWFC-ART has shown consistency in programs with high dimensional input domains.
Our simulation and empirical studies show that SWFC-ART reduces the computational overhead of FSCS-ART from quadratic to log-linear order while maintaining the failure-detection effectiveness of FSCS-ART, and remaining consistent in high dimensional input domains.
We recommend using SWFC-ART in practical software testing scenarios, where real-life programs often have high dimensional input domains and  low failure rates.
\end{abstract}


\end{frontmatter}




\section{Introduction}
\label{introduction}

Software testing~\cite{Myers:2004:AST:983238} is a fundamental software quality assurance activity.
Random testing (RT)~\cite{Duran1981} is a popular black-box software testing technique that randomly selects and executes a subset of test cases from the software's input domain.
Benefits of using RT include that it does not require explicit information, other than the inputs it takes, about the software under test~\cite{whenToUseRT}, and that it can easily be automated~\cite{hamlet1994random}.
RT has been used to test many real-life software packages, including:
Windows NT applications~\cite{10.5555/1267102.1267108};
embedded software systems~\cite{Regehr2005};
database systems~\cite{bati2007genetic, slutz1998massive};
\nrs{Android applications~\cite{Muangsiri2017};}
Java Just-In-Time Compilers~\cite{yoshikawa2003random};
.NET error detection~\cite{pacheco2008finding};
security assessment~\cite{10.1145/1379022.1375607};
Mac OS robustness assessment~\cite{10.1145/1145735.1145743};
graphical user interfaces~\cite{Daboczi2003}; and
UNIX utility programs~\cite{10.1145/96267.96279, millerfuzz1998}.

Although RT has been reported to be effective, a large body of related research has highlighted continuing questions about its actual effectiveness~\cite{Myers:2004:AST:983238, 5010257}.
\nrs{For example, RT does not take advantage of non-failure-revealing test cases, which may still have information about program behavior, and should not be discarded without careful inspection~\cite{Myers:2004:AST:983238}.}
Moreover, RT may not achieve satisfactory failure-detection effectiveness and code coverage~\cite{chen2013code}, which may make it unsuitable when there are constraints on the number of test cases that can be executed.

\begin{figure}[!t]
    \centering
    \includegraphics[width=\linewidth]{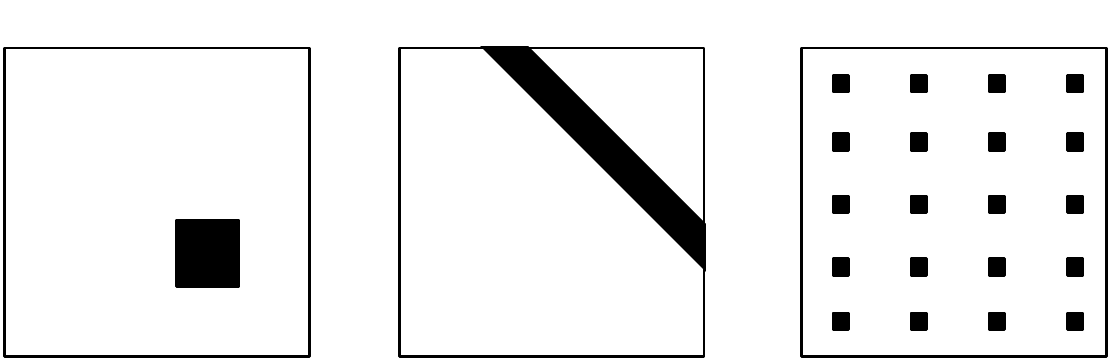}
    \caption{Block, strip and point failure patterns}
    \label{failure_patterns}
\end{figure}

An important finding in software testing has been the fact that inputs that cause programs to fail (failure-causing inputs), more often than not, form contiguous regions within the input domain of a program~\cite{10.1145/1295074.1295091, Arcuri2011, 10.1145/566171.566203, Bishop, Chan1996, Finelli1991, Ammann1988, White1980}.
\nrs{A family of testing techniques called Adaptive Random Testing (ART)~\cite{Chen2004art, Huang2019, Chen2019} is based on the idea that, if RT's basic technique is slightly modified, such that test cases are more evenly distributed, the chances of encountering failure-causing inputs can be significantly increased.}
\nrs{Three broad categories of patterns of these failure-causing inputs have been identified \cite{Chan1996}, as shown in Fig. \ref{failure_patterns}:
\textit{block}, \textit{strip} and \textit{point} patterns.}

An objective of ART is to minimize the number of test cases executions needed before revealing program failure.
\nrs{Theoretical support for ART includes not repeating testing where failure is unlikely to occur, covering every aspect of the program (coverage), and exploring as much variety (diversity) in inputs as possible \cite{10.1016/j.jss.2009.02.022, Ammann1988}.
ART has been shown to be more effective than RT in real-life software testing scenarios
---
such as resource-constrained testing~\cite{DBLP:conf/serp/MayerS06} and beta testing~\cite{Chen2007c}
---
achieving better code coverage~\cite{6449335, 4601538, 10.1016/j.ins.2011.01.025}, and using fewer test cases to find failure~\cite{Chen2004art}.}
ART can require up to 50\% fewer test cases to find the first failure than RT~\cite{10.1145/1363102.1363107}.
Recently, ART has been gaining traction as a viable approach for testing real-life programs and systems, such as:
testing deep neural networks~\cite{8944083};
detection of cross-site-scripting (XSS) attacks~\cite{Lv2019};
exposing SQL database vulnerabilities~\cite{Zhang2019}; and
testing object-oriented programs~\cite{ciupa2008}.

Several ART approaches exist, based on different strategies and motivations~\cite{Huang2019, Chen2019}.
\nrs{The Fixed-Size-Candidate-Set version of ART (FSCS-ART) was the first proposed, and is the most-widely-used ART strategy, best known for its simplicity and failure-detection effectiveness \cite{Huang2019}.}
FSCS-ART follows the ART principle that if a test case does not reveal a failure, then nearby inputs are also unlikely to do so.
Subsequent test cases, therefore, should be selected \textit{far away} from the previously executed, non-failure-causing test cases.
FSCS-ART achieves the concept of \textit{far away} through computing distances among test cases.

Unfortunately, FSCS-ART incurs a quadratic time complexity when generating test cases, due to the brute-force strategy for determining the nearest neighbors of \textit{candidate test cases} (see Section \ref{fscs-art})
---
the distance between each candidate test case and \textit{all} executed test cases must be calculated before the nearest neighbor can be identified.
\nrs{Furthermore, this computational cost rises sharply as the number of program input parameters (dimensions of the input domain, $D$) increases. }
\nrs{These two issues can be collectively referred to as the \textit{double-tier efficiency problem} of FSCS-ART.
Given that most real-world programs have high dimensional input domains~\cite{Lin2009} and low failure rates~\cite{Arcuri2011}
---
which means that many test case executions may be required before finding a failure
---
the FSCS-ART double-tier efficiency problem needs to be addressed. }

\nrs{
The work reported in this paper addresses the FSCS-ART double-tier efficiency problem.
By identifying FSCS-ART as an instance of the nearest neighbor search (NNS) problem, we hypothesize that a solution to the FSCS-ART efficiency problem may lie in addressing the NNS mechanism:
If the NNS mechanism can be scalable, and consistent with dataset size and dimensions, it may alleviate the double-tier efficiency problem.
Furthermore, \textit{approximate} NNS (ANNS) should be able to significantly alleviate the computational overheads of distance calculations, especially in high dimensional input domains~\cite{Indyk1998}.
In software testing, NNS has been used to find the most similar test cases in regression testing \cite{Ali2019}, test case prioritization (TCP) \cite{8453081}, and model-based testing \cite{cartaxo2011use}.
It has also been used to find the most diverse  (opposite to similar) test cases in ART \cite{Mao2019} and software product lines \cite{10.1145/2866614.2866627}.
ANNS has also been successfully applied to enhance the efficiency in other areas of software testing, including TCP \cite{8453081}, test suite reduction \cite{8812048} and prediction of test flakiness \cite{bertolino2020know}.}

\nrs{In this paper, we introduce an approach based on Hierarchical Navigable Small World graphs (HNSWGs), a technology that has outperformed \textit{tree}, \textit{hashing}, and other \textit{graph-based} NNS strategies on a wide variety of datasets~\cite{Li2019, Aumuller2017, Ponomarenko2014}.
HNSWGs represent an excellent potential solution for solving the computational overheads and high dimensionality problem of FSCS-ART.
The proposed method, referred to as FSCS-ART by Hierarchical Navigable Small World Graphs (abbreviated as SWFC-ART), stores previously executed, non-failure-causing test cases in an HNSWG data structure that is efficient for NNS queries, especially for high-dimensional datasets.
HNSWGs are built on navigable small world graphs with a controllable hierarchy for approximate k-nearest neighbor searches~\cite{Malkov2018}, making them suitable for alleviating the exhaustive distance computations burden of FSCS-ART. }
We evaluated SWFC-ART in a series of simulations and empirical studies, examining its efficiency and failure-detection effectiveness.
Our proposed method reduces the computational overhead of FSCS-ART from quadratic to \textit{log-linear} order while maintaining the failure-detection effectiveness of FSCS-ART, and remaining consistent in high dimensional input domains.

The rest of the paper is organized as follows:
\nrs{Section \ref{background} introduces some background information for NNS, FSCS-ART, the current state-of-the-art for FSCS-ART (KDFC-ART), and HNSWGs.}
The proposed method is explained in Section \ref{method}.
Section \ref{experimental-studies} describes the experimental setup used to evaluate the proposed method.
Section \ref{experimental-results} provides the experimental results and discussion.
Section \ref{threats-to-validity} describes potential limitations and threats to the validity of our work.
Related work is described in Section \ref{related-work}.
\nrs{The paper concludes with Section \ref{conclusion}, which also discusses some future work.}

\section{Background}
\label{background}

\subsection{Fixed-Size-Candidate-Set Adaptive Random Testing}
\label{fscs-art}

The Fixed-Size-Candidate-Set implementation of ART (FSCS-ART) uses two sets of test cases:
the candidate set ($C$), containing $k$ elements; and
the executed set ($E$) that contains the previously executed, non-failure-causing test cases.
Initially, both $E$ and $C$ are empty.
\nrs{The first test case is randomly\footnote{\nrs{Using a uniform probability distribution.}} generated and executed.}
If a failure is not found, the executed test case is added to $E$, and $k$ new candidate test cases are randomly generated and added to $C$.
The nearest neighbor in set $E$ for each $c_{i} \in C$ ($1\leq i\leq k $) is determined by calculating the distance between $c_i$ and all elements in $E$.
Finally, the best candidate test case ($c_{best}$) is selected as the candidate whose nearest neighbor is farthest away---
\nrs{following the \textit{min-max} strategy of FSCS-ART}.
This process can be described mathematically as follows (where $\delta$ is a similarity measure for test cases):
\setlength{\arraycolsep}{0.0em}
 \begin{align}
    \forall c \in C, \min_{e\in E} \delta(c_{best},e) \geq \min_{e\in E} \delta(c,e)
 \end{align}
The \textit{Euclidean distance} is typically used as $\delta$ for numeric programs~\cite{Huang2019, Chen2004art}:
FSCS-ART chooses the test case from $C$ that is most distant from previously executed, \nrs{non-failure-causing} test cases.

A core part of the FSCS-ART algorithm is finding the nearest neighbors of candidate test cases.
\nrs{Once the nearest neighbor of each candidate is found, $c_{best}$ can be determined in constant time. }
If a candidate test case ($c_i \in C $) is considered a query point, and the executed test case set ($E$) is the dataset, then the whole process becomes an instance of the NNS problem.
The NNS problem has been extensively studied in computer science, including in areas such as geographic information systems, artificial intelligence, pattern recognition, clustering, and outlier detection~\cite{ONeil2017}.

\begin{defn}
The NNS problem can be formally defined as follows:
Given a $d$-dimensional input domain  $D$ (\nrs{also called \textit{vector space} or \textit{input space}}),
and a distance function $\delta: D \times D \rightarrow \mathbb{R}$,
for a finite set $X = \{x_1, x_2, x_3, \dots, x_n\}$,
where $X \subset D$,
an \textit{effective probability search method} is needed to find the $x_i  \in X$ which is closest to $q \in D$ (according to $\delta$).
Each $x_i \in X$ and $q \in D$ are $d$-dimensional vectors.
\setlength{\arraycolsep}{0.0em}
\begin{align}
    \nrs{NNS(q) = arg \min_{x \in X} \delta(q, x)}
\end{align}
For FSCS-ART, $X$ is a set of executed test cases ($E$), $x_i$ is one executed test case (also called \textit{test input}), and $q$ is a candidate test case. An \textit{effective probability search method} is not guaranteed to identify the exact nearest neighbor for a given candidate.
Although the original FSCS-ART uses \textit{exact} NNS, it has been found that an ANNS can also be employed that maintains the FSCS-ART failure-detection effectiveness~\cite{Mao2019}.
\end{defn}

FSCS-ART uses a brute-force NNS~\cite{Arya1993} for each element $c \in C$, with the distances between $c$ and \textit{each} element of $E$ calculated to find its nearest neighbor.
The complexity of the brute-force NNS is $O\,(d \cdot n)$~\cite{Indyk1998}, and because there are $k$ elements in $C$, one iteration of FSCS-ART has a complexity of $O\,(k \cdot d \cdot n)$.
As the algorithm iterates $n$ times, the total time complexity becomes $O\,( n \cdot k \cdot d \cdot n) = O\,(k \cdot d \cdot n^{2})$.
This quadratic time complexity can take a prohibitive amount of time when testing programs with high dimensionality and low failure rates.

Due to its simplicity, failure-detection effectiveness and popularity for testing numeric, non-numeric and object-oriented programs (especially after the various distance metrics proposed in these domains by ART researchers), most studies of the application of ART in testing real-life software packages have used FSCS-ART.
\nrs{As noted by Huang et al.~\cite{Huang2019}, of 15 studies employing ART to test software packages from different application domains, 14 used FSCS-ART.}
In spite of this, however, the extent of research into improving FSCS-ART is less than that for partition-based ART strategies, which are less often used in real-life testing scenarios~\cite{Chen2019}.

\nrs{
\subsection{State-of-the-art: KDFC-ART}
\label{state_of_the_art_subsection}
}
\nrs{
A state-of-the-art FSCS-ART overhead reduction strategy called KDFC-ART~\cite{Mao2019} stores previously executed, non-failure-causing test cases in a tree-based data structure to perform efficient NNS.
LimBal-KDFC
---
the most efficient of the three KDFC-ART variants
---
incorporates limited backtracking and a semi-balancing strategy to perform ANNS, and appears to effectively address high-dimensionality computational challenges.
Its worst-case time complexity is $O\,(k \cdot d^2\cdot n \log n$)
---
where $k$ is the candidate set size;
$n$ is the number of generated test cases; and
$d$ is the input domain's dimensionality).}

\nrs{Previous studies that have shown that tree-based approaches can perform NNS in low dimensional ($d \leq 5)$ input spaces with $O\,(\log n)$ complexity.
However, in worst-case situations, this complexity can become $O\,(d \cdot  n^{1-1/d})$~\cite{Lee1977}.
LimBal-KDFC, a tree-based search method, is therefore expected also to suffer from the impact of this phenomenon.
}

\subsection{Hierarchical Navigable Small World Graphs}
\label{hnsw}

Graph-based approaches map vectors of a dataset into a graph data structure, and perform greedy traversals to find the nearest neighbor of a query point~\cite{Malkov2018}.
These approaches have been shown to out-perform both tree-based and hashing-based techniques~\cite{Chavez2010, Arya1993, Wang2015, Aoyama2011, Paredes2008,  Hajebi2011, Wang2012, Jiang2016}.
However, these techniques face power-law scaling of the number of steps with the size of the dataset, and may potentially get stuck in local minima~\cite{CarettaCartozo2009, Dong2011}.

To solve this problem, researchers have studied the construction of small world graphs (SWGs) instead of regular connected graphs.
The {\em small world} phenomenon is related to the \textit{Milgram Experiment}~\cite{Milgram1967}, which showed that most social entities are linked through a small number of connections (average of $6$).
\nrs{Watts~\&~Strogatz~\cite{Watts1998} showed that, due to their high clustering and small path lengths, some real-life networks, called small world networks, can lie between \textit{regular} and \textit{connected} networks.}
\nrs{These networks use a few long-range links as well as regular short-range links.}
Short-range links provide local connectivity by joining nodes with their neighbors.
Long-range links are responsible for global connectivity,  joining more distant nodes~\cite{Mehlhorn2013}.
Kleinberg~\cite{Kleinberg2000, Kleinberg} showed that if long-range links are introduced with a probability $r^{-\alpha}$
---
where $r$ is the distance between two distant nodes,
\nrs{and $\alpha$ is a fixed clustering coefficient}
---
\nrs{then the number of steps needed to reach the target node by a greedy search scales down to poly-logarithmic order.}
\nrs{The value of $\alpha$ can be set to the dimensionality of the vector space. }
\nrs{Based on this idea, many NNS and ANNS algorithms have been developed \cite{Lifshits2009, Karbasi2015, Beaumont2007, Beaumont} that have reduced the greedy routing complexity from power-law to poly-logarithmic scaling.}
Small world properties can be incorporated into a graph during its construction~\cite{Malkov2015}
---
\nrs{this has been used by NNS and ANNS, showing small world properties~\cite{Malkov2012, Malkov2014}.}

Hierarchical Navigable Small World graphs (HNSWGs)~\cite{Malkov2018}  aim to further reduce the complexity of SWGs.
HNSWGs are constructed by separating links into different layers based on their length:
This means that only a fixed number of the connections for each element are evaluated (independently of the graph size), which allows for logarithmic scaling.
Each element is assigned a layer level $l$, which denotes the highest layer it can belong to.
The NNS is initiated from the top layer (which has the longest links), and continues until a local minimum is reached \nrs{at that layer}.
The search then goes to the next lower layer, proceeding from the local minimum found in the upper layer.
This process continues until the bottom layer.

Because there is a fixed number of connections at each layer, if the layer level $l$ is set with exponentially decaying probability, then the overall NNS complexity scales down to logarithmic order.
The HNSWG structure is similar to probabilistic skip-lists~\cite{Vitter1990}, with proximity graphs replacing linked-lists.

\begin{figure}[!t]
    \centering
    \subfloat[]{\includegraphics[width=0.49\linewidth]{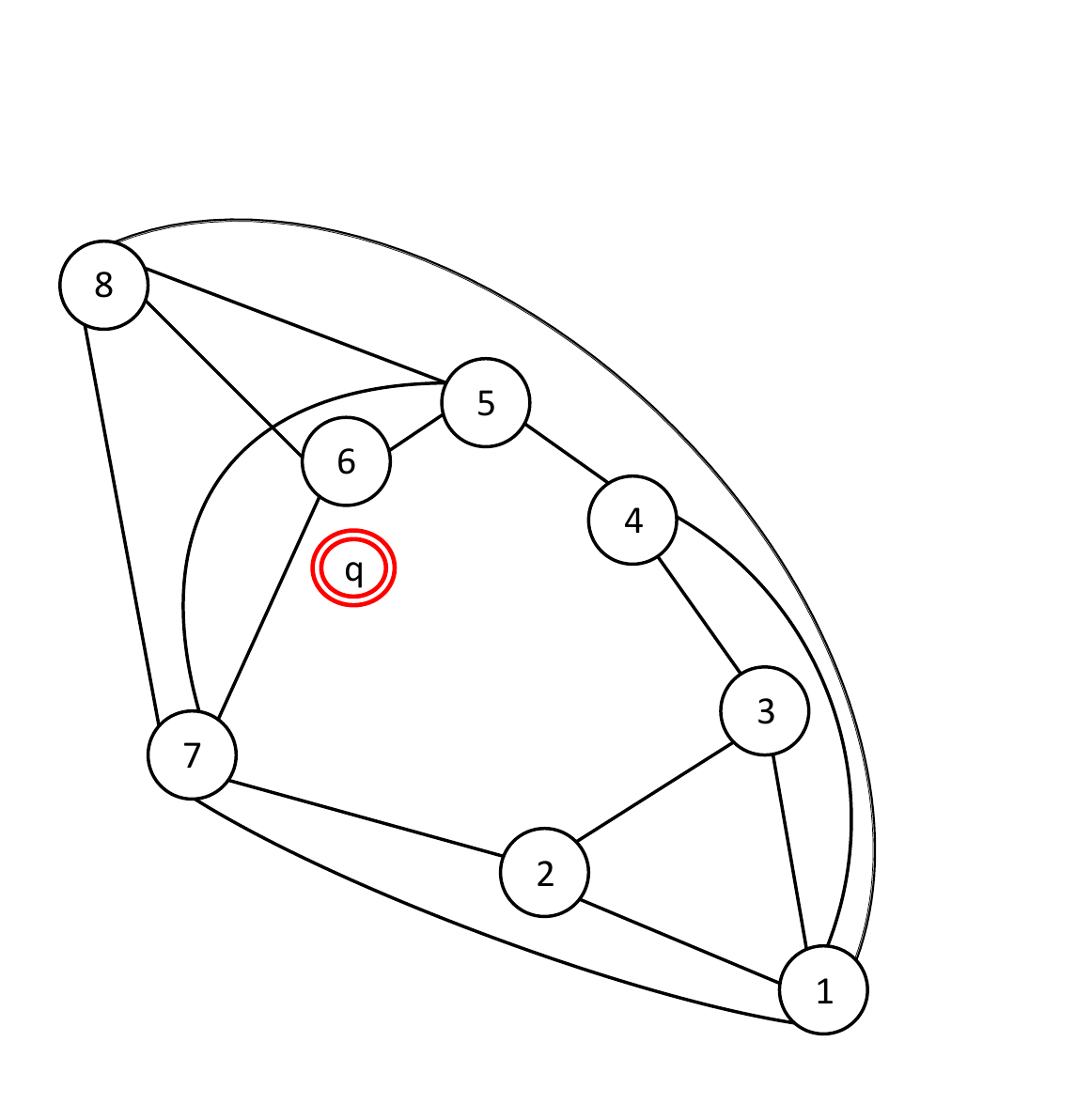}
    \label{simple_nsw}}
    \hfil
    \subfloat[]{\includegraphics[width=0.49\linewidth]{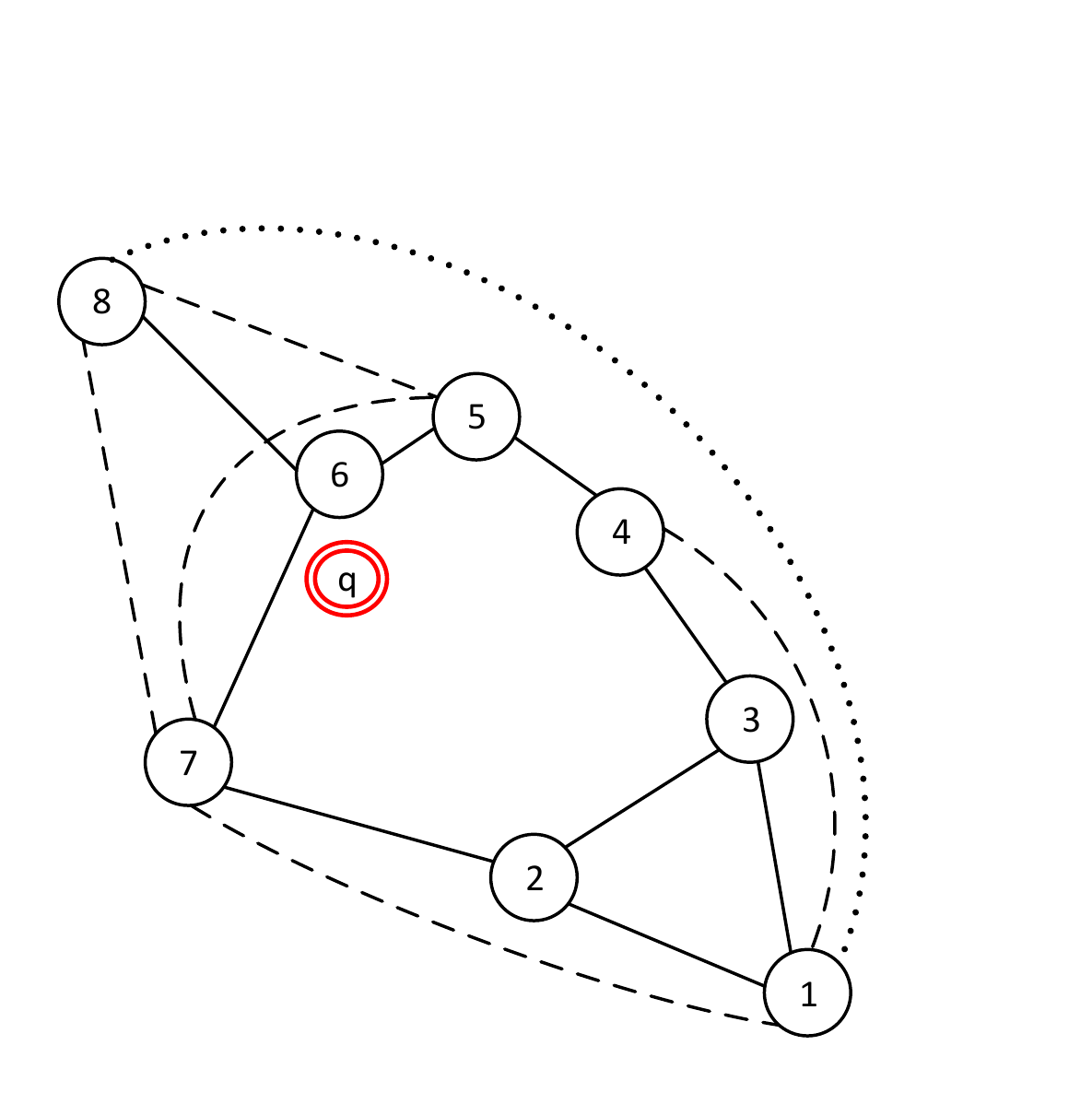}
    \label{converted_hnsw}}
    \caption{HNSWG Structure: (a) Basic NSWG; (b) Break-down of links}
    \label{nsw_to_hnsw}
\end{figure}

\begin{figure}[!t]
    \centering
    \includegraphics[width=2in]{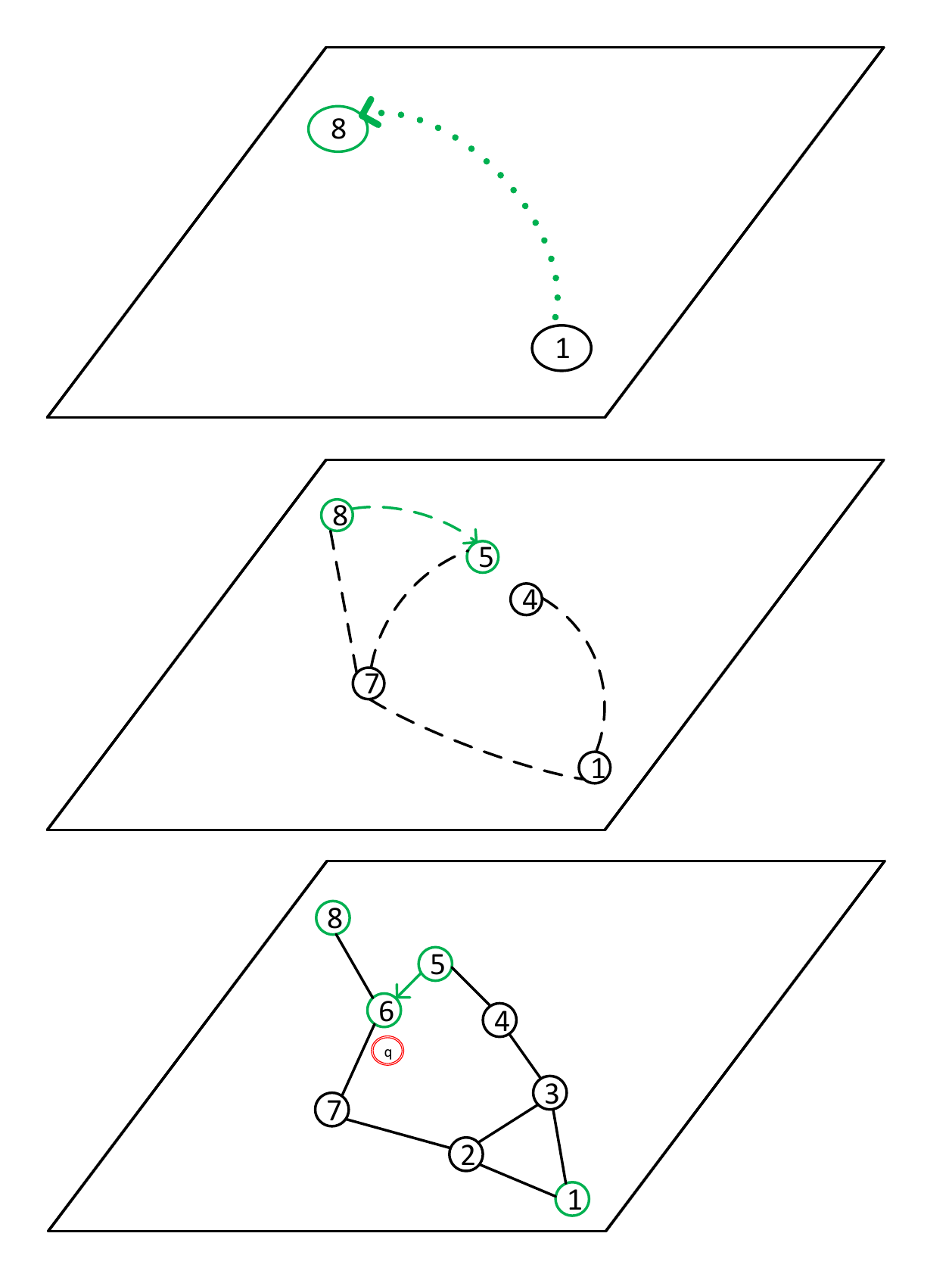}
    \caption{Layered Hierarchy Representation of Figure \ref{converted_hnsw}}
    \label{hnsw_layered}
\end{figure}

\subsubsection{Example}
Fig.~\ref{simple_nsw} presents a sample navigable small world graph (NSWG) where each node is connected to its neighboring nodes, and there are also some long-range links to more distant nodes.
For example, Node 1 has bidirectional short-range links with its neighboring nodes (Nodes 2 \& 3), and long-range links with Nodes 4, 7, and 8.
This NSWG is converted to an HNSWG by grouping links into three categories:
Long-, medium-, and short-range
---
which are represented in Fig. \ref{converted_hnsw} by dotted, dashed, and solid lines, respectively.
The long- and medium-range links
---
where a node is connected to nodes other than its neighbors
---
are responsible for the small world properties in the graph.
The short-range links connect nodes to some of their neighbors, making an approximate Delaunay graph~\cite{Malkov2012, Malkov2014}.
As shown in Fig. \ref{hnsw_layered}, links in the HNSWG are separated into three virtual layers (hierarchies) according to their length,  with long-, medium-, and short- links categorized into the top, middle, and bottom layers, respectively.

\subsubsection{NNS in HNSWG}
An NNS for query point $q$ (highlighted with a double circle in Figs. \ref{nsw_to_hnsw} \& \ref{hnsw_layered}) starts from the top layer by selecting the node with the most links (the ``maximum degree node'') as an entry point (Node 1).
The entry point is updated each time an element is inserted into the graph.
The nearest neighbor of $q$ in the top layer is determined (Node 8), and the search proceeds to the middle layer, restarting from the local minimum found in the top layer.
The nearest neighbor is revised, with Node 5 now identified as the best possible solution.
Finally, the search proceeds to the final (bottom) layer and again attempts to refine the nearest neighbors, resulting in Node 6 now being identified as the final nearest neighbor for $q$.
According to this process, the nearest neighbor of $q$ is found in three steps, compared with the eight distance calculations that would have had to be performed in a brute-force (exhaustive) search.

\section{Method}
\label{method}

This section introduces our proposed method, SWFC-ART (a \textbf{S}mall-\textbf{W}orld-graph-based approach for \textbf{F}ixed-size-\textbf{C}andidate-set \textbf{A}daptive \textbf{R}andom \textbf{T}esting), which uses an HNSWG to store $E$, and to efficiently find nearest neighbors for each $c_i \in C$.

\begin{figure}[!t]
    \centering
    \includegraphics[width=\linewidth]{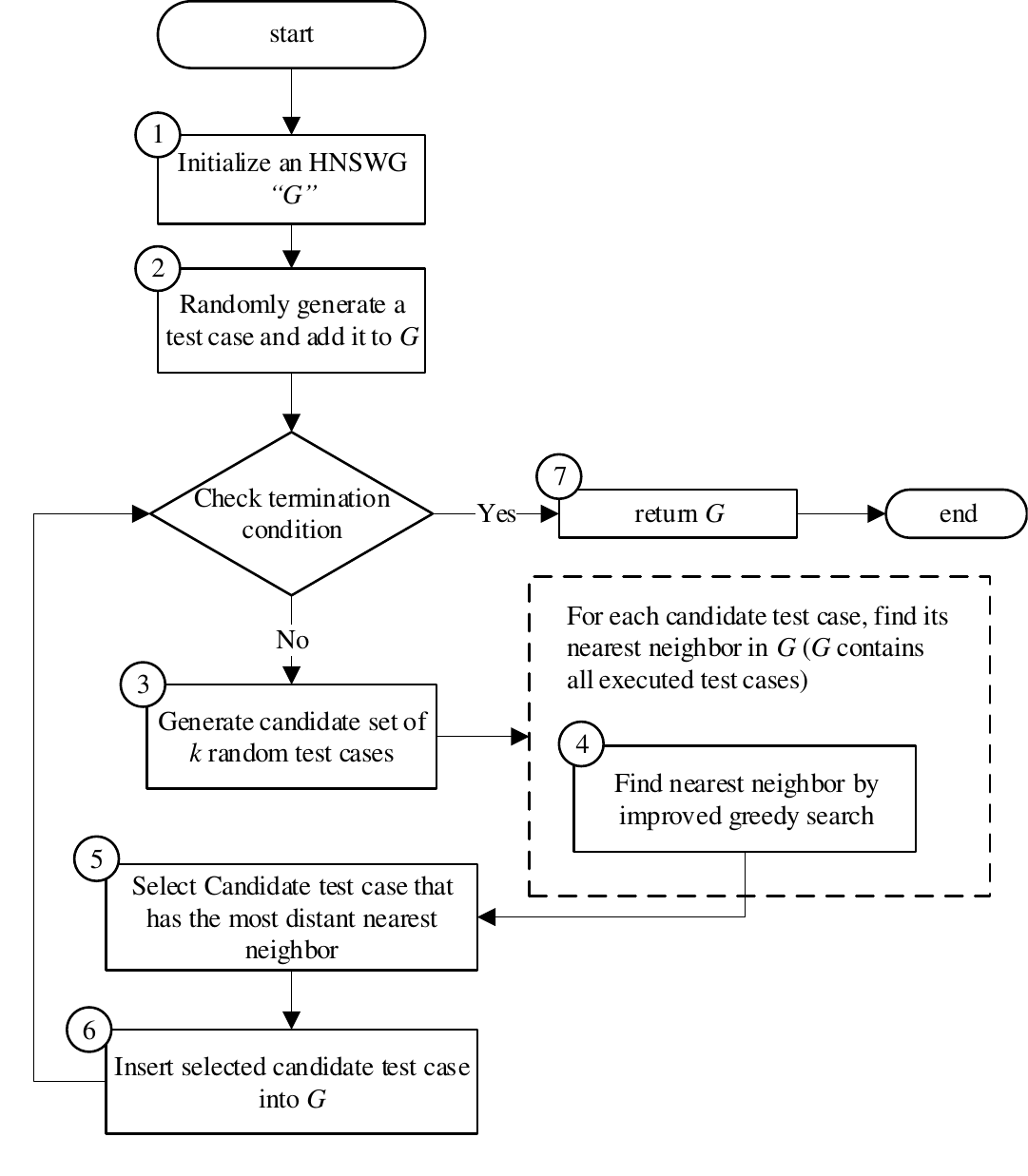}
    \caption{\nrs{SWFC-ART Framework}}
    \label{flowchart}
\end{figure}

\subsection{Framework}
\label{architecture}


The SWFC-ART method can be divided into seven major steps, as shown in Fig.~\ref{flowchart}.
In the first step, an HNSWG $G$, which will map all executed test cases to its nodes, is initialized.
The initialization phase requires a number of parameters (discussed in Section \ref{parameter_optimization}), including:
the graph size, the number of nearest neighbors to be \nrs{ searched for} in each layer, and the number of nearest neighbors to be connected for each inserted node.
Once the graph is initialized, the first test case is randomly generated, inserted into the $G$ (Step 2), and used to test the software under test \textit{(SUT)}.
If the method has not terminated, Step 3 involves randomly generating $k$ candidate test cases.
In Step 4, the multilayered $G$ is traversed to find the nearest neighbor for each candidate.
Step 5 determines the best candidate (the one whose nearest neighbor is most distant).
\nrs{In Step 6, the best candidate is inserted into $G$, and used as the next test case for the SUT.}
This process repeats until a termination criterion is reached.
Possible termination criteria include:
finding a failure;
executing a specific number of test cases;
running the algorithm for a specific time limit;
or any other specified criterion.
If a termination criterion is satisfied, the algorithm terminates and returns $G$ with all executed test cases as its nodes (Step 7).

\subsection{SWFC-ART}
\label{algorithmic-description}

\nrs{SWFC-ART is a modified form of the FSCS-ART algorithm~\cite{Chen2004art}, storing previously executed, non-failure-causing test cases in an HNSWG~\cite{Malkov2018}, instead of the arrays and trees used by FSCS-ART and LimBal-KDFC, respectively.}

\subsubsection{Algorithm}

\begin{algorithm}[!t]
\renewcommand{\thealgorithm}{\arabic{section}.\arabic{subsection}.\arabic{subsubsection}}
 \label{swfc_art_algorithm}
 \caption{SWFC-ART}
 \begin{algorithmic}[1]
 \renewcommand{\algorithmicrequire}{\textbf{Input:}}
 \renewcommand{\algorithmicensure}{\textbf{Output:}}

 \REQUIRE
\begin{enumerate}[nosep]
 \item Size of candidate test case set: $k$
 \item Program input domain: $D$
 \item Distance function: $\delta$
 \item Nearest neighbors to be searched for at each layer: $ef$
 \item Number of connections for each inserted element: $M$
 \item Size of dynamic list for enhancing the accuracy of nearest neighbor search: $efConst$
 \item Layers controller: $m_{l}$
 \item Base size of the graph: $b$
\end{enumerate}
 \ENSURE  HNSWG object $hnswg$ consisting of executed test cases as its nodes
  \STATE $d \gets$ dimensionality of $D$ \label{1_1}
  \STATE \nrs{Globally-stored $ep$}
  \STATE \nrs{Initialize $G$ ($d$, $\delta$, $b$, $M$, $efConst$, $ef$)} \label{1_7}
  \STATE $\nrs{t_{init}} \gets$ randomly generate a test case \label{1_8}
  \STATE \nrs{Call \texttt{\textbf{procedure} G.Insert} ($t_{init}$, $M$, $efConst$, $ep$, $m_{l}$)}  \label{1_9}
  \WHILE{termination condition is not satisfied}\label{1_10}
  \STATE $C \gets $ Randomly generate $k$ test cases \label{1_11}
    \FOR{each $c$ in $C$}\label{1_12}
     \STATE \nrs{Call \texttt{\textbf{procedure} G.NNS} ($c$, $ef$, $ep$)}
     \label{1_13}
    \ENDFOR\label{1_14}
    \STATE Select $c_{best}$ from $C$\label{1_15}
    \STATE \nrs{Call \texttt{\textbf{procedure} G.Insert} ($c_{best}$, $M$, \textit{efConst}, $ep$, $m_{l}$)}\label{1_16}
    \IF  {\nrs{$G.currentSize()$} \textbf{equals}  $b$}\label{if_block_start}
      \STATE \nrs{$b \gets 2 \cdot b$}\label{1_18}
      \STATE \nrs{re-calculate $efConst$}\label{1_19}
      \STATE \nrs{ $G_{temp} \gets G.items()$}\label{1_20}
      \STATE \nrs{Re-initialize $G$ ($d$, $\delta$, $b$, $M$, $efConst$, $ef$)} \label{1_21}
      \STATE \nrs{Call \texttt{\textbf{procedure} G.Insert} ($G_{temp}$, $M$, \textit{efConst}, $ep$, $m_{l}$)}\label{1_23}
    \ENDIF\label{if_block_end}
  \ENDWHILE\label{1_25}
 \RETURN \nrs{$G$}\label{1_26}
 \end{algorithmic}
\end{algorithm}

SWFC-ART takes \nrs{eight} inputs:
1) $k$ (the size of the candidate test set);
2) $D$ (the SUT's input domain);
3) $\delta$ (the distance function);
4) $ef$ (the size of the dynamic list for the number of nearest neighbors to be searched for in each layer);
5) $M$ (the number of connections for an inserted test case in each layer of the HNSWG);
6) \textit{efConst} (the size of the dynamic list for enhancing the accuracy of returned nearest neighbors
---
although this parameter is the same as $ef$, a different value is used during HNSW construction, and therefore we call it $efConst$ (efConstruction);
7) $m_{l}$ (a non-zero integer to control the number of layers with exponentially decaying probability);
and
\nrs{8) $b$ (the initial base size of the graph, representing the number of nodes that it can accommodate).
The parameters are discussed further in Section \ref{parameter_optimization}.
The algorithm returns an HNSWG ($G$) whose nodes are the executed test cases, the number of which corresponds to the {\em F-measure} (Section \ref{evaluation-metrics}).}

\nrs{The algorithm begins by calculating the dimensionality $d$ of the input domain $D$ (line 1).
The entry point $ep$ (line 2) is stored globally, and updated each time an element is inserted into $G$
---
this differentiates HNSWGs from NSWGs, where the entry point is randomly chosen on each search iteration.
On line 3, $G$ has been initialized by specifying the parameter values.
In the next phase, a randomly-generated test case from the SUT's input domain is executed, and inserted into $G$ (line \ref{1_8}).
If no termination criterion has been satisfied, then $k$ test cases are randomly generated, and put into $C$ (line \ref{1_11}).
The nearest neighbor of each candidate ($c \in C$) is determined by calling the \texttt{NNS procedure} (Section \ref{nns-procedure}), and the candidate with the maximum distance from its nearest neighbor ($c_{best}$) is selected as the next test case (line \ref{1_15}).
This selected test case is inserted into $G$ by calling the \texttt{Insert procedure} (Section \ref{insert-procedure}).
The \texttt{if block} (Lines \ref{if_block_start}-\ref{if_block_end}) maintains the dynamic size of $G$, which is doubled if the number of stored test cases  reaches the limit ($b$) (causing $efConst$ to be re-calculated).}

The \texttt{NNS} and \texttt{Insert} procedures called by SWFC-ART require the \texttt{Searcher} procedure to identify the nearest neighbors on each layer of the HNSWG.
Because these procedures have been comprehensively explained by Malkov \& Yashunin~\cite{Malkov2018}, the following is only a general overview.

\subsubsection{Insert procedure}
\label{insert-procedure}

\begin{algorithm}[!t]
\floatname{algorithm}{Procedure}
\renewcommand{\thealgorithm}{\arabic{section}.\arabic{subsection}.\arabic{subsubsection}}
 \caption{ \texttt{ hnswg.Insert} ($t$, $M$, $efConst$, $ep$, $m_{l}$) }
 \begin{algorithmic}[1]
 \label{insert_procedure}
 \renewcommand{\algorithmicrequire}{\textbf{Input:}}
 \renewcommand{\algorithmicensure}{\textbf{Output:}}
  \STATE $l \gets \lfloor -\ln{(unif(0...1))} \cdot m_{l}  \rfloor $
  \STATE $L \gets$ level of $ep$ \quad // entry point to top layer
  \STATE $W \gets \phi$
  \\ \textit{Phase I}
  \FOR {$l_c \gets {L . . . l}$}
    \STATE $W \gets$ \texttt{\textbf{procedure} Searcher}($t$, $ep$, $ef=1$, $l_c$)
  \ENDFOR
  \\ \textit{Phase II}
  \FOR {$l_c \gets {l . . . 2,1,0}$}
    \STATE $W \gets$ Call \texttt{\textbf{procedure} Searcher}($t$, $ep$, $efConst$, $l_c$)
   \STATE bidirectionally connect $M$ ($M_0$ if $l_c = 0$) elements from $W$ to $t$
   \STATE shrink connections if needed
  \ENDFOR
  \IF{$l > maxLayer$}
    \STATE $ep \gets t$  \quad // update entry point
    \STATE $maxLayer \gets l$  \quad // update maximum layer level
  \ENDIF
 \RETURN updated $G$ with inserted $t$
 \end{algorithmic}
\end{algorithm}

The \texttt{Insert procedure}
takes a test case $t$, entry point $ep$, and three integer value parameters ($efConst$, $M$ and $m_{l}$), and returns an updated $G$  reflecting the insertion of $t$.

\nrs{For each $t$, a maximum layer $l$ is randomly selected with an exponentially decaying probability distribution (normalized by $m_L$) (line 1).
$L$ represents the layer of entry point node $ep$, which is the top layer of $G$.}
$W$, which is initially empty, stores the nearest neighbors of $t$ in each layer (line 3).
The NNS for $t$ consists of two phases:
In Phase I (lines 4-6), the search moves from the top layer $L$, to $t$'s layer $l$, identifying exactly one nearest neighbor in each layer.
In Phase II (lines 7-11), each layer below $l$ is searched for $efConst$ nearest neighbors, with a goal of increasing the accuracy of the greedy search at lower layers (line 8).
The best $M$ nodes from $W$ are linked to $t$.
If the current layer is the ground layer ($l_0$), then $t$ is linked to \nrs{ $M_0$ neighboring nodes (Section \ref{M_parameter})}.
The maximum number of connections for an element 
\nrs{is} kept to within a fixed limit, thus maintaining logarithmic complexity (line 10).
The entry point and maximum layer level are updated (lines 12-15), and the updated $G$ is returned (line 16).

\subsubsection{NNS procedure}
\label{nns-procedure}

\begin{algorithm}[!t]
\floatname{algorithm}{Procedure}
\renewcommand{\thealgorithm}{\arabic{section}.\arabic{subsection}.\arabic{subsubsection}}
 \caption{ \texttt{ hnswg.NNS} ($t$, $ef$, $ep$) }
 \label{NNS}
 \begin{algorithmic}[1]
 \renewcommand{\algorithmicrequire}{\textbf{Input:}}
 \renewcommand{\algorithmicensure}{\textbf{Output:}}
 \STATE $L \gets$ level of $ep$ \quad // entry point to top layer
  \STATE $W \gets \phi$  \quad // List of currently found nearest neighbors
  \\ \textit{Phase I}
  \FOR {$l_c \gets {L . . . 1}$}
    \STATE $W \gets$ Call \texttt{\textbf{procedure} Searcher}($t$, $ep$, $ef=1$, $l_c$)
  \ENDFOR
  \\ \textit{Phase II}
  \STATE $W \gets$ Call \texttt{\textbf{procedure} Searcher}($t$, $ep$, $ef$, $l_0$)
  \STATE Sort $W$ in ascending order
 \RETURN first element of $W$ as nearest neighbor of $t$
 \end{algorithmic}
\end{algorithm}

The \texttt{NNS procedure} is similar to the \texttt{Insert procedure}, with the exception that Phase I spans from the top layer to the second last layer.
\nrs{Similar to \texttt{Insert}, exactly {\em one} nearest neighbor is identified for $t$ in Phase I (line 3-5).}
In Phase II, only the bottom layer ($l_0$) is searched for $ef$ nearest neighbors (line 6).
\nrs{Finally, $W$ is sorted according to  $\delta(W[i], \, t) < \delta\,(W[i+1], \,t)$, and the first element is returned as the nearest neighbor of $t$ (line 8).}

\subsubsection{Searcher procedure}
\label{searcher-procedure}

\begin{algorithm}[!t]
\floatname{algorithm}{Procedure}
\renewcommand{\thealgorithm}{\arabic{section}.\arabic{subsection}.\arabic{subsubsection}}
 \caption{ \texttt{Searcher} ($t$, $ep$, $ef$, $l$) }
 \label{searcher}
 \begin{algorithmic}[1]
 \renewcommand{\algorithmicrequire}{\textbf{Input:}}
 \renewcommand{\algorithmicensure}{\textbf{Output:}}
  \STATE $ d_{init} \gets \delta (t, ep) $  \quad // initial distance from entry point to test case
  \STATE $W \gets ep$
  \FOR {each $e \in neighborhood\,(ep)$ at layer $l$}
    \IF{$\delta(t,e)<d_{init}$}
      \STATE $ W  \gets W \bigcup e $;
      \STATE greedily search neighborhood of $e$ recursively
    \ENDIF
  \ENDFOR
  \STATE sort $W$ in ascending order
 \RETURN first $ef$ elements of $W$ as closest neighbors of $t$  at layer $l$
 \end{algorithmic}
\end{algorithm}

Both \texttt{NNS} and \texttt{Insert} procedures call the \texttt{Searcher procedure} (a type of greedy search)
to search for $ef$ nearest neighbors in a given layer.
\texttt{Searcher} searches for the nearest neighbors of test case $t$ in layer $l$, given entry point $ep$, and returns $ef$ nearest neighbors.
Initially, the entry points are taken as temporary nearest neighbors (line 1), and stored in $W$.
Next, the neighborhood of each entry point is recursively searched, in a greedy manner, for other nearest neighbors (lines 3-8), with any identified closer neighbors added to $W$.
\nrs{The first $ef$ elements of $W$ (that are at a minimum distance from $t$) are then returned.}

\subsection{Parameter Optimization}
\label{parameter_optimization}

SWFC-ART is controlled by a number of parameters: $k$, $ef$, $efConst$, $M$, $b$ and $\delta$.
\nrs{If all the parameter values are set to the minimum possible, then the HNSWG is not used, and the algorithm's effectiveness can become similar to that of RT.}
Using the optimal parameter values, as explained in the following, is therefore critical to the success of SWFC-ART.

\subsubsection{Number of candidate test cases ($k$)}
In most ART studies, the size of the candidate test set (the number of test cases randomly generated in each iteration), $k$, is usually set to $10$~\cite{Chen2004art}.

\subsubsection{Graph size ($b$)}
The size of the HNSWG needs to be given in advance, with larger graphs incurring more construction time (lowering efficiency).
However, while this parameter has no apparent impact on the failure-detection effectiveness, it can affect efficiency.
\nrs{The actual number of nodes (executed test cases) in the final graph corresponds to the F-measure (the number of test cases executed before finding the first failure), which can depend on the failure rate of the SUT (which is unknown in advance).}
In ART, test cases are usually generated incrementally until a termination criterion is reached.
One approach to deal with this would be to assign the maximum available size (as supported by the hardware and software platform) to the HNSWG, but this can incur a very heavy construction cost, especially for software with high failure rates.
Because our analysis of varying $b$ between $10^{2}$ and $2 \times 10^{7}$ showed that the graph construction time for $10^{2}$ to $10^{4}$ remained relatively stable, but then increased significantly for larger sizes, we initially set $b$ to $10^{4}$, and double this any time additional nodes are needed.
\nrs{In practice, testers may set $b$ according to their own specific needs.}

\subsubsection{Distance function ($\delta$)}
Because the Euclidean distance has been used in many FSCS-ART studies~\cite{Chen2004art, Huang2019}, especially for numeric programs, we adopted it in our simulations and experiments.

\subsubsection{Size of dynamic list ($ef$)}
The size of the dynamic list ($ef$) controls the number of a candidate's closest neighbors that are \nrs{searched for} in layers higher than its own layer.
Because SWFC-ART employs ANNS, the identified nearest neighbor may not be the {\em actual} nearest neighbor.
There is a tension between the efficiency and effectiveness for the $ef$ value:
Increasing $ef$ incurs an additional time cost, but also increases the chances of finding actual nearest neighbors.
It should be at least equal to the number of desired nearest neighbors for a candidate, and, since ART seeks only one nearest neighbor for each candidate, the minimum value of $ef$ can be $1$.
With $ef=1$, our analysis showed NNS accuracy of $90\%$ in all dimensions and failure rates under study.
Increasing $ef$ to $2$ incurred a little additional time cost, but also raised the accuracy to 98\%.
Because the scope of our study was to increase efficiency while keeping effectiveness at a comparable level to the state-of-the-art, and $ef=2$ shows effectiveness similar to that of FSCS-ART and LimBal-KDFC (Section \ref{experimental-results}), we did not increase $ef$ beyond $2$.
Testers may choose to increase $ef$ if they are interested in further enhancing the effectiveness.

\subsubsection{Number of links ($M$)}
\label{M_parameter}
This parameter controls the number of connections made to an inserted element:
More connections increase the failure-detection effectiveness, but compromise the efficiency.
Following Malkov~et~al.~\cite{Malkov2012}, who recommended that
\nrs{a newly-inserted element}
should be connected to at least its $M = 3 \cdot d$ closest neighbors (where $d$ is the dimensionality of dataset), we set $M = 3 \cdot d$.
\nrs{On the ground layer ($l_0$), a separate parameter $M_0$ has been used.
Setting $M_0\, = \, M$ reduces the NNS accuracy.
$M_0 = 2 \cdot M$ is the recommended choice, because higher values can lead to performance degradation and excessive memory usage \cite{Malkov2018}.}

\subsubsection{Construction parameter ($efConst$)}
The construction parameter ($efConst$) specifies the number of nearest neighbor candidates used during graph construction.
As only $M$ closest candidates are connected to the inserted element, $efConst \geq M$.
In the \texttt{Insert} procedure, a candidate has to be searched for only one nearest neighbor in layers above its own layer, but this is increased to $efConst$ for lower layers to improve the NNS accuracy.
The value of $efConst$ is logarithmic to the size of the dataset and is very similar to the $w$ parameter described by Malkov~et~al.~\cite{Malkov2014}:
$A \cdot \log(N)$ (where $N$ is the size of the dataset, and $A$ can be any natural number).

\subsection{Illustration}
\label{illustration}

\begin{figure*}
    \centering
    \begin{subfigure}[b]{0.24\textwidth}
        \includegraphics[width=\textwidth]{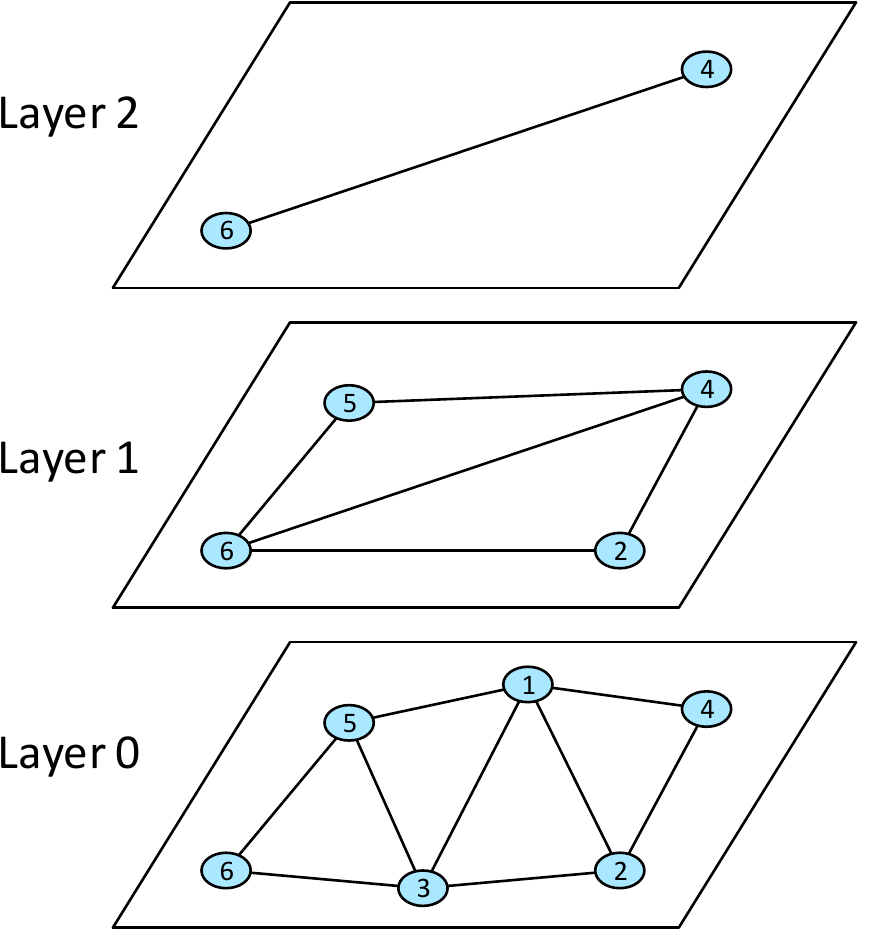}
        \caption{Initial HNSWG}
        \label{fig:Initial-HNSWG}
    \end{subfigure}
    ~ 
    \begin{subfigure}[b]{0.24\textwidth}
        \includegraphics[width=\textwidth]{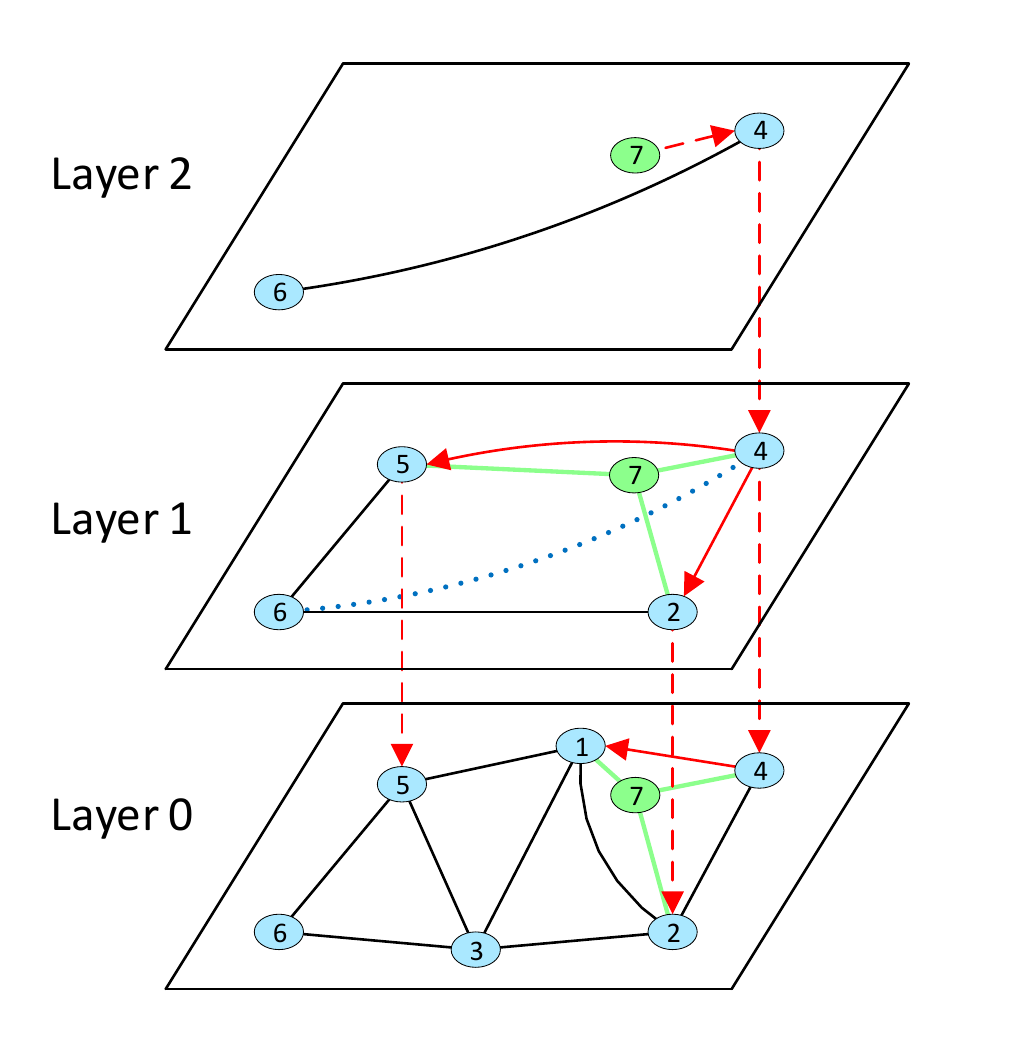}
        \caption{\nrs{Insertion of Node 7}}
        \label{fig:insertion-node7}
    \end{subfigure}
    ~ 
    \begin{subfigure}[b]{0.24\textwidth}
        \includegraphics[width=\textwidth]{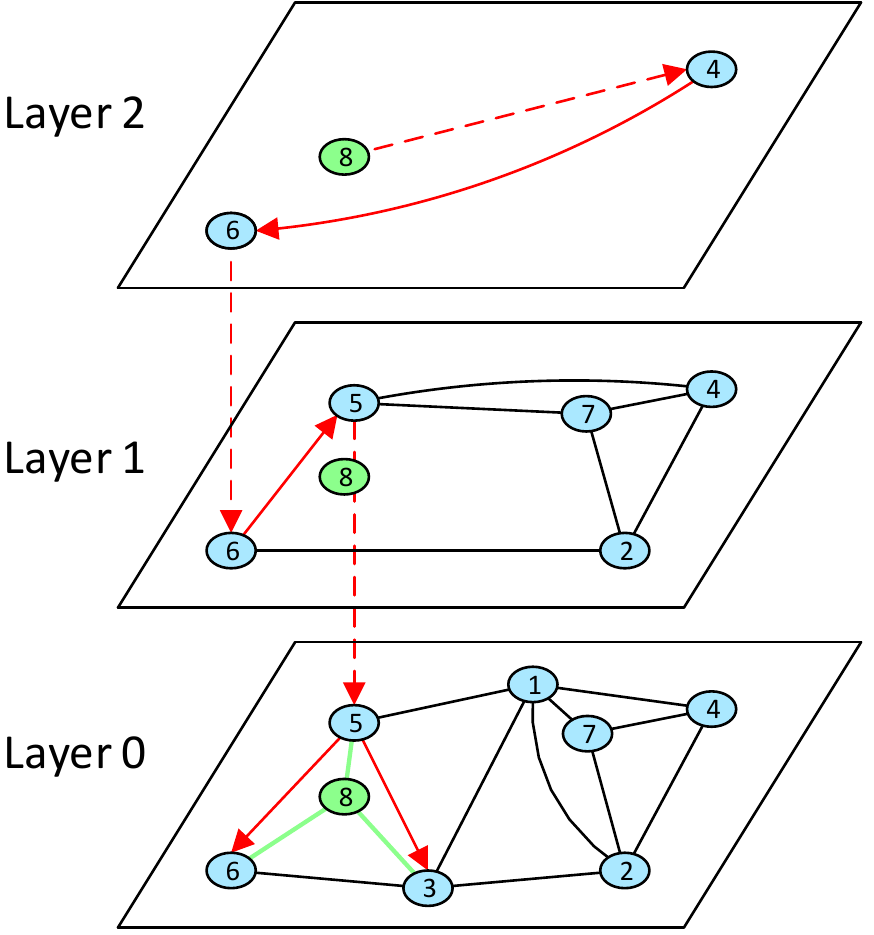}
        \caption{Insertion of Node 8}
        \label{fig:insertion-node8}
    \end{subfigure}
    \begin{subfigure}[b]{0.24\textwidth}
        \includegraphics[width=\textwidth]{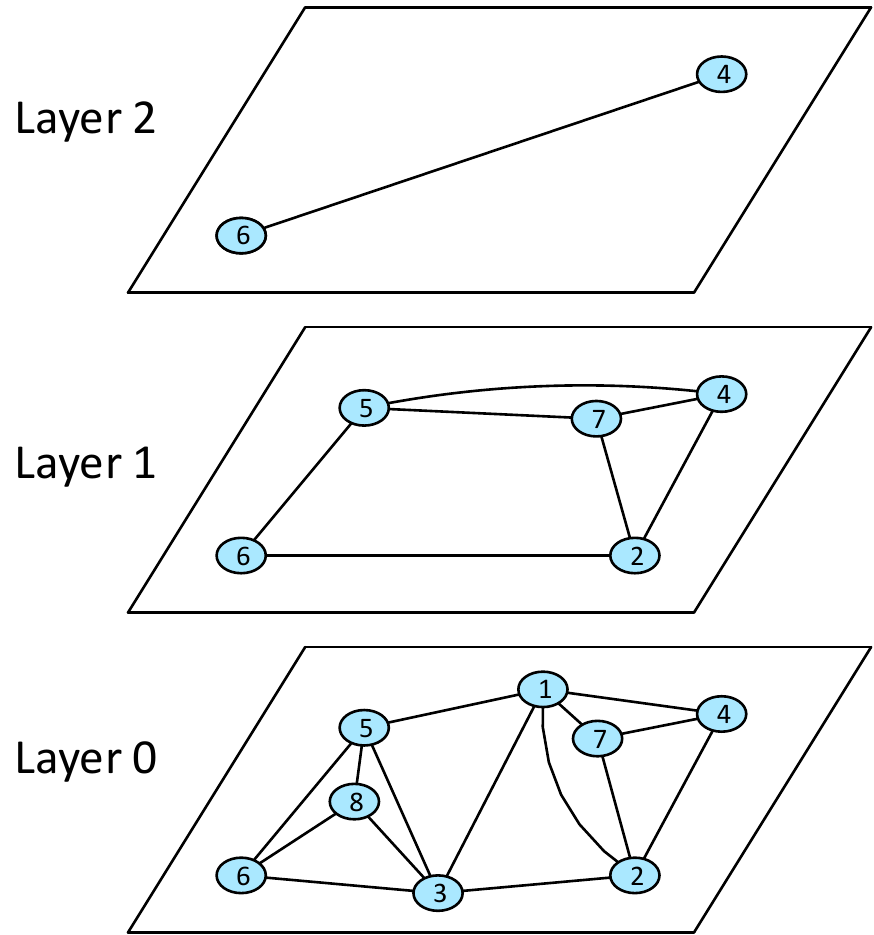}
        \caption{Updated HNSWG}
        \label{fig:final-HNSWG}
    \end{subfigure}
    \caption{Sample illustration of HNSWG node insertion}
    \label{hnswg_illustration}
\end{figure*}

Fig. \ref{hnswg_illustration} demonstrates an example of inserting two new nodes into an HNSWG.
\nrs{Without loss of generality, the parameter values in this example were set as follows:}
$ef=1$;
$M=3$;
$M_0 =2 \cdot M = 6$; and
$efConst = 3$.
\nrs{(The value of $m_l$ is not important for demonstration purposes.)}
The initial HNSWG is shown in Fig.~\ref{fig:Initial-HNSWG}, with long-range links in the top layer (Layer 2), and short-range links in the ground layer (Layer 0).
Node 4 is the entry point to the graph.

Fig \ref{fig:insertion-node7} shows the process of inserting Node 7 into the HNSWG.
\nrs{Suppose that Node 7 is inserted into Layer 1, which means that it will be inserted into this, and all lower layers.}
The search starts from the top layer (Layer 2), with Node 4 as the entry point.
(As a reminder:
For layers above the insertion layer, the search identifies the $ef$ closest elements; and
for the insertion and lower layers, the search identifies the $efConst$ closest elements.)
In Layer 2, all neighbors of Node 4 are examined, and the \textit{one} ($ef=1$) closest to the inserted node is identified
---
in this case, Node 4.
Node 4 is then used as the entry point for Layer 1, where its neighbors (Nodes 2 and 5) are examined, and then their neighbors are examined.
This continues until all the neighborhood has been examined.
In this example, because $efConst=3$, Nodes 2, 4, and 5, are identified as the closest elements, and are connected directly to Node 7.
They will then be used as entry points to Layer 0.
At this stage, because Node 4 now has four connections, which exceeds the limit ($M=3$), its closest $M$ neighbors are identified and any remaining connections are discarded:
The connection between Nodes 4 and 6 is therefore removed (\nrs{represented by the dotted blue line in Layer 1 of Fig.~\ref{fig:insertion-node7}}).
Finally, in Layer 0, the neighbors of the entry points (Nodes 2, 4, and 5) are examined, and the $efConst=3$ closest elements are identified and connected to Node 7:
Nodes 1, 2, and 4.

Fig.~\ref{fig:insertion-node8} shows the insertion of Node 8.
\nrs{Suppose that Layer 0 is the insertion layer for Node 8}
The search again begins from the top layer (Layer 2), from the entry point Node 4.
In Layer 2, all neighbors of Node 4 are examined, and the \textit{one} ($ef=1$) closest to the inserted node is identified
---
Node 6.
Node 6 is then used as the entry point for Layer 1, where examination of the various connected neighbors results in Node 5 being identified as the \textit{one} closest node to Node 8, and thus being used as the entry point to Layer 0. In Layer 0, the neighbors of Node 5 are examined, and the \textit{three} ($efConst = 3$) closest elements are identified and connected to Node 8:
Nodes 3, 5, and 6.

Fig.~\ref{fig:final-HNSWG} shows the updated HNSWG after insertion of both Nodes 7 and 8. The \texttt{NNS Procedure} is similar to the \texttt{Insert Procedure}, with the slight differences that $ef=1$ nearest neighbors are searched for in \textit{all} layers except the ground layer ($l_0)$, where $efConst=3$ nearest neighbors are identified (as explained in Section \ref{nns-procedure}); and that the node is not inserted into the HNSWG (bidirectional connecting does not take place).

\subsection{Complexity Analysis}
\label{complexity-analysis}

The time complexity of SWFC-ART can be considered in two parts:
the NNS; and
the HNSWG construction complexity.
For each candidate test case, the algorithm searches for $ef$ or $efConst$ nearest neighbors on each layer, of which there are a maximum of $m_{l}$ in the graph).
\nrs{Because $ef$, $efConst$, and $m_{l}$ are constant values (and thus do not depend on the size of the dataset), the overall NNS complexity scales down to $O\,(\log n)$ for one candidate test case, and $O\,(k\cdot \log n)$ for $k$ candidates.}
The HNSWG is constructed through sequential insertion of test cases, for each of which $M$ nearest neighbors are connected in each layer.
The complexity of inserting one test case into the HNSWG therefore becomes equal to the search complexity $O\,(\log n)$.
The total graph construction complexity for the sequential insertion of $n$ test cases scales to $O\,(n\cdot\log n)$.
\nrs{For $k$ candidates and a $d$-dimensional input domain the overall complexity becomes log-linear (also called \textit{linearithmic} or \textit{quasilinear}): $O\,(k\cdot d\cdot n\cdot\log n)$.}
The storage complexity of SWFC-ART depends on the number of links (both long- and short-range):
For four billion elements (nodes or test cases), four-byte unsigned integers can be used to store the HNSWG connections.
The typical memory requirement for one HNSWG object is about 60-450 bytes, which has been confirmed by simulation analysis~\cite{Malkov2018}.

\section{Experimental Studies}
\label{experimental-studies}


Our study aimed at solving the double-tier efficiency problem of FSCS-ART, an ART version known for its failure-detection effectiveness and application in real-life programs.
In addition to FSCS-ART, LimBal-KDFC (Section \ref{state_of_the_art_subsection}) was also selected as a baseline for comparison.

\subsection{Research Questions}
\label{research-questions}

The double-tier efficiency problem conceptualizes two efficiency issues of the FSCS-ART algorithm.
\nrs{The first issue relates to the growing executed test set size when failure has not yet been revealed:
This is a {\em scalability} issue.
The second issue relates to the computational load associated with dimensionality increases for any size of test set:}
This is a {\em consistency} issue.
\nrs{In addition to examining the effect of these issues, we also wanted to investigate the impact (similarities and differences) of the ANNS strategies of LimBal-KDFC and SWFC-ART on the failure-detection effectiveness of FSCS-ART.}
Therefore, the following research questions were designed to guide our experiments:
\begin{enumerate}[leftmargin=1cm, nosep]
    \item[\textbf{RQ1:}] Does SWFC-ART successfully solve the double-tier efficiency problem? (Efficiency)
    \item[\textbf{RQ2:}] How effective is SWFC-ART at revealing failures? (Effectiveness)
    \item[\textbf{RQ3:}] How evenly does SWFC-ART distribute test cases? (Test case distribution)
\end{enumerate}

\subsection{Evaluation Metrics}
\label{evaluation-metrics}

\subsubsection{Efficiency metrics}
\nrs{Because a goal of this study was to reduce the computational cost associated with FSCS-ART generating test cases, the {\em test case generation time} ($T_G$) was adopted an efficiency metric.
The \textit{test case execution time} ($T_E$) was also recorded.
$T_G$ includes the time taken to generate a fixed number of test cases, with lower times indicating better efficiency;
while $T_E$ is the total time taken by a program to execute the generated test cases.}

\subsubsection{Failure-detection effectiveness metrics}
The {\em F-measure} is defined as the expected number of test case executions required by a method to find the first failure~\cite{Chen2004art}, with
lower F-measure values (fewer test cases to find a failure) corresponding to better effectiveness.
The F-measure was used as the failure-detection effectiveness metric in our study.
If the failure rate, $\theta$, of an SUT is defined as the ratio of the failure-causing inputs to the total size of the SUT input domain,
then the theoretical \textit{F-measure} of ($F_{RT}$) (with replacement) is $1 / \theta$.
Because ART aims to improve on the failure-detection effectiveness of RT, a measure of the extent of this improvement, known as the {\em F-ratio} ($F_{ART} / F_{RT}$), is also used in this paper.

\subsubsection{Test case distribution metrics}
\nrs{\textit{Discrepancy} refers to the differences of point densities in different sub-domains of the software input domain ($D$)
---
larger sub-domains should have more test cases than smaller ones.
In an ideal situation, discrepancy values should be zero, indicating that the test cases ($E$) are evenly distributed.
The input domain can have an infinite number of sub-domains~\cite{liu2011adaptive};
its Monte Carlo approximation can be obtained by~\cite{Chen2007a}:}
\begin{equation}
    Discrepancy  =  \max_{i=1 \cdots m} \left \lvert \frac{|E_i|}{|E|} - \frac{|D_i|}{|D|}  \right \rvert
\end{equation}

\nrs{where $D_1$, $D_2$, $D_3$, $\cdots$, $D_m$ are hyper-rectangular sub-domains of $D$ whose size and location are randomly defined with uniform probability \cite{Chow2013}};
\nrs{$E_1$, $E_2$, $E_3$, $\cdots$, $E_m$ are the subsets of $E$ falling in each sub-domain, respectively}; and
\nrs{$m$ is the number of randomly defined sub-domains.
A value of $m$ that is too low causes unreliable approximation, but a value that is too high incurs significant overheads for discrepancy calculation \cite{Ackah-Arthur2019}  ($1000$ is a commonly-used value~\cite{Ackah-Arthur2019, Liu2010, Chen2007a}).}

\subsection{Simulations and Subject Programs}
\label{simulations-and-subject-programs}

\begin{table*}[!t]
    \centering
    \caption{Details of the subject programs}
    \label{tbl_programs}
    \resizebox{\textwidth}{!}{%
    \setlength{\extrarowheight}{1pt}
    \begin{tabular}{|c||c||c|c||c||c||c||c|}
        \hline
        \multirow{2}{*}{Program} &
        \multirow{2}{*}{$d$} &
        \multicolumn{2}{c||}{Input Domain ($D$)} &
        \multirow{2}{*}{\makecell{Size\\ (LOC)} } &
        \multirow{2}{*}{Fault Types} &
        \multirow{2}{*}{\makecell{Total\\ Faults}} &
        \multirow{2}{*}{$\theta$} \\ \cline{3-4}
        &  & from & to &  &  &  &  \\ \hline\hline

        bessj0 & 1 & -300000 & 3000000 & 28 & AOR,ROR,SVR,CR & 5 & 0.001373 \\ \hline

        airy & 1 & -5000 & 5000 & 43 & CR & 1 & 0.000716 \\ \hline

        asinh &  1 &  -10000 &  10000 &  360 & AOR,ROR & 2 & 0.0001001 \\ \hline

        erfcc & 1 & -30000 & 30000 & 14 & AOR,ROR,SVR,CR & 4 & 0.000574 \\ \hline

        probks & 1 & -50000 & 50000 & 22 & AOR,ROR,SVR,CR & 4 & 0.000387 \\\hline

        tanh & 1 & -500 & 500 & 18 & AOR,ROR,SVR,CR & 4 & 0.001817 \\ \hline

        bessj & 2 & (2, -1000) & (300, 15000) & 99 & AOR,ROR,CR & 4 & 0.001298 \\ \hline
        gammq & 2 & (0, 0) & (1700, 40) & 106 & ROR,CR & 4 & 0.000830 \\ \hline
        sncndn & 2 & (-5000, -5000) & (5000, 5000) & 64 & SVR,CR & 5 & 0.001623 \\ \hline

        binomial  &  2   & (0, 0) &  (128, 128) & 501 &  CR  &  1 &  0.0001341  \\ \hline

        plgndr & 3 & (10, 0, 0) & (500, 11, 1) & 36 & AOR,ROR,CR & 5 & 0.000368 \\ \hline
        golden & 3 & (-100, -100, -100) & (60, 60, 60) & 80 & ROR,SVR,CR & 5 & 0.000550 \\ \hline
        cel & 4 & \makecell{(0.001, 0.001,\\ 0.001, 0.001)} & \makecell{(1, 300,\\ 10000, 1000)} & 49 & AOR,ROR,CR & 3 & 0.000332 \\ \hline
        el2 & 4 & (0, 0, 0, 0) & (250, 250, 250, 250) & 78 & AOR,ROR,SVR,CR & 9 & 0.000690 \\ \hline

        period  & 4 &  \makecell{(-10000, -10000, \\ -10000, -10000)}   &  \makecell{(10000, 10000, \\ 10000, 10000)}  &  1128  &  CR  &  1  &  NA  \\ \hline

        calDay & 5 & (1, 1, 1, 1, 1800) & (12, 31, 12, 31, 2020) & 37 & SDL & 1 & 0.000632 \\ \hline
        complex & 6 & \makecell{(-20, -20, -20,\\ -20, -20, -20)} & \makecell{(20, 20, 20,\\ 20, 20, 20)} & 68 & SVR & 1 & 0.000901 \\ \hline
        pntLinePos & 6 & \makecell{(-25, -25, -25,\\ -25, -25, -25)} & \makecell{(25, 25, 25,\\ 25, 25, 25)} & 23 & CR & 1 & 0.000728 \\ \hline
        triangle & 6 & \makecell{(-25, -25, -25,\\ -25, -25, -25)} & \makecell{(25, 25, 25,\\ 25, 25, 25)} & 21 & CR & 1 & 0.000713 \\ \hline
         line & 8 & \makecell{(-10, -10, -10, -10,\\ -10, -10, -10, -10)} & \makecell{(10,  10, 10, 10,\\ 10, 10, 10, 10)} & 86 & ROR & 1 & 0.000303 \\ \hline
        pntTrianglePos & 8 & \makecell{(-10, -10, -10, -10, \\-10,  -10, -10, -10)} & \makecell{(10, 10, 10, 10,\\ 10, 10, 10, 10)} & 68 & CR & 1 & 0.000141 \\ \hline
        twoLinesPos & 8 & \makecell{(-15, -15, -15, -15,\\ -15, -15, -15, -15)} & \makecell{(15, 15, 15, 15,\\ 15, 15, 15, 15)} & 28 & CR & 1 & 0.000133 \\ \hline
        nearestDistance & 10 & \makecell{(1, 1, 1, 1, 1,\\ 1, 1, 1, 1, 1)} & \makecell{(15, 15, 15, 15, 15,\\ 15, 15, 15, 15, 15)} & 26 & CR & 1 & 0.000256 \\ \hline
        calGCD & 10 & \makecell{(1, 1, 1, 1, 1,\\ 1, 1, 1, 1, 1)} & \makecell{(1000, 1000, 1000,\\ 1000, 1000, 1000,\\ 1000, 1000, 1000, 1000)}  & 24 & AOR & 1 & NA \\ \hline
        select & 11 & \makecell{(1, 1, 1, 1, 1, 1, \\ 1, 1, 1, 1, 1)} & \makecell{(10, 100, 100,\\ 100, 100, 100, 100,\\ 100, 100, 100, 100)}  & 117 & RSR,CR & 2 & NA \\ \hline
        tcas & 12 & \makecell{(0, 0, 0, 0, 0, 0,\\ 0, 0, 0, 0, 0, 0)} & \makecell{(1000, 1, 1, 50000,\\ 1000, 50000, 3, \\1000, 1000, 2, 2, 1)} & 182 & CR & 1 & NA \\ \hline

        matrixProcessor  &  12  &  \makecell{($-10^4$, $-10^4$, $-10^4$,\\ $-10^4$, $-10^4$, $-10^4$,\\ $-10^4$, $-10^4$,  $-10^4$,\\ $-10^4$, $-10^4$, $-10^4$)}  &  \makecell{($10^4$, $10^4$, $10^4$,\\ $10^4$, $10^4$, $10^4$,\\ $10^4$, $10^4$,  $10^4$,\\ $10^4$, $10^4$, $10^4$)}  &  462  &  CR  &  1  &  NA \\ \hline

        java.util.Arrays  &  15  &  \makecell{($-10^4$, $-10^4$, $-10^4$,\\ $-10^4$, $-10^4$, $-10^4$,\\ $-10^4$, $-10^4$,  $-10^4$,\\ $-10^4$, $-10^4$, $-10^4$,\\ $-10^4$, $-10^4$, $-10^4$)}  &  \makecell{($10^4$, $10^4$, $10^4$,\\ $10^4$, $10^4$, $10^4$,\\ $10^4$, $10^4$,  $10^4$,\\ $10^4$, $10^4$, $10^4$,\\ $10^4$, $10^4$, $10^4$)}  &  1357  &  CR  & 1  &  NA  \\ \hline
    \end{tabular}
    }
\end{table*}

\nrs{To answer RQ1, the $T_G$ values of FSCS-ART, LimBal-KDFC, and SWFC-ART were recorded for test suites of sizes 500, 1000, 2000, 5000, 10,000, 15,000 and 20,000, in 2-, 3-, 4-, 5-, 10-, and 15-dimensional input domains. }
The $T_G$ and $T_E$ for the real-life programs (discussed below) were also recorded.

The failure-detection effectiveness of ART methods depends on several factors, including the shape of the failure region, \nrs{ the failure rate ($\theta$), and the dimensionality ($d$) of the input domain ($D$)}~\cite{Chen2007c}.
\nrs{It is common practice in ART studies investigating failure-finding effectiveness according to the {\em F-measure} ($F_{ART}$) to use both simulations and empirical studies.}
We have followed this tradition in our study to answer RQ2.

Generally speaking, the failure-causing inputs of a software tend to cluster into \textit{block}, \textit{strip} or \textit{point} failure patterns~\cite{Chan1996}.
In our simulations,
\nrs{the block patterns were created by randomly generating a hyper-cube in $D$ whose hyper-volume and length of each side equalled $\theta$ and $\sqrt[d]{\theta}$, respectively.}
\nrs{The strip patterns were simulated by randomly selecting points on adjacent borders of $D$, joining them, and expanding the strip magnitude until the hyper-volume became equal to $\theta$}.
\nrs{Strips generated in corners of the input domain were discarded due to their unrealistic thickness.}
The point failure pattern was simulated by randomly generating 25 small, non-overlapping, block failure patterns, with the total hyper-volume of all the blocks being appropriate for the given $\theta$.
Simulations were performed for all three failure pattern types, with $d = \{2, 3, 4, 5, 10\}$ and  $\theta = \{0.01, 0.005, 0.002, 0.001, 0.0005, 0.0002, 0.0001\}$.

To answer RQ3, 100, 1000, and 10,000 test cases were generated in 2-, 3-, 4-, 5-, 10-, and 15-dimensional hyper-cube input domains using FSCS-ART, LimBal-KDFC, and SWFC-ART.
Each dimension of the hyper-cubes was continuous, ranging from $-5000$ to $5000$.

We also used 28 programs, of different sizes and dimensions, in our empirical studies.
\nrs{Faults were seeded into the programs using the following mutation operators~\cite{Jia2011}:
constant replacement (CR);
arithmetic operator replacement (AOR);
return statement replacement (RSR);
scalar variable replacement (SVR);
statement deletion (SDL);
and relational operator replacement (ROR).}
Table \ref{tbl_programs} summarizes details about their $d$, $D$, size (in terms of lines of code), fault\footnote{According to the IEEE~\cite{5733835}, a \textit{fault} (\textit{defect} or \textit{bug}) is an oversight of a programmer. When \textit{fault} is confronted during program execution, \textit{failure} is said to have occurred i.e. software behaves unexpectedly.}
types, number of seeded faults, and $\theta$.
\nrs{(The ``NA'' for $\theta$ in some cases represents situations where the failure rate was not calculated.)}
The first $12$ subject programs have been commonly used in ART research, and are from \textit{Numerical Recipes} ~\cite{10.5555/1403886} and \textit{ACM Collected Algorithms} ~\cite{collected_algorithms_acm}.
The programs \textit{calDay}, \textit{complex} and \textit{line} are from Ferrer~et~al.~\cite{Ferrer2012}.
The programs \textit{pntLinePos},  \textit{pntTrianglePos}, \textit{twoLinesPos} and \textit{triangle} were written as exercises from the textbook \textit{Introduction to Java Programing and Data Structures}~\cite{Y.DanielLiang2017}.
The \textit{nearestDistance} program takes five points in 2-dimensional space and returns the two points that are nearest to each other~\cite{Mao2019}.
The \textit{calGCD} program takes $10$ integers and returns their greatest common divisor.
The \textit{select} program~\cite{May2007} returns the $k$-th largest element from an unordered array.
The \textit{tcas} program is an aircraft collision avoidance system, from Siemens~\cite{Do2005}.
The \textit{asinh}, \textit{binomial}, and \textit{period} programs are from Walkinshaw \& Fraser~\cite{7927980}.
\nrs{The program \textit{matrixProcessor}\footnote{https://github.com/ritish78/NumericMatrixProcessor} manipulates matrices according to specified matrix operations, and was written as an exercise from \textit{JetBrains Academy}~\cite{matrix_processor}.}
\nrs{The \textit{java.util.Arrays}~\cite{java_util_arrays, 10.1145/1868281.1868289} is an array manipulation library in the Java API that contains various array helper functions (such as for sorting and searching).}

\subsection{Data Collection and Statistical Analysis}
\label{data-collection}

The simulations were conducted by continuously generating test cases, using one of the testing strategies under study, until a test case fell \textit{inside} the failure region.
In the experiments with real programs, failure-revealing test inputs were identified when the output of the fault-seeded program differed from the output of the original.
The number of test cases generated and executed before a failure was found (the F-measure) was recorded.

\nrs{All trials were run $S$ times to ensure that mean values had a 95\% confidence level and 5\% accuracy range, according to the central limit theorem~\cite{freund1988modern,Chen2004art}.
In the simulations, $S$ was set to 10,000 for calculating the F-measure, and set to 1000 for $T_G$.}
In the empirical studies, $S$ was also set to 10,000 trials.
The sample sizes were confirmed to be large enough to obtain results with the desired confidence level and accuracy.

\nrs{We used the unpaired two-tailed \textit{Wilcoxon rank-sum test}~\cite{Wilcoxon1945} (reciprocal of \textit{Mann-Whitney U test}~\cite{mann1947}) to analyze the significance of differences between the SWFC-ART and FSCS-ART data, and between the SWFC-ART and LimBal-KDFC data.}
\nrs{For two random samples, the Wilcoxon rank-sum test returns a $z$-statistic which is then converted into a $p$-value (probability value).
For a 95\% confidence interval (or 5\% significance level), a $z$-static $\geq 1.96$, or a $p$-value $\leq$ $0.01$, means that there is sufficient evidence to reject the \textit{null hypothesis} ($H_{0}$)~\cite{Ackah-Arthur2019}.
The $H_{0}$ states that there is no significant difference between the observed values of the two samples~\cite{Wilcoxon1945}. }
The \textit{effect size}~\cite{Sullivan2012} is used to calculate the impact of the results of the experiment on an evaluation metric.
The effect size for the Wilcoxon rank-sum test was calculated as~\cite{pallant_2016}:
\begin{equation}
    r = \frac{|z|}{\sqrt{n_{1} + n_{2} }}
\end{equation}
where $z$ is the $z$-statistic returned by the rank-sum test and $n_1$ and $n_2$ are the sample sizes.
Cohen~\cite{jacobcohen1988} identified effect sizes as
{\em large} for $r=0.5$;
{\em medium} for $r=0.3$; and
{\em low} for $r=0.1$.

\subsection{Experimental Environment}
\label{experimental-environment}

Java 1.8.0\_221 was the programming language used to develop and run the simulations and
experiments\footnote{We have released the SWFC-ART source code, and made it available online: https://github.com/ashfaq92/swfc-art}.
Two machines were used to conduct the study, both of which ran under the Microsoft Windows 10 Pro 64-bit operating system.
\begin{itemize}
    \item Machine 1: Acer Aspire V3-572G, Intel\textregistered \quad Core\texttrademark  i5-5200U CPU @ 2.2GHz, 2 Cores, 4 Logical Processors, 12GB RAM.
    \item Machine 2: Dell OptiPlex 7050, Intel\textregistered \quad  Core\texttrademark  i7-7700 CPU @ 3.60GHz, 4 Cores, 8 Logical Processors, 16GB RAM.
\end{itemize}
\nrs{The simulations and most of the studies with the subject programs were conducted on Machine 1. }
However, due to the huge size and prohibitive time required, \textit{java.util.Arrays} and \textit{period} were tested on Machine 2.
The experimental parameters were set as described in Section \ref{parameter_optimization}.

\section{Experimental Results}
\label{experimental-results}

\subsection{Simulations}
\label{simulations}

\subsubsection{Computational Efficiency}
\label{sim-efficiency}

\begin{table*}[!t]
\centering
\caption{Wilcoxon Rank-Sum Tests and Effect Size Analyses of Test Case Generation Times for FSCS-ART, LimBal-KDFC and SWFC-ART}
\label{tbl_efficiency}
\begin{tabular}{|c||c||c|c|c||c|c||c|c|}
\hline
\multirow{2}{*}{$d$} &
\multirow{2}{*}{$n$} &
\multicolumn{3}{c||}{Mean running time (ms)} & \multicolumn{2}{c||}{\makecell{FSCS-\\ART} vs \makecell{SWFC-\\ART}} & \multicolumn{2}{c|}{\makecell{LimBal-\\KDFC} vs \makecell{SWFC-\\ART}} \\ \cline{3-9}
 &  & FSCS-ART & LimBal-KDFC & SWFC-ART & $p$-value & effect size & $p$-value & effect size \\ \hline
\multirow{7}{*}{2} & 500 & 21.86 & \textbf{6.38} & 28.81 & 0.0000 & 0.8573 & 0.0000 & 0.8658 \\ \cline{2-9}
 & 1000 & 99.60 & \textbf{14.20} & 57.24 & 0.0000 & 0.8658 & 0.0000 & 0.8658 \\ \cline{2-9}
 & 2000 & 406.98 & \textbf{33.88} & 127.23 & 0.0000 & 0.8658 & 0.0000 & 0.8658 \\ \cline{2-9}
 & 5000 & 2347.86 & \textbf{92.00 }& 367.43 & 0.0000 & 0.8658 & 0.0000 & 0.8658 \\ \cline{2-9}
 & 10000 & 8466.74 & \textbf{202.73} & 818.66 & 0.0000 & 0.8658 & 0.0000 & 0.8658 \\ \cline{2-9}
 & 15000 & 18367.03 &\textbf{ 328.03} & 1525.58 & 0.0000 & 0.8658 & 0.0000 & 0.8658 \\ \cline{2-9}
 & 20000 & 32235.35 &\textbf{ 470.50} & 2145.45 & 0.0000 & 0.8658 & 0.0000 & 0.8658 \\ \hline \hline
\multirow{7}{*}{3} & 500 & 26.15 & \textbf{9.87} & 30.37 & 0.0000 & 0.8535 & 0.0000 & 0.8658 \\ \cline{2-9}
 & 1000 & 113.63 & \textbf{22.03} & 67.66 & 0.0000 & 0.8654 & 0.0000 & 0.8658 \\ \cline{2-9}
 & 2000 & 484.51 & \textbf{50.80} & 150.57 & 0.0000 & 0.8658 & 0.0000 & 0.8658 \\ \cline{2-9}
 & 5000 & 2776.46 & \textbf{146.15} & 434.76 & 0.0000 & 0.8658 & 0.0000 & 0.8658 \\ \cline{2-9}
 & 10000 & 10190.99 & \textbf{332.34} & 990.50 & 0.0000 & 0.8658 & 0.0000 & 0.8658 \\ \cline{2-9}
 & 15000 & 22372.77 & \textbf{546.68} & 1842.28 & 0.0000 & 0.8658 & 0.0000 & 0.8658 \\ \cline{2-9}
 & 20000 & 39157.61 & \textbf{788.84} & 2550.49 & 0.0000 & 0.8658 & 0.0000 & 0.8658 \\ \hline \hline
\multirow{7}{*}{4} & 500 & 30.82 & \textbf{15.15} & 33.90 & 0.0000 & 0.8220 & 0.0000 & 0.8658 \\ \cline{2-9}
 & 1000 & 136.47 & \textbf{35.20} & 74.88 & 0.0000 & 0.8658 & 0.0000 & 0.8658 \\ \cline{2-9}
 & 2000 & 568.14 & \textbf{84.39} & 173.33 & 0.0000 & 0.8658 & 0.0000 & 0.8658 \\ \cline{2-9}
 & 5000 & 3209.42 & \textbf{255.38} & 504.40 & 0.0000 & 0.8658 & 0.0000 & 0.8658 \\ \cline{2-9}
 & 10000 & 11958.25 & \textbf{603.19} & 1141.58 & 0.0000 & 0.8658 & 0.0000 & 0.8624 \\ \cline{2-9}
 & 15000 & 26301.38 &\textbf{ 1028.92} & 2136.11 & 0.0000 & 0.8658 & 0.0000 & 0.8641 \\ \cline{2-9}
 & 20000 & 46700.79 &\textbf{ 1537.45} & 2975.18 & 0.0000 & 0.8658 & 0.0000 & 0.8658 \\ \hline \hline
\multirow{7}{*}{5} & 500 & 35.50 & \textbf{23.55} & 40.17 & 0.0000 & 0.8322 & 0.0000 & 0.8658 \\ \cline{2-9}
 & 1000 & 156.48 & \textbf{56.48} & 87.91 & 0.0000 & 0.8658 & 0.0000 & 0.8657 \\ \cline{2-9}
 & 2000 & 666.15 & \textbf{140.33} & 200.98 & 0.0000 & 0.8658 & 0.0000 & 0.8641 \\ \cline{2-9}
 & 5000 & 3718.41 & \textbf{449.71} & 605.54 & 0.0000 & 0.8658 & 0.0000 & 0.8568 \\ \cline{2-9}
 & 10000 & 13815.72 & \textbf{1096.25} & 1367.65 & 0.0000 & 0.8658 & 0.0000 & 0.8641 \\ \cline{2-9}
 & 15000 & 30474.20 &\textbf{ 1904.76} & 2448.46 & 0.0000 & 0.8658 & 0.0000 & 0.8606 \\ \cline{2-9}
 & 20000 & 54599.66 & \textbf{2879.91} & 3399.76 & 0.0000 & 0.8658 & 0.0000 & 0.8639 \\ \hline \hline
\multirow{7}{*}{10} & 500 & \textbf{60.04} & 114.19 & 69.03 & 0.0000 & 0.8602 & 0.0000 & 0.8658 \\ \cline{2-9}
 & 1000 & 263.29 & 366.03 & \textbf{165.59} & 0.0000 & 0.8658 & 0.0000 & 0.8658 \\ \cline{2-9}
 & 2000 & 1063.11 & 982.16 & \textbf{380.31 }& 0.0000 & 0.8658 & 0.0000 & 0.8658 \\ \cline{2-9}
 & 5000 & 6116.69 & 3216.35 & \textbf{1129.49 }& 0.0000 & 0.8658 & 0.0000 & 0.8658 \\ \cline{2-9}
 & 10000 & 23420.62 & 8056.34 & \textbf{2654.56} & 0.0000 & 0.8658 & 0.0000 & 0.8658 \\ \cline{2-9}
 & 15000 & 52877.05 & 14336.57 & \textbf{4923.99} & 0.0000 & 0.8658 & 0.0000 & 0.8658 \\ \cline{2-9}
 & 20000 & 95485.52 & 21384.65 & \textbf{6948.82} & 0.0000 & 0.8658 & 0.0000 & 0.8658 \\ \hline \hline
\multirow{7}{*}{15} & 500 & \textbf{81.78} & 146.45 & 97.22 & 0.0000 & 0.8658 & 0.0000 & 0.8645 \\ \cline{2-9}
 & 1000 & 350.05 & 629.85 & \textbf{236.66} & 0.0000 & 0.8608 & 0.0000 & 0.8658 \\ \cline{2-9}
 & 2000 & 1374.82 & 2266.66 & \textbf{583.09} & 0.0000 & 0.8658 & 0.0000 & 0.8658 \\ \cline{2-9}
 & 5000 & 8049.52 & 8243.52 & \textbf{1796.36} & 0.0000 & 0.8658 & 0.0000 & 0.8658 \\ \cline{2-9}
 & 10000 & 31545.61 & 21661.91 & \textbf{4229.87} & 0.0000 & 0.8658 & 0.0000 & 0.8658 \\ \cline{2-9}
 & 15000 & 72274.64 & 38793.33 & \textbf{8074.45} & 0.0000 & 0.8658 & 0.0000 & 0.8658 \\ \cline{2-9}
 & 20000 & 137040.00 & 58535.86 & \textbf{11391.41} & 0.0000 & 0.8658 & 0.0000 & 0.8658 \\ \hline
\end{tabular}
\end{table*}


\begin{figure*}[!t]
    \centering
    \subfloat[2-$d$ input domain]{\includegraphics[width=0.49\linewidth]{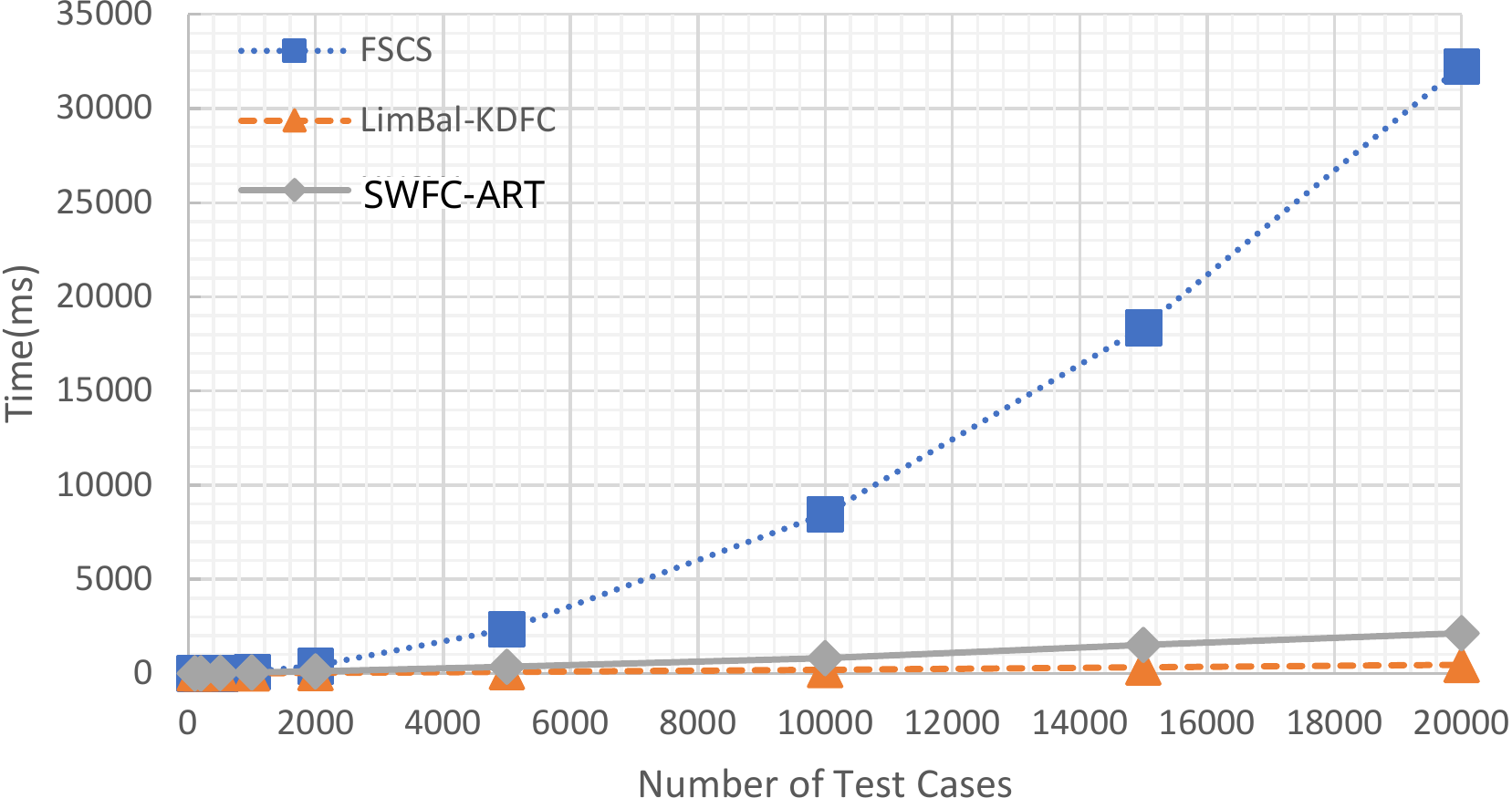}
    \label{efficiency_2d}}
    \hfil
    \subfloat[3-$d$ input domain] {\includegraphics[width=0.49\linewidth]{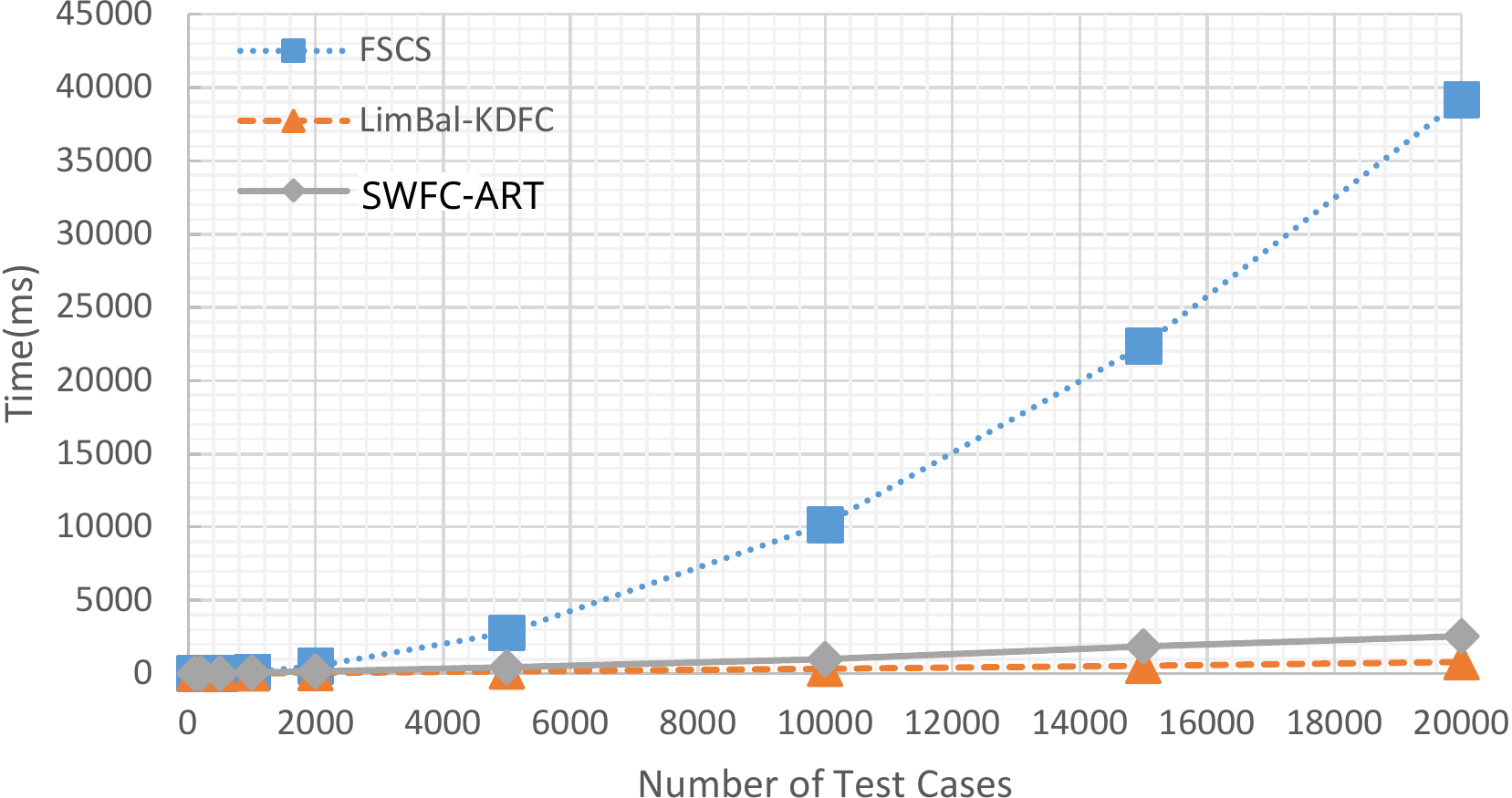}
    \label{efficiency_3d}}
    \hfil
    \subfloat[4-$d$ input domain]{\includegraphics[width=0.49\linewidth]{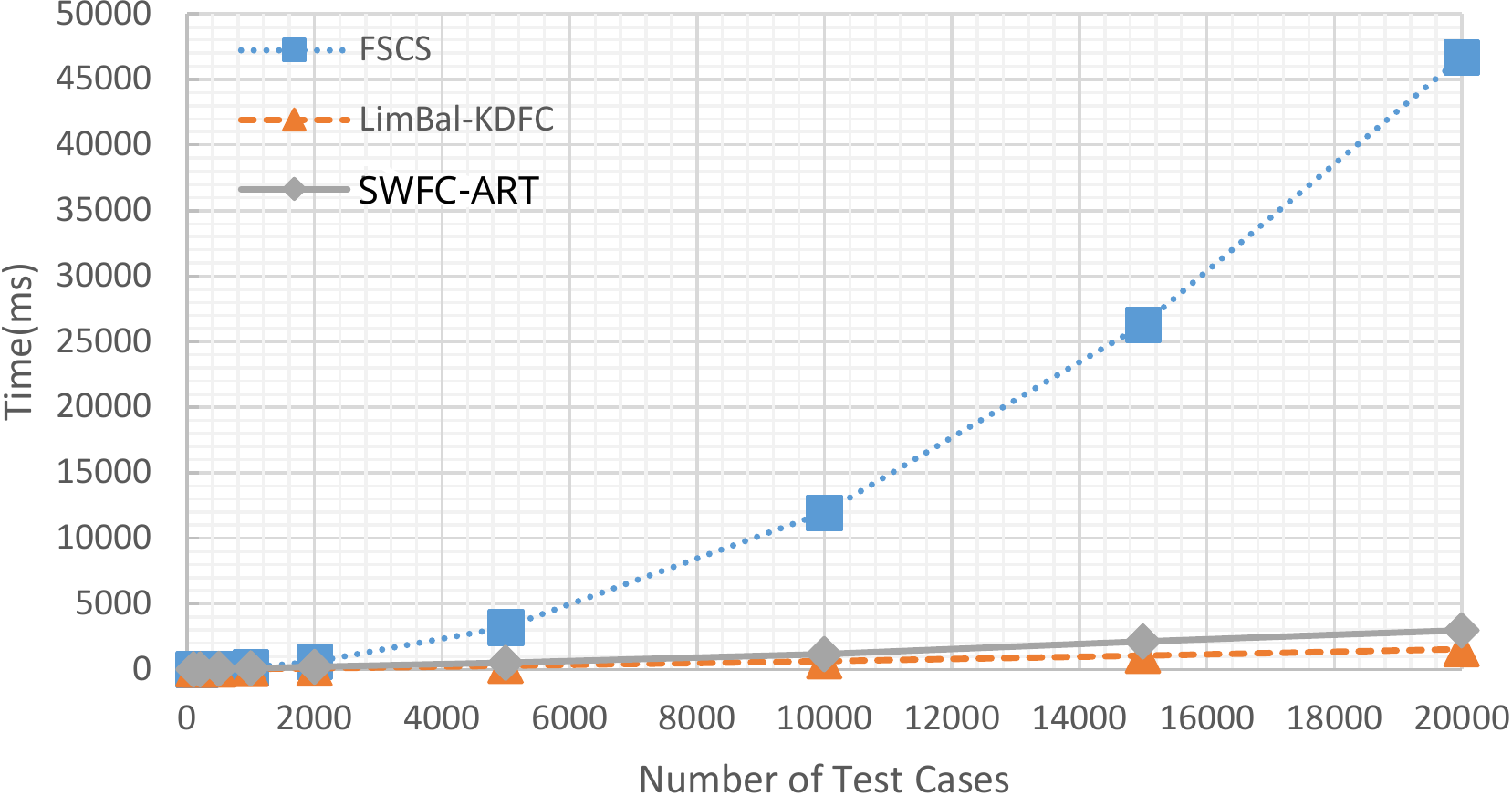}
    \label{efficiency_4d}}
    \hfil
    \subfloat[5-$d$ input domain]{\includegraphics[width=0.49\linewidth]{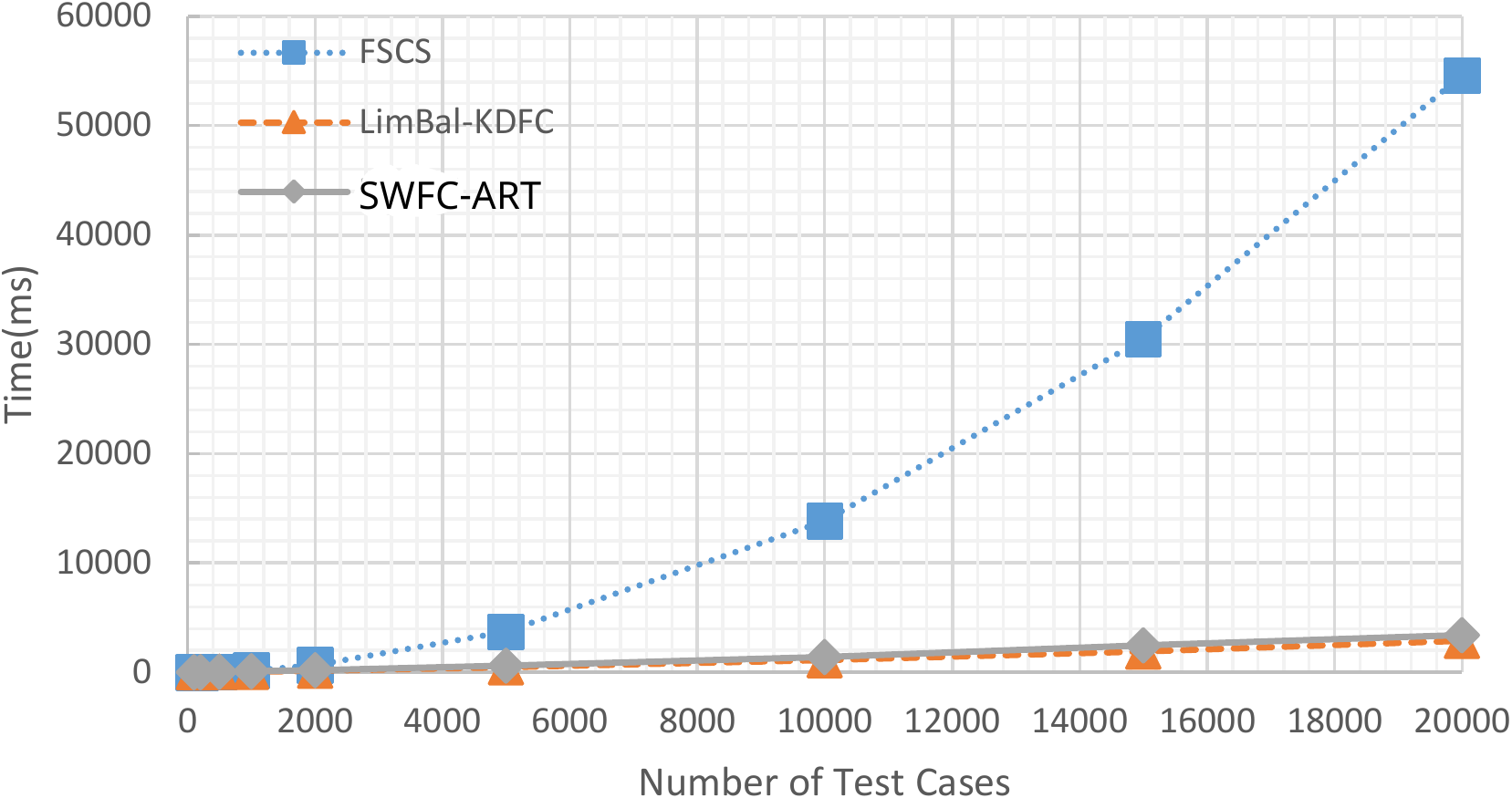}
    \label{efficiency_5d}}
    \hfil
    \subfloat[10-$d$ input domain]{\includegraphics[width=0.49\linewidth]{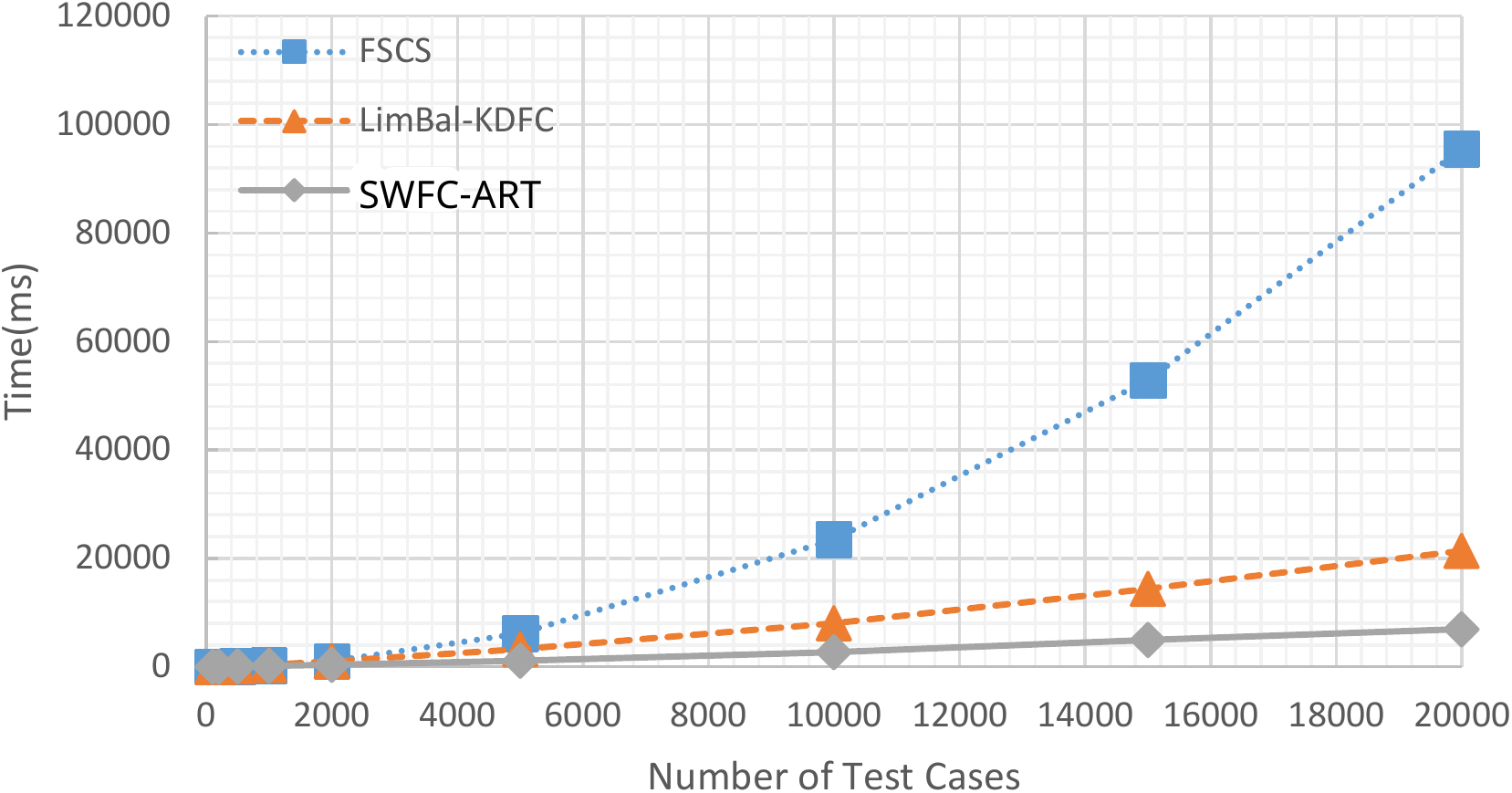}
    \label{efficiency_10d}}
    \hfil
    \subfloat[15-$d$ input domain]{\includegraphics[width=0.49\linewidth]{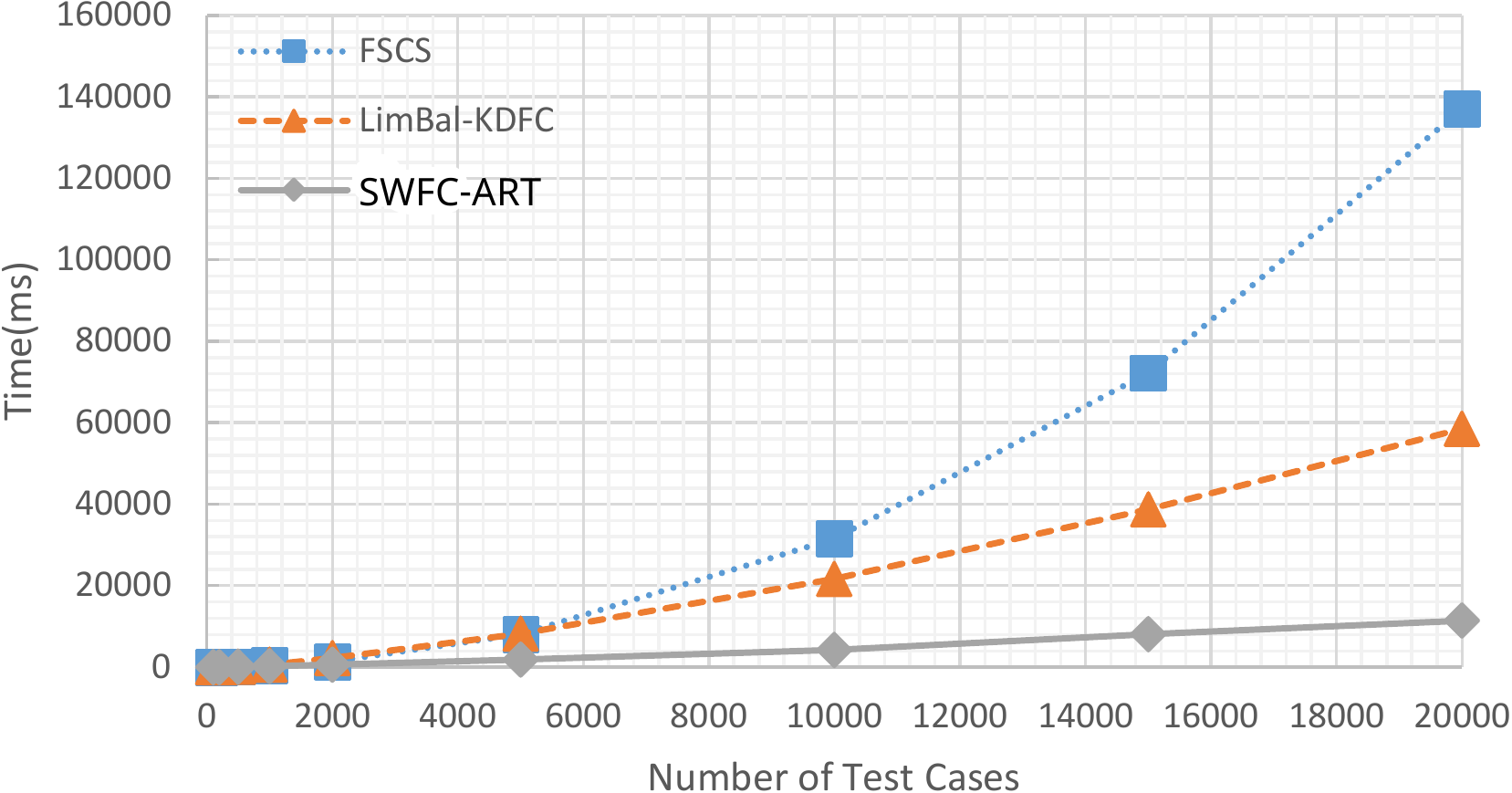}
    \label{efficiency_15d}}
    \caption{Test case generation times for FSCS-ART, LimBal-KDFC and SWFC-ART, for different input domain dimensions}
    \label{efficiency}
\end{figure*}

\begin{figure*}[!t]
    \centering
    \subfloat[$n=500$]{\includegraphics[width=0.49\linewidth]{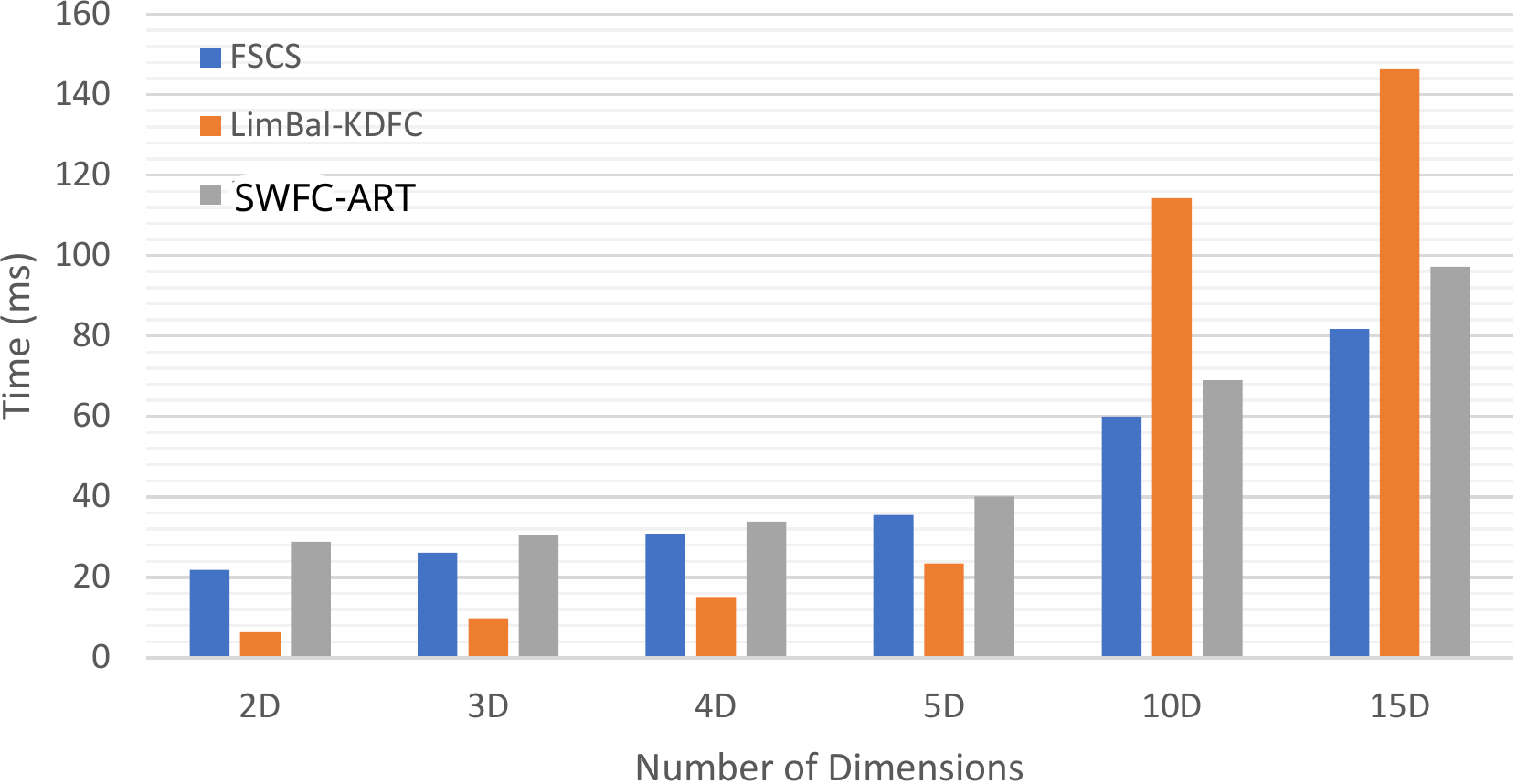}
    \label{efficiency_500}}
    \hfil
    \subfloat[$n=1000$]{\includegraphics[width=0.49\linewidth]{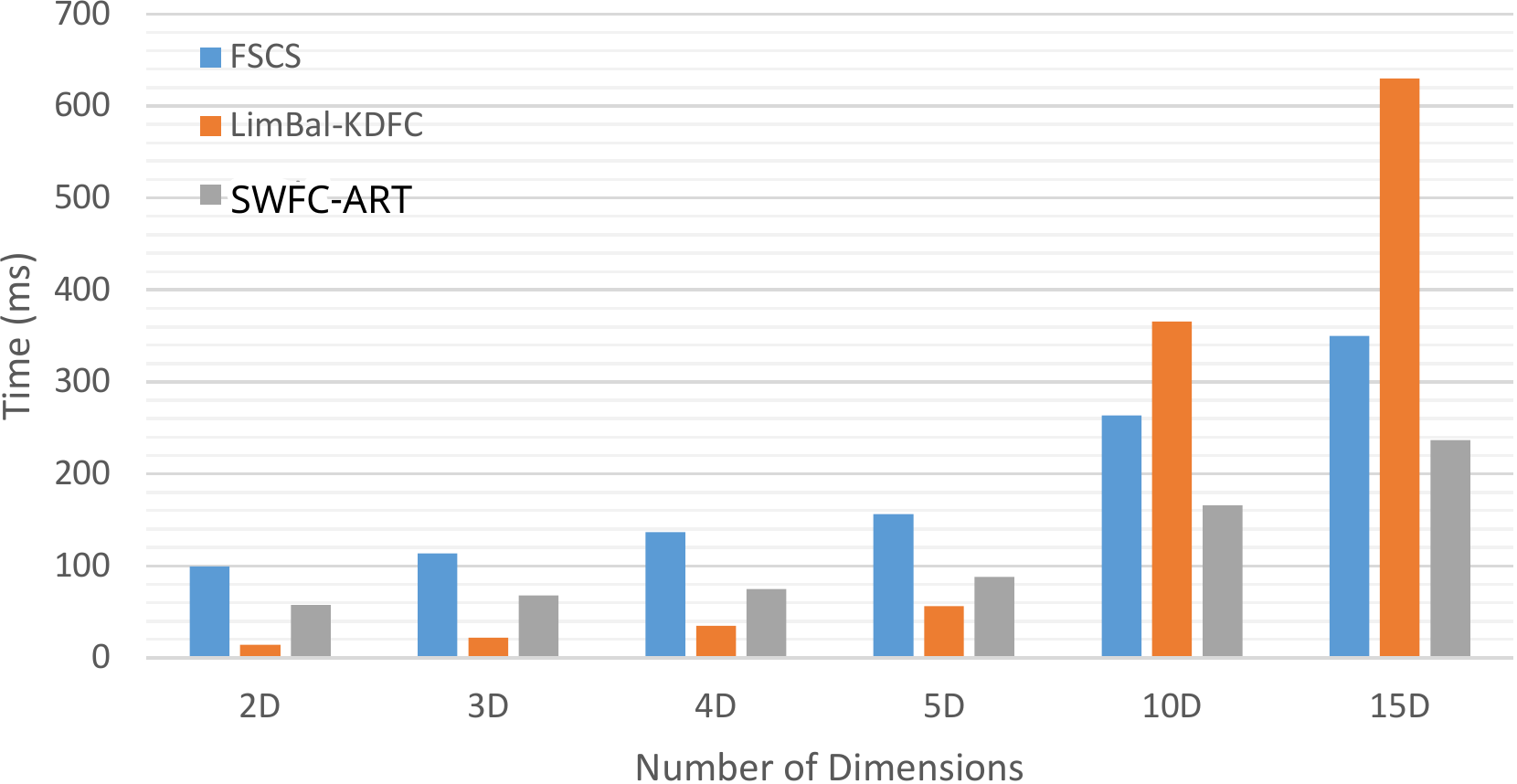}
    \label{efficiency_1000}}
    \hfil
    \subfloat[$n=2000$]{\includegraphics[width=0.49\linewidth]{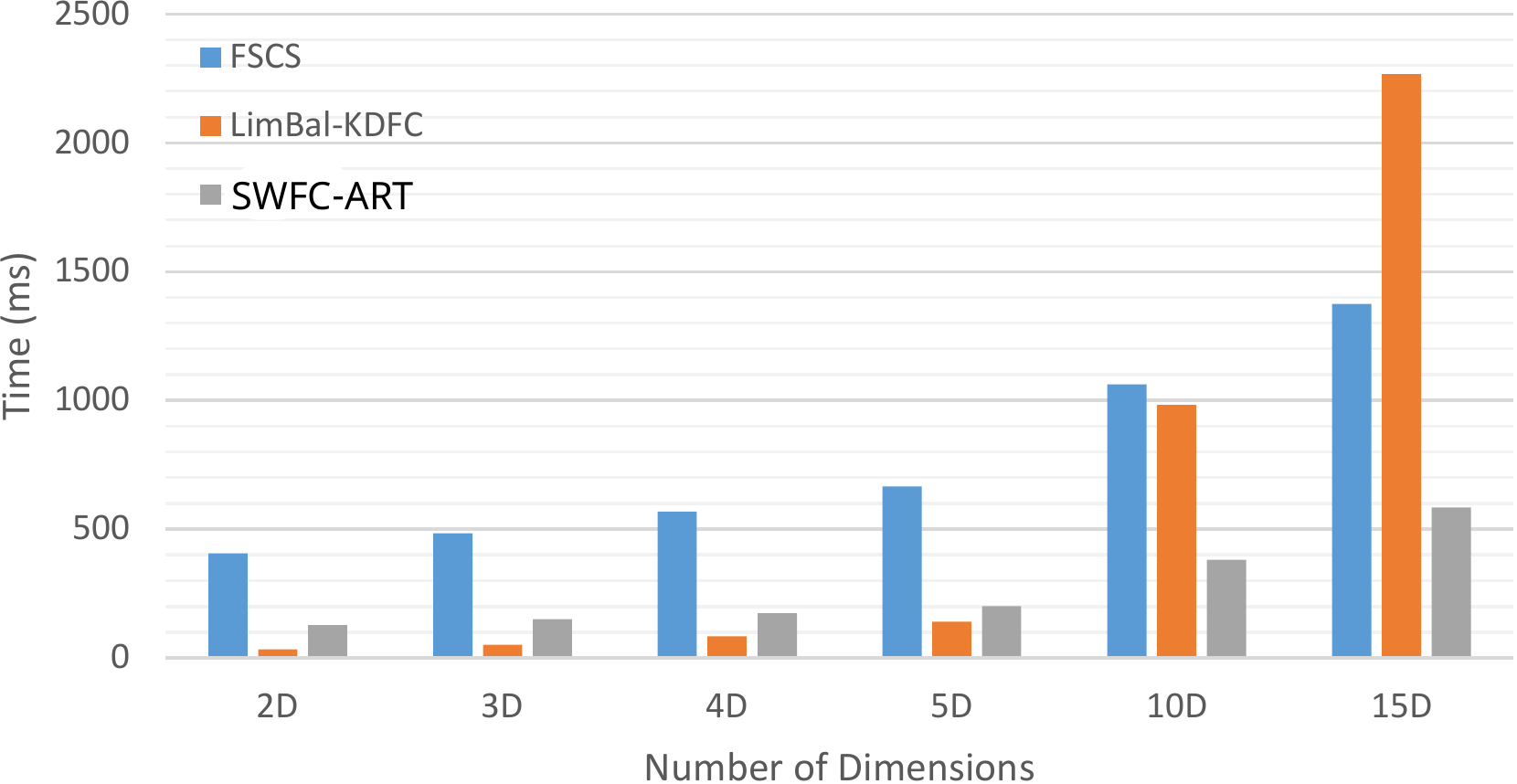}
    \label{efficiency_2000}}
    \hfil
    \subfloat[$n=5000$]{\includegraphics[width=0.49\linewidth]{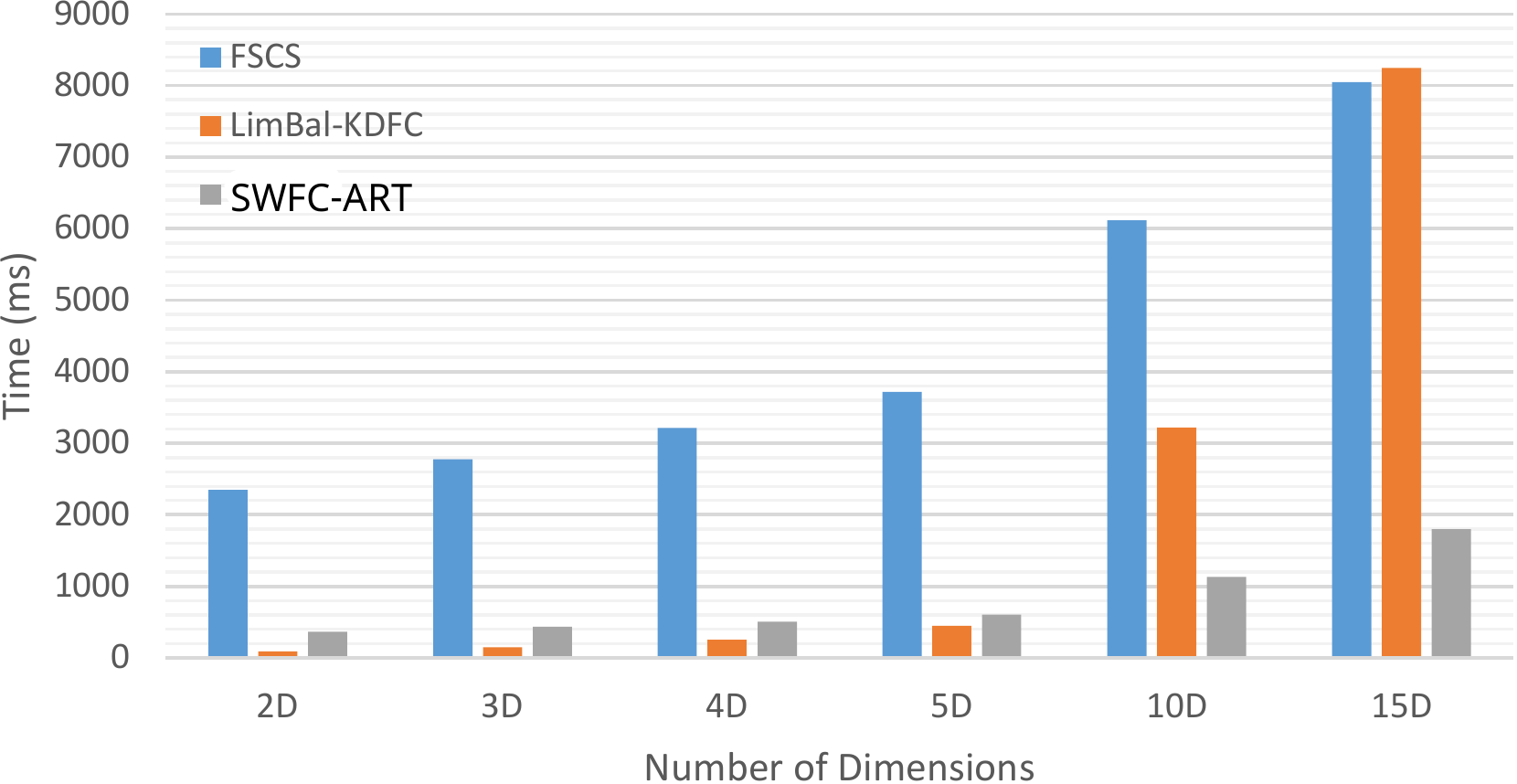}
    \label{efficiency_5000}}
    \hfil
    \subfloat[$n=10,000$]{\includegraphics[width=0.49\linewidth]{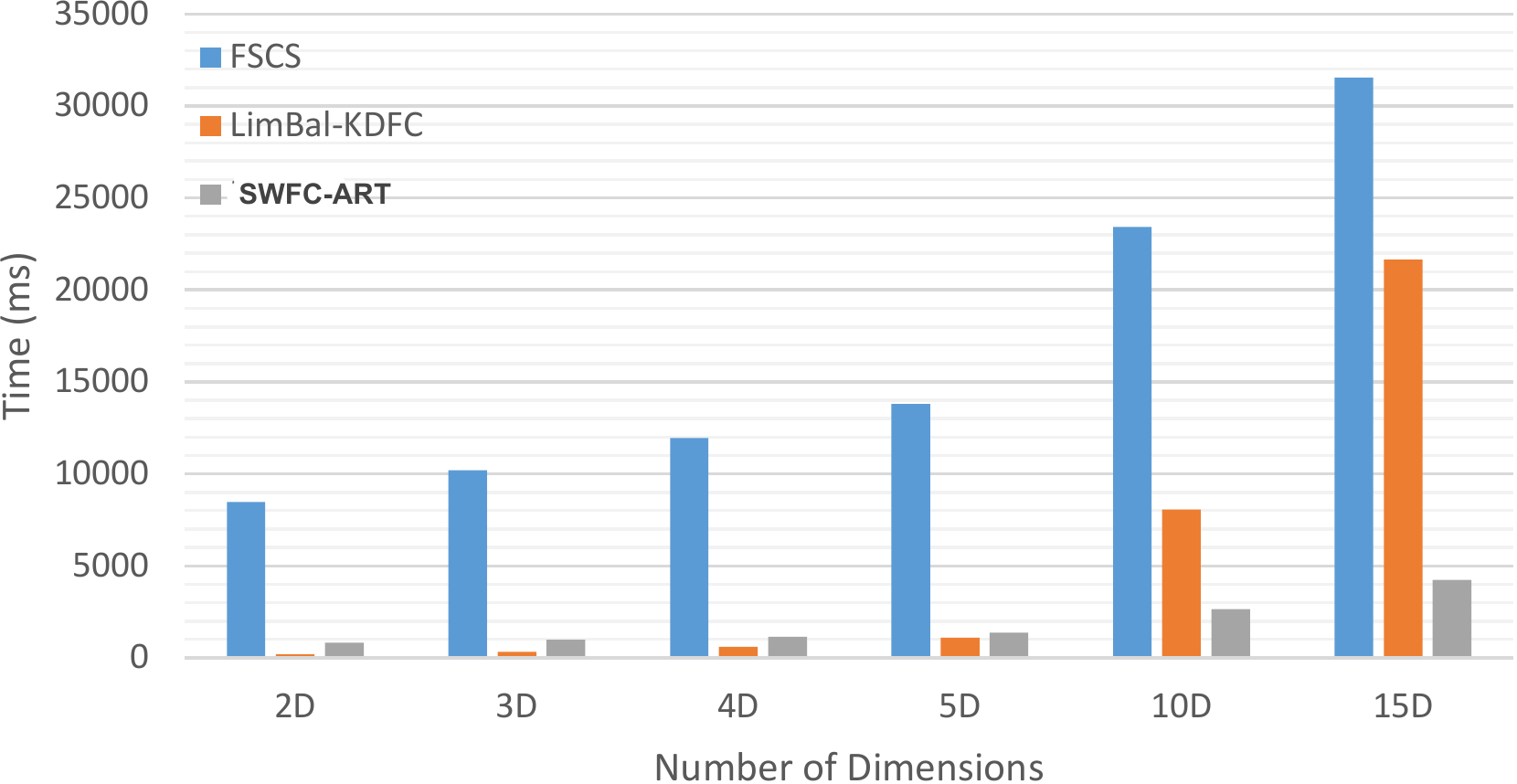}
    \label{efficiency_10000}}
    \hfil
    \subfloat[$n=15,000$]{\includegraphics[width=0.49\linewidth]{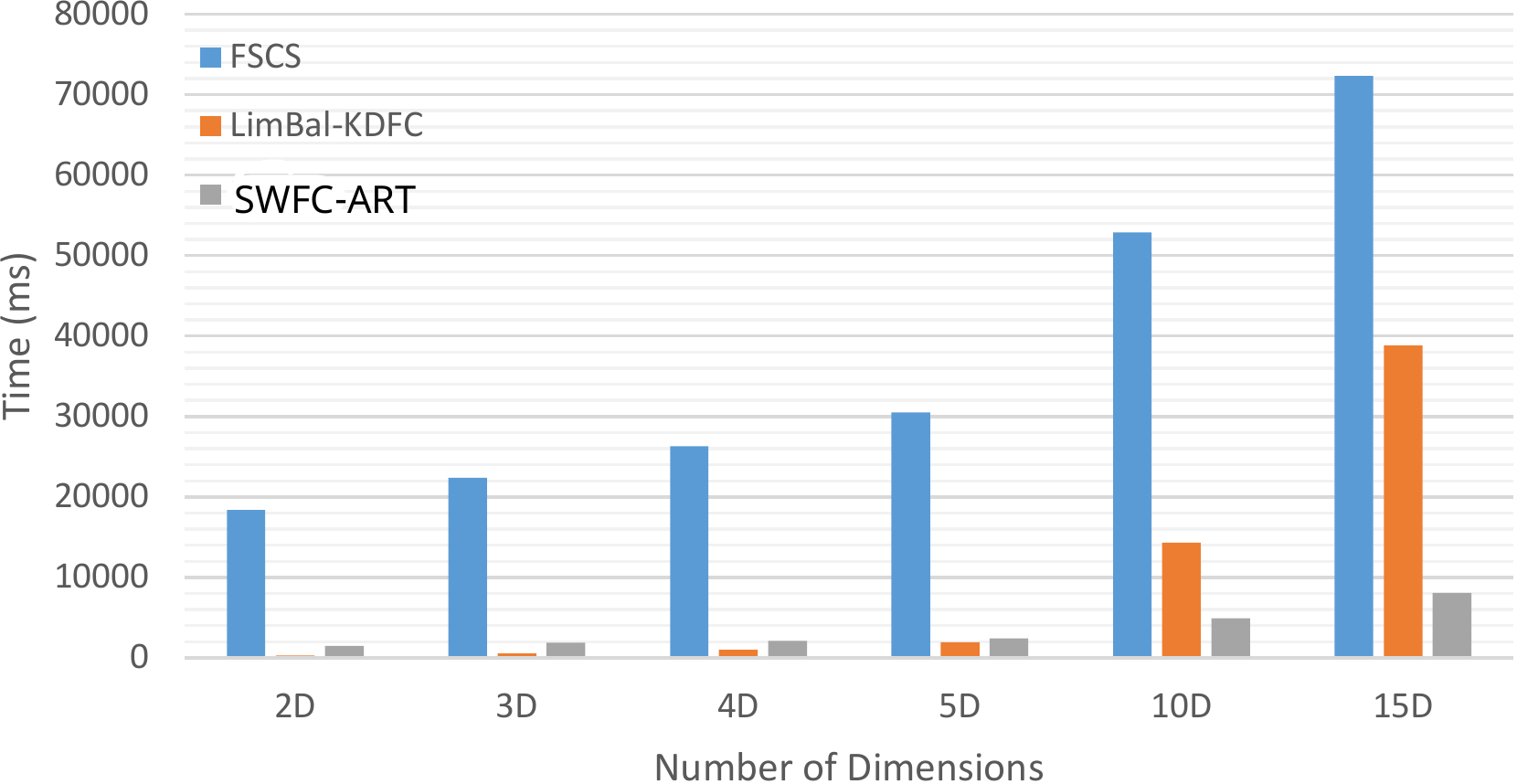}
    \label{efficiency_15000}}
    \hfil
    \subfloat[$n=20,000$]{\includegraphics[width=0.49\linewidth]{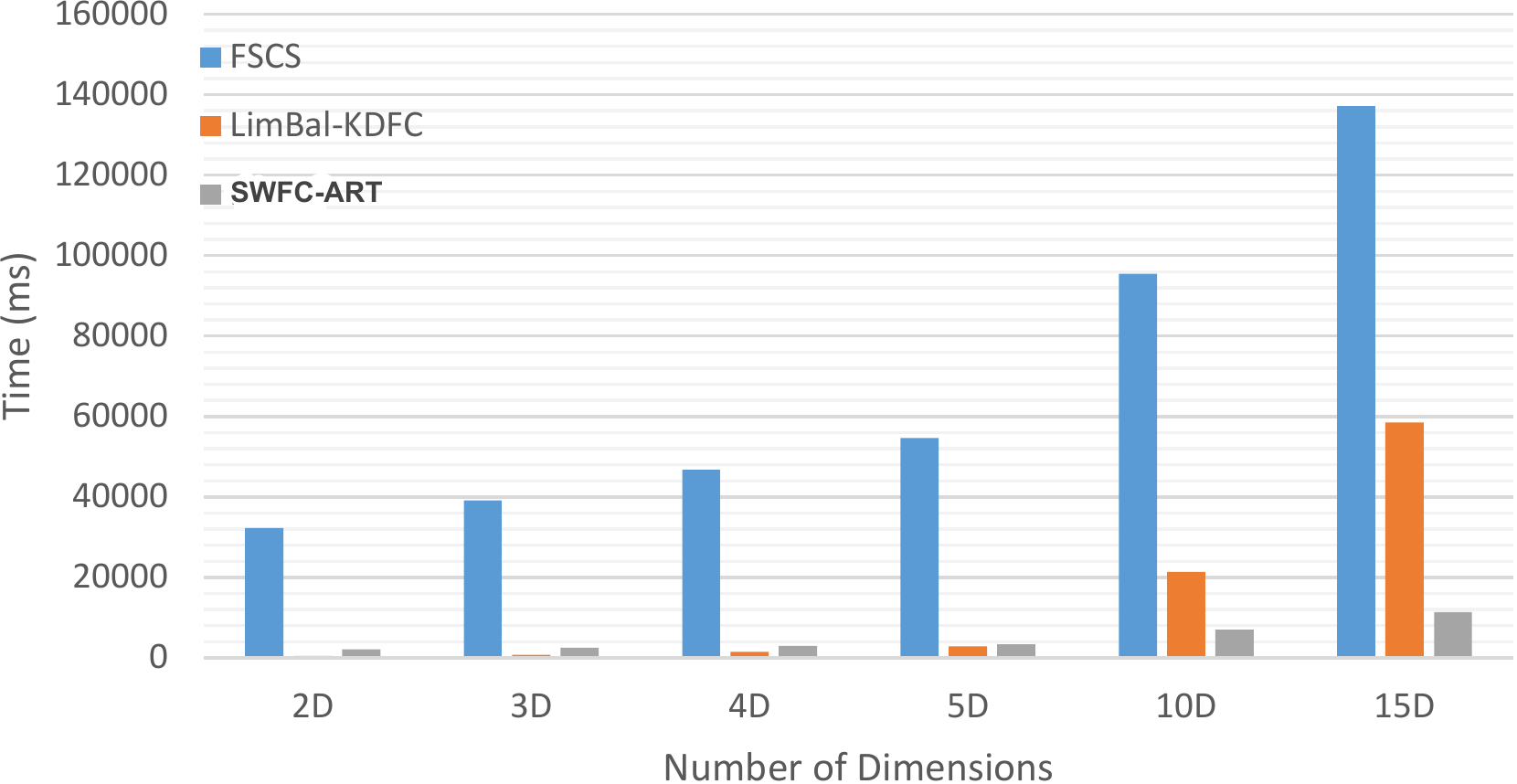}
    \label{efficiency_20000}}
    \hfil
    \subfloat[Trend-lines for $n=20,000$]{\includegraphics[width=3in]{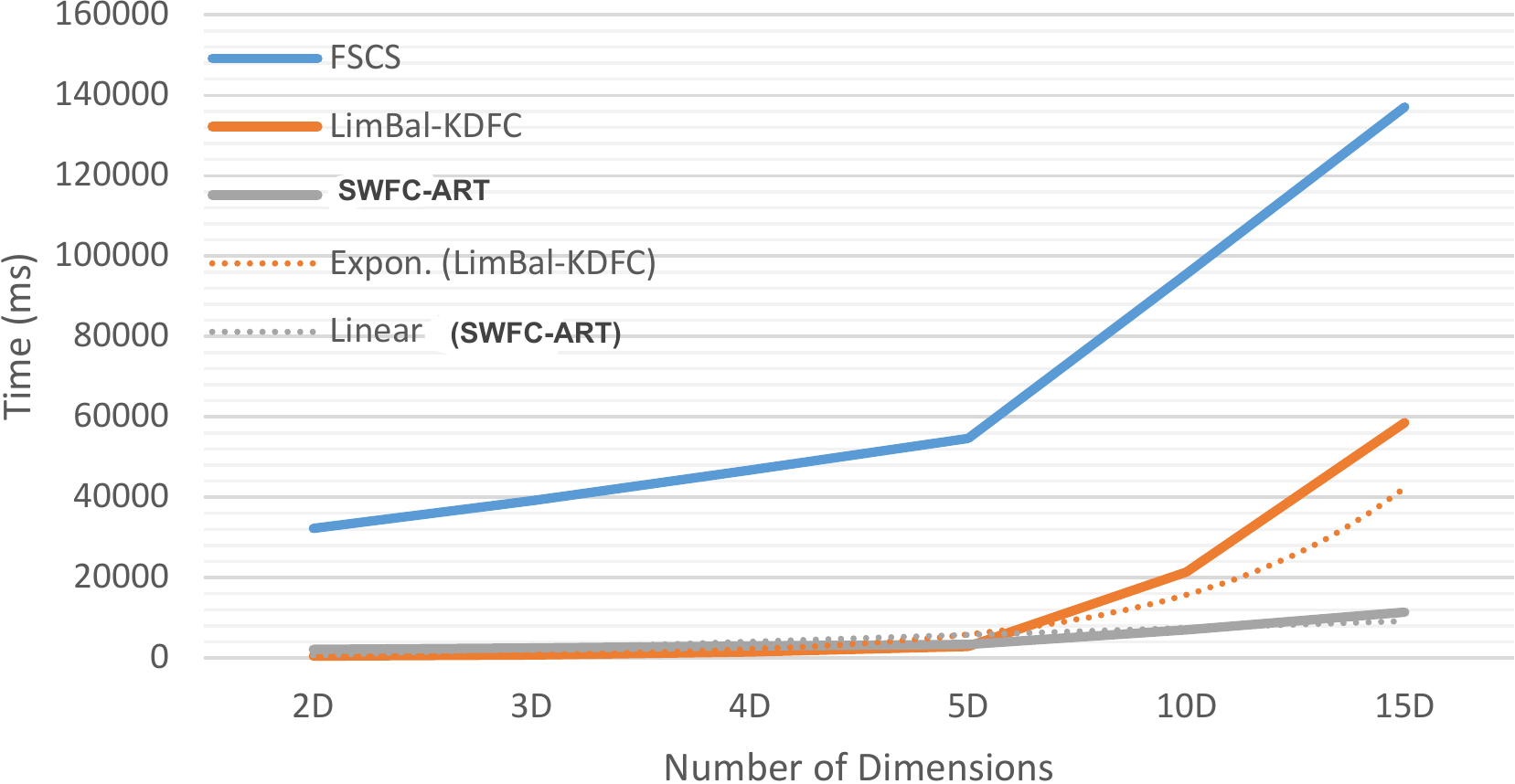}
    \label{efficiency_20000_line}}
    \caption{Time taken to generate a fixed number of test cases, in various dimensions}
    \label{efficiency_E}
\end{figure*}

Table \ref{tbl_efficiency} presents the efficiency results, with $d$ and $n$ denoting the dimensionality of the SUT input domain and the number of test cases generated, respectively.
All $p$-values are less than $0.05$, and all effect sizes are greater than $0.5$, which means that the SWFC-ART test case generation times are significantly different from those of FSCS-ART and LimbBal-KDFC, with large effect sizes.
\nrs{These results are next discussed from the perspective of the double-tier efficiency problem: }

\textit{Efficiency Tier-1 ($T_G$ trends when changing ``$n$'' within a fixed ``$d$''):}
\nrs{The overall trend for generating $n$ test cases, as shown in Fig.~\ref{efficiency}, is that the FSCS-ART time complexity grows in quadratic order while the complexities of SWFC-ART and LimBal-KDFC grow in log-linear order, $O\,(n\cdot \log n)$.}

For a 2-dimensional (2-$d$) input domain (Fig.~\ref{efficiency_2d}), FSCS-ART outperforms SWFC-ART when only a few test cases are generated
---
when $n \leq 500$, for example, FSCS-ART performs 7\% faster than SWFC-ART.
However, the advantage of SWFC-ART starts becoming apparent when larger numbers of tests are generated
---
when $n=20,000$, for example, SWFC-ART shows a 93\% improvement over FCSC-ART.
\nrs{Although both LimBal-KDFC and SWFC-ART have log-linear growth, LimBal-KDFC has a flatter slope, making it the most efficient method in 2-$d$ input domains.}

A similar trend to the 2-$d$ observations continues until $d=5$ (Fig.~\ref{efficiency_5d}), where, when $n > 500$, SWFC-ART is considerably more efficient than FSCS-ART, and LimBal-KDFC remains the most efficient method.
A close look at Figs.~\ref{efficiency_2d}, \ref{efficiency_3d}, \ref{efficiency_4d} and \ref{efficiency_5d}, however, shows that the $T_G$ of LimBal-KDFC starts to rise, and the gap between SWFC-ART and LimBal-KDFC decreases with the increasing dimensionality of the input domain
---
when generating 20,000 test cases, for example, the performance difference between SWFC-ART and LimBal-KDFC decreases from 78\% to 15\%, from 2-$d$ to 5-$d$ input domains.
When $d=5$, LimBal-KDFC and SWFC-ART appear to have the same performance
---
when $d > 5$ (Figs.~\ref{efficiency_10d} and \ref{efficiency_15d}), LimBal-KDFC, which had the best efficiency in low dimensions, is now outperformed by SWFC-ART.

Interestingly, when $d=10$ and $n \leq 1000$, FSCS-ART (with quadratic complexity) performs better than LimBal-KDFC.
However, SWFC-ART maintains its scalability and efficiency, and outperforms the other methods
---
at $d=10$, for example, when generating 20,000 test cases, SWFC-ART performs 92\% and 67\% faster than FSCS-ART and LimBal-KDFC, respectively.
When $d=15$, the LimBal-KDFC performance is worse than FSCS-ART until $n = 5000$, but the SWFC-ART performance remains consistent
---
when generating 20,000 test cases, SWFC-ART performs 91\% and 80\% faster than FSCS-ART and LimBal-KDFC, respectively.


In summary, the performance improvement of SWFC-ART over FSCS-ART is always consistent (greater than 90\% for $n$=20,000, irrespective of the input domain dimensionality).
LimBal-KDFC, however, shows inconsistency, including that the rate of improvement of SWFC-ART over LimBal-KDFC when $d>5$ is greater than that of LimBal-KDFC over SWFC-ART when $d\leq5$.

\textit{Efficiency Tier-2 ($T_G$ trends when changing ``$d$'' for fixed ``$n$''):}
Fig.~\ref{efficiency_E} shows the time taken to generate $n$ test cases in 2-, 3-, 4-, 5-, 10- and 15-dimensional input domains.
SWFC-ART is the worst performer when generating 500 test cases in a 2-$d$ input domain (Fig.~\ref{efficiency_500}).
As the dimensionality increases, the LimBal-KDFC's $T_G$ increases at a higher rate than the other two methods.
When generating 20,000 test cases (Fig.~\ref{efficiency_20000}), although FSCS-ART has the highest $T_G$, its rate of increase, as the dimensionality increases, is lower than than that of LimBal-KDFC.
Overall, although SWFC-ART has poor relative performance when generating a small number of test cases in low dimensional input spaces,
it gradually becomes faster and more consistent when the number of test cases and input domain dimensionality increase.

\nrs{In general, the time taken to generate $n$ test cases while moving from $d=2$ to $d=4$, by FSCS-ART, LimBal-KDFC, and SWFC-ART, increases $1.37$, $2.5$ and $1.32$ times, respectively;}
when moving from $d=5$ to $d=10$, they increase $1.68$, $5.68$ and $1.83$ times, respectively.
The time taken by LimBal-KDFC to generate $n$ test cases rises at the fastest rate of all three methods.
Fig.~\ref{efficiency_20000_line} summarizes the $T_G$ trends for $n=20,000$ for the three methods:
When $d \leq 5$, all methods show monotonous growth, but when $d > 5$, they all appear to encounter the \textit{curse of dimensionality}~\cite{Domingos2012, Bellman1957, Bellman1954}, with $T_G$ values increasing sharply for FSCS-ART and LimBal-KDFC, but SWFC-ART remaining consistent.

\subsubsection{Failure-detection effectiveness}
\label{sim-effectiveness}

\begin{table*}[!t]
    \centering
    \begin{threeparttable}
        \caption{F-ratios, Wilcoxon Rank-Sum Tests, and Effect Size Analyses for FSCS-ART, LimBal-KDFC and SWFC-ART for Block Failure Patterns}
        \label{tbl_effectiveness_block}
        \begin{tabular}{|c||c||c|c|c||c|c||c|c|}
            \hline
            \multirow{2}{*}{$d$} & \multirow{2}{*}{$\theta$} & \multicolumn{3}{c||}{F-ratio (\%)} & \multicolumn{2}{c||}{\makecell{FSCS-\\ART} vs \makecell{SWFC-\\ART}} & \multicolumn{2}{c|}{\makecell{LimBal-\\KDFC} vs \makecell{SWFC-\\ART}}
            \\ \cline{3-9}
             &  &  FSCS-ART & LimBal-KDFC & SWFC-ART & $p$-value & effect size &  $p$-value & effect size  \\ \hline\hline
            \multirow{7}{*}{2} & 0.0100 & 69.68 & \textbf{67.34} & 68.66 & 0.2724 & 0.0078 & 0.0236 & 0.0160 \\ \cline{2-9}
             & 0.0050 & \textbf{66.08} & 66.15 & 66.86 & 0.3867 & 0.0061 & 0.5118 & 0.0046 \\ \cline{2-9}
             & 0.0020 & \textbf{64.08} & 65.21 & 64.51 & 0.7632 & 0.0021 & 0.2016 & 0.0090 \\ \cline{2-9}
             & 0.0010 & \textbf{63.80} & 63.85 & 64.29 & 0.5442 & 0.0043 & 0.4965 & 0.0048 \\ \cline{2-9}
             & 0.0005 & 64.07 & 64.75 & \textbf{63.21} & 0.2354 & 0.0118 &  0.0444 & 0.0200 \\ \cline{2-9}
             & 0.0002 & 64.16 & \textbf{62.79} & 63.10 & 0.1556 & 0.0141 & 0.8199 & 0.0022 \\ \cline{2-9}
             & 0.0001 & \textbf{61.53} & 62.55 & 62.98 & 0.2057 & 0.0126 & 0.8351 &  0.0020 \\ \hline \hline
            \multirow{7}{*}{3} & 0.0100 & 85.75 &\textbf{ 83.65} & 85.54 & 0.6644 & 0.0031 & 0.0517 & 0.0138 \\ \cline{2-9}
             & 0.0050 & 80.99 & \textbf{80.18} & 81.82 & 0.5225 & 0.0045 & 0.0816 & 0.0123 \\ \cline{2-9}
             & 0.0020 & \textbf{76.97} & 77.65 & 77.49 & 0.7567 & 0.0022 & 0.3305 & 0.0069 \\ \cline{2-9}
             & 0.0010 & 75.46 & \textbf{74.79} & 75.88 & 0.2838 & 0.0076 & 0.0858 & 0.0121 \\ \cline{2-9}
             & 0.0005 & 73.73 & \textbf{73.53} & 73.88 & 0.4099 & 0.0058 & 0.8224 & 0.0016 \\ \cline{2-9}
             & 0.0002 & 72.72 & \textbf{71.45} & 72.27 & 0.9138 & 0.0008 & 0.2490 & 0.0082 \\ \cline{2-9}
             & 0.0001 & \textbf{71.36} & 73.24 & 71.88 & 0.8894 & 0.0014 & 0.2246 & 0.0121 \\ \hline \hline
            \multirow{7}{*}{4} & 0.0100 & 106.86 & \textbf{105.25 }& 106.30 & 0.2040 & 0.0090 & 0.8604 & 0.0012 \\ \cline{2-9}
             & 0.0050 & 100.79 & \textbf{98.86 }& 100.37 & 0.7294 & 0.0024 & 0.2443 & 0.0082 \\ \cline{2-9}
             & 0.0020 & 94.19 &\textbf{ 91.87} & 93.66 & 0.3998 & 0.0060 & 0.1958 & 0.0091 \\ \cline{2-9}
             & 0.0010 & 90.99 & \textbf{88.82} & 90.13 & 0.8465 & 0.0014 & 0.1374 & 0.0105 \\ \cline{2-9}
             & 0.0005 & 86.77 & \textbf{86.55} & 87.78 & 0.6803 & 0.0029 & 0.4614 & 0.0052 \\ \cline{2-9}
             & 0.0002 & 84.01 & \textbf{82.83 }& 84.11 & 0.1276 & 0.0196 & 0.6135 & 0.0051 \\ \cline{2-9}
             & 0.0001 & \textbf{80.34} & 82.00 & 83.44 & 0.0482 & 0.0255 & 0.1773 & 0.0135 \\ \hline \hline
            \multirow{7}{*}{5} & 0.0100 & 133.93 & \textbf{127.96 }& 129.21 & 0.0261 & 0.0157 & 0.8760 & 0.0011 \\ \cline{2-9}
             & 0.0050 & 125.75 & \textbf{118.50} & 122.12 & 0.0550 & 0.0136 & 0.0916 & 0.0119 \\ \cline{2-9}
             & 0.0020 & 116.38 & \textbf{109.38 }& 111.52 & 0.0038 & 0.0205 & 0.0816 & 0.0123 \\ \cline{2-9}
             & 0.0010 & 107.97 & \textbf{105.29} & 105.82 & 0.1404 & 0.0104 & 0.6252 & 0.0035 \\ \cline{2-9}
             & 0.0005 & 105.38 & \textbf{98.58} & 102.45 & 0.0187 & 0.0166 & 0.0124 & 0.0177 \\ \cline{2-9}
             & 0.0002 & 100.55 & \textbf{96.54} & 97.79 & 0.0146 & 0.0173 & 0.5037 & 0.0047 \\ \cline{2-9}
             & 0.0001 & 96.49 & \textbf{92.39} & 94.21 & 0.0886 & 0.0120 & 0.2567 & 0.0080 \\ \hline \hline
            \multirow{7}{*}{10} & 0.0100 & 405.86 & 392.27 & \textbf{350.56 }& 0.0000 & 0.0738 & 0.0000 & 0.0582 \\ \cline{2-9}
             & 0.0050 & 365.57 & 339.91 & \textbf{305.96} & 0.0000 & 0.0841 & 0.0000 & 0.0659 \\ \cline{2-9}
             & 0.0020 & 313.77 & 268.07 & \textbf{259.12} & 0.0000 & 0.0989 & 0.0011 & 0.0324 \\ \cline{2-9}
             & 0.0010 & 290.71 & 236.69 & \textbf{227.99} & \textbf{0.0000} & \textbf{0.1105} & 0.0000 & 0.0430 \\ \cline{2-9}
             & 0.0005 & 266.85 & \textbf{203.16} & 213.68 & 0.0000 & 0.0980 & 0.1833 & 0.0133 \\ \cline{2-9}
             & 0.0002 & 242.48 & \textbf{180.95} & 195.57 & 0.0000 & 0.0933 & 0.0032 & 0.0295 \\ \cline{2-9}
             & 0.0001 & 220.75 & \textbf{169.04} & 180.67 & \textbf{0.0030} & \textbf{0.1381} & 0.0064 & 0.0273 \\ \hline
        \end{tabular}
    \end{threeparttable}
\end{table*}

The simulation results for the block failure pattern are reported in Table \ref{tbl_effectiveness_block}.
For the 2-$d$ input domain, \textit{all} $p$-values are much greater than $0.05$ and all effect sizes are less than $0.1$, meaning that no significant difference exists between the F-ratios of FSCS-ART, LimBal-KDFC, and SWFC-ART.
A similar trend can be seen for the 3- and 4-$d$ input domains.
The trend is also present in the 5-$d$ input domain, except when $\theta=0.002$, for which the $p$-value for FSCS-ART and SWFC-ART is $0.0038$
---
although the effect size is is still less than $0.5$ ($0.0205$)
---
which means that there is insufficient evidence to conclude whether or not the F-ratios are different.
In the 10-$d$ input domain, the $p$-values are less than $0.05$, meaning that the F-ratios are significantly different, with SWFC-ART outperforming FSCS-ART.
However, the effect sizes are still far less than $0.05$, which means that even if the samples are different, there is still only a negligible effect on the F-ratios. 
In summary, there was insufficient evidence to reject $H_0$ and thus we conclude that the F-ratio results for all three methods are similar for the block failure pattern.


\begin{table*}[!t]
    \centering
    \begin{threeparttable}
        \caption{F-ratios, Wilcoxon Rank-Sum Tests, and Effect Size Analyses for FSCS-ART, LimBal-KDFC and SWFC-ART for Strip Failure Patterns}
        \label{tbl_effectiveness_strip}
        \begin{tabular}{|c||c||c|c|c||c|c||c|c|}
            \hline
            \multirow{2}{*}{$d$} & \multirow{2}{*}{$\theta$} & \multicolumn{3}{c||}{F-ratio (\%)} & \multicolumn{2}{c||}{\makecell{FSCS-\\ART} vs \makecell{SWFC-\\ART}} & \multicolumn{2}{c|}{\makecell{LimBal-\\KDFC} vs \makecell{SWFC-\\ART}}
            \\ \cline{3-9}
             &  &  FSCS-ART & LimBal-KDFC & SWFC-ART & $p$-value & effect size &  $p$-value & effect size  \\ \hline\hline
            \multirow{7}{*}{2} & 0.0100 & 93.29 & 91.90 & \textbf{91.59} & 0.1827 & 0.0094 & 0.6709 & 0.0030 \\ \cline{2-9}
             & 0.0050 & 94.48 & \textbf{94.25} & 94.37 & 0.4042 & 0.0059 & 0.7006 & 0.0027 \\ \cline{2-9}
             & 0.0020 & 97.85 &\textbf{ 96.14 }& 97.76 & 0.6088 & 0.0036 & 0.0933 & 0.0119 \\ \cline{2-9}
             & 0.0010 & 98.25 & 99.85 & \textbf{96.85} & 0.4100 & 0.0058 & 0.0296 & 0.0154 \\ \cline{2-9}
             & 0.0005 & 98.86 & 98.34 & \textbf{95.36} & 0.1472 & 0.0144 & 0.5097 & 0.0065 \\ \cline{2-9}
             & 0.0002 & 97.62 & \textbf{95.91} & 101.04 & 0.4311 &  0.0078 & 0.2896 & 0.0105 \\ \cline{2-9}
             & 0.0001 & 100.38 & 101.53 & \textbf{98.21} & 0.4917 & 0.0068 & 0.1760 & 0.0135 \\ \hline\hline
            \multirow{7}{*}{3} & 0.0100 & 97.01 & 97.84 & \textbf{96.53} & 0.5924 & 0.0038 & 0.7784 & 0.0020 \\ \cline{2-9}
             & 0.0050 & \textbf{98.09} & 98.65 & 98.96 & 0.6399 & 0.0033 & 0.7847 & 0.0019 \\ \cline{2-9}
             & 0.0020 & 99.63 & 99.64 & \textbf{98.83 }& 0.9175 & 0.0007 & 0.8003 & 0.0018 \\ \cline{2-9}
             & 0.0010 & 99.94 & \textbf{98.92} & 99.45 & 0.5357 & 0.0044 & 0.5505 & 0.0042 \\ \cline{2-9}
             & 0.0005 & \textbf{98.99} & 99.93 & 99.92 & 0.5154 & 0.0046 & 0.4547 & 0.0053 \\ \cline{2-9}
             & 0.0002 & 99.23 & 100.86 & \textbf{98.82} & 0.6649 & 0.0031 & 0.0176 & 0.0168 \\ \cline{2-9}
             & 0.0001 & 101.24 & 101.31 & \textbf{101.16} & 0.8457 & 0.0019 & 0.5195 & 0.0064 \\ \hline\hline
            \multirow{7}{*}{4} & 0.0100 & 99.74 & \textbf{98.96} & 101.09 & 0.1633 & 0.0099 & 0.0233 & 0.0160 \\ \cline{2-9}
             & 0.0050 & 99.62 &\textbf{ 99.23} & 99.86 & 0.9521 & 0.0004 & 0.2178 & 0.0087 \\ \cline{2-9}
             & 0.0020 &\textbf{ 98.52} & 100.58 & 100.78 & 0.0752 & 0.0126 & 0.3637 & 0.0064 \\ \cline{2-9}
             & 0.0010 & \textbf{98.40} & 99.93 & 99.99 & 0.7849 & 0.0019 & 0.8062 & 0.0017 \\ \cline{2-9}
             & 0.0005 &\textbf{ 99.87} & 100.22 & 99.91 & 0.4659 & 0.0052 & 0.6741 & 0.0030 \\ \cline{2-9}
             & 0.0002 & \textbf{98.68} & 101.92 & 98.98 & 0.4159 & 0.0105 & 0.0936 & 0.0168 \\ \cline{2-9}
             & 0.0001 & 101.15 & 99.39 & \textbf{97.10} & 0.1709 & 0.0176 & 0.0658 & 0.0184 \\ \hline\hline
            \multirow{7}{*}{5} & 0.0100 & 100.48 & \textbf{98.75} & 101.88 & 0.2224 & 0.0086 & 0.0512 & 0.0138 \\ \cline{2-9}
             & 0.0050 & \textbf{99.68} & 99.87 & 99.77 & 0.8492 & 0.0013 & 0.4507 & 0.0053 \\ \cline{2-9}
             & 0.0020 & 99.41 & \textbf{98.36} & 100.22 & 0.4515 & 0.0053 & 0.0414 & 0.0144 \\ \cline{2-9}
             & 0.0010 & \textbf{98.66} & 100.89 & 100.55 & 0.3531 & 0.0066 & 0.9837 & 0.0001 \\ \cline{2-9}
             & 0.0005 & 100.96 & 99.71 & \textbf{98.77} & 0.2659 & 0.0079 & 0.7164 & 0.0026 \\ \cline{2-9}
             & 0.0002 & 100.02 & 99.17 & \textbf{99.01} & 0.3432 & 0.0067 & 0.9036 & 0.0009 \\ \cline{2-9}
             & 0.0001 & 99.26 & \textbf{98.57} & 100.08 & 0.8683 & 0.0017 & 0.3254 & 0.0098 \\ \hline\hline
            \multirow{7}{*}{10} & 0.0100 & \textbf{99.26} & 101.28 & 103.13 & 0.2442 & 0.0116 & 0.0966 & 0.0166 \\ \cline{2-9}
             & 0.0050 & 102.04 & \textbf{101.12} & 101.21 & 0.9038 & 0.0012 & 0.1216 & 0.0154 \\ \cline{2-9}
             & 0.0020 & 99.38 & \textbf{97.64} & 100.69 & 0.6301 & 0.0048 & 0.3151 & 0.0100 \\ \cline{2-9}
             & 0.0010 & \textbf{94.90} & 101.24 & 96.10 & 0.2681 & 0.0143 & 0.0621 & 0.0187 \\ \cline{2-9}
             & 0.0005 & 100.31 & 101.25 & \textbf{100.19} & 0.7399 & 0.0033 & 0.3629 & 0.0090 \\ \cline{2-9}
             & 0.0002 & \textbf{100.52} & 100.60 & 101.03 & 0.2239 & 0.0157 & 0.9277 & 0.0009 \\ \cline{2-9}
             & 0.0001 & 101.70 & 99.68 & \textbf{99.24} & 0.6137 & 0.0331 & 0.6475 & 0.0046 \\ \hline
        \end{tabular}%
    \end{threeparttable}
\end{table*}

Table \ref{tbl_effectiveness_strip} shows the strip failure pattern simulation results.
For all dimensions, the $p$-values are greater than $0.05$ and the effect sizes are less than $0.1$, which again means that there is insufficient evidence to reject the null hypothesis.

\begin{table*}[!t]
    \centering
    \begin{threeparttable}
        \caption{F-ratios, Wilcoxon Rank-Sum Tests, and Effect Size Analyses for FSCS-ART, LimBal-KDFC and SWFC-ART for Point Failure Patterns}\label{tbl_effectiveness_point}
        \begin{tabular}{|c||c||c|c|c||c|c||c|c|}
            \hline
            \multirow{2}{*}{$d$} & \multirow{2}{*}{$\theta$} & \multicolumn{3}{c||}{F-ratio (\%)} & \multicolumn{2}{c||}{\makecell{FSCS-\\ART} vs \makecell{SWFC-\\ART}} & \multicolumn{2}{c|}{\makecell{LimBal-\\KDFC} vs \makecell{SWFC-\\ART}}
            \\ \cline{3-9}
             &  &  FSCS-ART & LimBal-KDFC & SWFC-ART & $p$-value & effect size &  $p$-value & effect size  \\ \hline\hline
            \multirow{7}{*}{2} & 0.0100 & \textbf{99.18} & 99.62 & 103.04 & 0.0054 & 0.0197 & 0.0280 & 0.0155 \\ \cline{2-9}
             & 0.0050 & 100.10 & 99.44 & \textbf{99.19 }& 0.5853 & 0.0039 & 0.8042 & 0.0017 \\ \cline{2-9}
             & 0.0020 & \textbf{96.31} & 98.37 & 97.68 & 0.3614 & 0.0065 & 0.8473 & 0.0013 \\ \cline{2-9}
             & 0.0010 & 97.79 & 98.38 & \textbf{97.75} & 0.4763 & 0.0050 & 0.5360 & 0.0043 \\ \cline{2-9}
             & 0.0005 & 98.31 & 96.50 & \textbf{96.46 }& 0.6352 & 0.0047 &  0.7385 & 0.0033 \\ \cline{2-9}
             & 0.0002 & 94.88 & 95.60 & \textbf{94.41} & 0.7026 & 0.0038 & 0.4861 &  0.0069 \\ \cline{2-9}
             & 0.0001 & \textbf{95.67} & 97.01 & 96.10 & 0.9466 & 0.0006 & 0.4030 & 0.0083 \\ \hline\hline
            \multirow{7}{*}{3} & 0.0100 & 112.00 & 112.25 & \textbf{111.65} & 0.6711 & 0.0030 & 0.5859 & 0.0039 \\ \cline{2-9}
             & 0.0050 & \textbf{106.91} & 107.81 & 108.71 & 0.4195 & 0.0057 & 0.9024 & 0.0009 \\ \cline{2-9}
             & 0.0020 & 105.68 & 104.67 & \textbf{104.63} & 0.3997 & 0.0060 & 0.8001 & 0.0018 \\ \cline{2-9}
             & 0.0010 & 102.32 & 102.37 & \textbf{102.19} & 0.6468 & 0.0032 & 0.8468 & 0.0014 \\ \cline{2-9}
             & 0.0005 & 101.63 & 101.58 & \textbf{100.29 }& 0.3700 & 0.0063 & 0.5001 & 0.0048 \\ \cline{2-9}
             & 0.0002 & \textbf{99.26} & 101.21 & 99.65 & 0.6180 & 0.0035 & 0.4311 & 0.0056 \\ \cline{2-9}
             & 0.0001 & 98.36 & \textbf{97.88} & 99.24 & 0.6133 & 0.0065 & 0.6882 & 0.0040 \\ \hline\hline
            \multirow{7}{*}{4} & 0.0100 & 129.63 & \textbf{127.25} & 128.92 & 0.5684 & 0.0040 & 0.1229 & 0.0109 \\ \cline{2-9}
             & 0.0050 & 125.38 & 124.26 & \textbf{123.47} & 0.8625 & 0.0012 & 0.8929 & 0.0010 \\ \cline{2-9}
             & 0.0020 & 117.24 & \textbf{116.33} & 116.41 & 0.5400 & 0.0043 & 0.9631 & 0.0003 \\ \cline{2-9}
             & 0.0010 & 114.61 & \textbf{113.50} & 113.83 & 0.8201 & 0.0016 & 0.5857 & 0.0039 \\ \cline{2-9}
             & 0.0005 &\textbf{ 107.00} & 108.62 & 110.34 & 0.4013 & 0.0108 & 0.6633 & 0.0044 \\ \cline{2-9}
             & 0.0002 & 107.79 & 105.65 & \textbf{105.60} & 0.4562 & 0.0096 & 0.6999 & 0.0039 \\ \cline{2-9}
             & 0.0001 & \textbf{105.59} & 106.37 & 106.20 & 0.6547 & 0.0057 & 0.7695 & 0.0029 \\ \hline\hline
            \multirow{7}{*}{5} & 0.0100 & 153.57 & 150.38 & \textbf{149.39 }& 0.0383 & 0.0146 & 0.2678 & 0.0078 \\ \cline{2-9}
             & 0.0050 & 145.70 & 141.07 & \textbf{140.73} & 0.0062 & 0.0193 & 0.8661 & 0.0012 \\ \cline{2-9}
             & 0.0020 & 134.72 & \textbf{130.30} & 130.71 & 0.0444 & 0.0142 & 0.7011 & 0.0027 \\ \cline{2-9}
             & 0.0010 & 129.35 & 126.02 & \textbf{125.73 }& 0.1885 & 0.0093 & 0.5049 & 0.0047 \\ \cline{2-9}
             & 0.0005 & 124.58 &\textbf{ 121.15} & 122.53 & 0.0752 & 0.0126 & 0.5975 & 0.0037 \\ \cline{2-9}
             & 0.0002 & 119.99 & \textbf{114.77 }& 119.38 & 0.2236 & 0.0086 & 0.0620 & 0.0132 \\ \cline{2-9}
             & 0.0001 & 115.45 & \textbf{111.16} & 114.53 & 0.6723 & 0.0054 & 0.0952 & 0.0167 \\ \hline\hline
            \multirow{7}{*}{10} & 0.0100 & 252.49 & 247.17 & \textbf{237.79} & 0.7246 & 0.0045 & 0.3984 & 0.0084 \\ \cline{2-9}
             & 0.0050 & 278.97 & 271.20 & \textbf{244.12} & 0.0000 & 0.0586 & 0.0000 & 0.0494 \\ \cline{2-9}
             & 0.0020 & 292.82 & 265.93 & \textbf{240.21} & 0.0000 & 0.0852 & 0.0000 & 0.0641 \\ \cline{2-9}
             & 0.0010 & 291.09 & 244.76 & \textbf{236.16} & 0.0000 & 0.0974 & 0.0038 & 0.0289 \\ \cline{2-9}
             & 0.0005 & 272.80 & 233.53 &\textbf{ 227.46} & 0.0000 & 0.0840 & 0.0441 & 0.0201 \\ \cline{2-9}
             & 0.0002 & 242.80 & \textbf{206.08 }& 209.57 & \textbf{0.0000 }& \textbf{0.1109} & 0.6922 & 0.0040 \\ \cline{2-9}
             & 0.0001 & 236.35 & \textbf{191.35 }& 192.74 & 0.0000 & 0.0194 & 0.5468 & 0.0060 \\ \hline
        \end{tabular}%
    \end{threeparttable}
\end{table*}

The point failure pattern simulation results, shown in Table \ref{tbl_effectiveness_point}, present similar trends to those seen in the block pattern results.
In low dimensional input domains ($d\leq5$), the $p$-values and effect sizes show that the F-ratios of all three methods are similar.
In the 10-$d$ input domain, the $p$-values show a significant difference, especially between SWFC-ART and FSCS-ART.
\nrs{The mean SWFC-ART F-ratios are better than those of FSCS-ART, and significantly better than LimBal-KDFC when $\theta$=$0.01$, $0.005$ and $0.001$;}
LimBal-KDFC outperforms SWFC-ART when $\theta$=$0.0002$ and $0.0001$.
However, the effect size values are not large enough to allow strong conclusions to be drawn.

In summary, we can conclude that the F-ratios of all three methods are similar, for all the failure rates, failure patterns, and input dimensions under study.
\nrs{Furthermore, the ANNS strategies (LimBal-KDFC and SWFC-ART) performed better than the exact NNS of FSCS-ART in high dimensional input spaces. }
The ANNS strategies employed by LimBal-KDFC and SWFC-ART were also significantly different from each other in high dimensions, while showing similar failure-detection effectiveness.

\subsubsection{Test Case Distribution}
\label{sim-tc-distribution}

\begin{table*}[!t]
    \centering
    \caption{Discrepancy in Various Dimensions}
    \label{tbl_discrepancy}
    \begin{tabular}{|c|c|c|c|c|c|c|c|c|}
        \hline
        \multirow{2}{*}{Test Cases} & \multirow{2}{*}{Method} & \multicolumn{7}{c|}{Discrepancy} \\ \cline{3-9}
         &  & 1-$d$ & 2-$d$ & 3-$d$ & 4-$d$ & 5-$d$ & 10-$d$ & 15-$d$ \\ \hline\hline
        \multirow{3}{*}{100} & FSCS-ART & 0.0750 & 0.1393 & \textbf{0.2159} & 0.2697 & \textbf{0.3112} & 0.3135 & 0.3108 \\ \cline{2-9}
         & LimBal-KDFC & 0.0628 &\textbf{ 0.1295} & 0.2250 & 0.2856 & 0.3138 & 0.3070 & 0.2886 \\ \cline{2-9}
         & SWFC-ART & \textbf{0.0562} & 0.1359 & 0.2420 & \textbf{0.2652} & 0.3206 & \textbf{0.2942} & \textbf{0.2759} \\ \hline\hline
        \multirow{3}{*}{1000} & FSCS-ART & 0.0181 & 0.0381 & 0.1036 & 0.1620 & 0.1899 & 0.2228 & 0.2090 \\ \cline{2-9}
         & LimBal-KDFC & \textbf{0.0163} & \textbf{0.0347} & 0.0961 & 0.1592 & 0.1884 & 0.2140 &\textbf{ 0.2055} \\ \cline{2-9}
         & SWFC-ART & 0.0168 & 0.0355 & \textbf{0.0930} & \textbf{0.1515} & \textbf{0.1739} & \textbf{0.2010} & 0.2106 \\ \hline\hline
        \multirow{3}{*}{10,000} & FSCS-ART & \textbf{0.0084} & 0.0163 & 0.0574 & 0.1091 & 0.1534 & 0.1987 & 0.1889 \\ \cline{2-9}
         & LimBal-KDFC & 0.0085 & \textbf{0.0133} & 0.0560 & 0.1098 & \textbf{0.1385} & 0.1883 & 0.1859 \\ \cline{2-9}
         & SWFC-ART & 0.0089 & 0.0151 & \textbf{0.0525} & \textbf{0.1033} & 0.1388 & \textbf{0.1795} & \textbf{0.1623} \\ \hline
    \end{tabular}
\end{table*}

Table \ref{tbl_discrepancy} shows the discrepancy values for 100, 1000 and 10,000 test cases, for all three methods.
Three important trends can be observed:
\begin{enumerate}
    \item
        When generating a specific number of test cases in a particular dimension, there is no significant difference in the discrepancy values for all three methods.
        \nrs{When there is a difference, the ANNS methods (SWFC-ART and LimBal-KDFC) are usually better than the exact NNS method (FSCS-ART), with FSCS-ART only having better discrepancy values three out of $63$ times.}
        Hence, it can be concluded that the test case distribution of methods employing ANNS is at least equal to the exact NNS methods.

    \item
        \nrs{All test case generation strategies appear to display degradation in distribution with increasing dimensionality of the input domain: }
        The SWFC-ART discrepancy values, for example, increase from $0.0562$ to $0.2759$ from 1-$d$ to 15-$d$, for 100 test cases.
        \nrs{However, SWFC-ART appears least affected by the \textit{curse of dimensionality}~\cite{Domingos2012, Bellman1957, Bellman1954}.}


        To generate 100 test cases, SWFC-ART shows better discrepancy values in 1, 4, 10, and 15 dimensions;
        \nrs{for 1000 test cases, SWFC-ART has better discrepancy in 3, 4, 5, and 10 dimensions; and}
        for 10,000 test cases, SWFC-ART has better discrepancy in 3, 4, 10, and 15 dimensions.

    \item
        All methods showed better discrepancy values as the number of generated test cases increased.
\end{enumerate}

\subsection{Experiments with Real-life Programs}
\label{empirical-studies}

\subsubsection{Computational Efficiency}
\label{emp-efficiency}

\begin{table*}[!t]
    \caption{Test Case Generation Time and Execution Time for Detecting Failures in the 23 Subject Programs}
    \label{tbl_emp_efficiency}
    \centering
    \begin{tabular}{|l||c||ccc||ccc|}
        \hline
        \multirow{2}{*}{Program} &
        \multirow{3}{*}{$d$} &
        \multicolumn{3}{c||}{$T_G$ (ms)} & \multicolumn{3}{c|}{$T_E$ (ms)} \\
         & & FSCS-ART & \makecell{LimBal-\\KDFC}  & \makecell{SWFC-\\ART} & FSCS-ART & \makecell{LimBal-\\KDFC}  & \makecell{SWFC-\\ART} \\ \hline \hline
        bessj0 & 1 & 22.14 & \textbf{4.16} & 20.33 & 1.47 & 1.36 & 1.45 \\
        airy  & 1 & 86.78 & 10.01 & \textbf{2.23} & 2.32 & 1.96 & 2.23 \\
        asinh   &  1 &  3.8e+9 &  \textbf{9.5e+7} &  4.6e+8 & 3e+6 &  2e+6 &  2e+6 \\
        erfcc  & 1 & 129.65 & \textbf{12.08} & 57.40 & 4.69 & 3.99 & 4.43 \\
        probks  & 1 & 270.41 & \textbf{19.87} & 90.66 & 78.41 & 76.93 & 78.64 \\
        tanh  & 1 & 13.59 & \textbf{3.61 }& 18.22 & 1.44 & 1.33 & 1.67 \\
        bessj  & 2 & 34.61 & \textbf{5.56} & 24.24 & 3.70 & 3.60 & 3.76 \\
        gammq  & 2 & 229.95 & \textbf{16.60 }& 71.80 & 5.26 & 4.21 & 4.72 \\
        Sncndn  & 2 & 87.48 & \textbf{10.11} & 42.29 & 2.86 & 2.28 & 2.63 \\
        binomial   & 2 &  4.1e+9 &  \textbf{1.3e+8} &  4.9e+8 &  1.4e+7 &  1.4e+7 &  1.4e+7 \\
        plgndr  & 3 & 651.81 &\textbf{ 31.86} & 136.44 & 21.34 & 18.11 & 19.14 \\
        golden & 3  & 784.13 & \textbf{51.98} & 160.40 & 25.43 & 23.47 & 24.27 \\
        cel & 4  & 702.66 & \textbf{39.14} & 139.56 & 9.63 & 7.41 & 8.09 \\
        el2 & 4  & 157.08 & \textbf{29.26 }& 66.21 & 4.48 & 3.91 & 4.28 \\
        period   & 4 &  3.9e+9 &  \textbf{1.3e+8} &  4.5e+8 &  1.0e+7 &  1.0e+7 &  1.0e+7 \\
        calDay & 5  & 496.08 & \textbf{51.18} & 120.65 & 10.55 & 10.24 & 10.68 \\
        complex  & 6 & 597.86 & \textbf{150.12} & 155.63 & 0.75 & 0.61 & 0.62 \\
        pntLinePos & 6  & 968.20 & \textbf{206.68} & 209.48 & 0.40 & 0.24 & 0.27 \\
        triangle  & 6 & 774.64 & \textbf{179.19} & 191.10 & 0.86 & 0.52 & 0.52 \\
        line  & 8 & 5097.56 & 1311.34 & \textbf{656.25} & 0.75 & 0.52 & 0.48 \\
        pntTrianglePos & 8 & 15699.36 & 2449.96 & \textbf{1311.27} & 6.94 & 4.16 & 3.82 \\
        twoLinesPos  & 8 & 42078.38 & 4656.06 & \textbf{2403.85} & 8.24 & 4.31 & 4.08 \\
        nearestDistance  & 10 & 668.51 & 552.93 & \textbf{232.98} & 0.66 & 0.61 & 0.56 \\
        calGCD  & 10 & 522.59 & 440.67 &\textbf{ 179.26} & 3.29 & 3.33 & 3.31 \\
        select  & 11 & 4508.39 & 2333.18 & \textbf{744.19} & 6.25 & 5.75 & 5.01 \\
        tcas  & 12 & 2310.45 & \textbf{92.20} & 348.51 & 1.46 & 0.95 & 0.90 \\
        matrixProcessor   & 12 &  1.5e+10 &  7.0e+9 &  \textbf{1.8e+9} &  8.0e+7 &  8.0e+7 &  7.0e+7 \\
        java.util.Arrays  & 15 &  1.9e+10 &  7.6e+9 &  \textbf{2.4e+9} &  2.0e+8 &  2.0e+8 &  2.0e+8 \\
        \hline
    \end{tabular}
\end{table*}

The time taken to test a program is divided into two parts: the test case generation time ($T_G$) and the execution time ($T_E$).
When testing the real-life programs, test cases were incrementally generated and executed until a failure was revealed.

As can be seen in Table \ref{tbl_emp_efficiency}, the $T_E$ values for all the programs under study were far less than the $T_G$ values, with the $T_E$ results for all methods being comparable.
\nrs{$T_G$ is, therefore, the main time cost for FSCS-ART, and any reduction in $T_G$ should have a positive impact on FSCS-ART efficiency, especially when $T_E < T_G$.  }

SWFC-ART has significantly lower $T_G$ results than FSCS-ART for all the programs except \textit{tanh}, with results for the programs \textit{airy}, \textit{period},  \textit{pntTrianglePos}, and \textit{twoLinesPos} being particularly dramatic (reductions of about 90\%).
SWFC-ART outperforms FSCS-ART by 80-90\% for the programs  \textit{asinh}, \textit{binomial}, \textit{matrixProcessor}, \textit{java.util.Arrays}, \textit{line}, \textit{tcas}, \textit{select}, and \textit{cel};
and 70-80\% for \textit{complex}, \textit{triangle}, \textit{CalDay}, \textit{pntLinePos}, \textit{plgndr}, and \textit{golden}.
Overall, SWFC-ART reduced the FSCS-ART $T_G$ by more than 50\% for \textit{all} the programs under study except \textit{bessj}, \textit{bessj0} and \textit{tanh}
---
where the performance improvement was 29\%, 8\% and -34\%, respectively.


SWFC-ART performs a little worse than  LimBal-KDFC in (programs with) low dimensions, with five or less input parameters, but has better performance in the high dimensional programs ($d>5$)
---
\textit{matrixProcessor, \textit{java.util.Arrays}}, \textit{select}, \textit{calGCD}, \textit{nearestDistance}, \textit{line}, \textit{twoLinesPos}, and \textit{PntTrianglePos} where SWFC-ART outperforms LimBalKDFC by 73\%, 68\%, 68\%, 59\%, 57\%, 49\%, 48\%, and 46\%, respectively.
SWFC-ART can also sometimes outperform LimBal-KDFC in low dimensional programs (\textit{airy});
and
LimBal-KDFC can sometimes perform better than SWFC-ART in some high dimensional programs (\textit{triangle}, and \textit{tcas}). For \textit{complex} and \textit{pntLinePos}, both methods show similar $T_G$.


\nrs{In summary, the results of the experimental studies with real-life programs are consistent with the simulations:
Both LimBal-KDFC and SWFC-ART outperform FSCS-ART in terms of computational efficiency, and SWFC-ART remains consistent in high dimensional programs.}

\subsubsection{Failure-detection effectiveness}
\label{emp-effectiveness}

\begin{table*}[!t]
\centering
 \caption{Wilcoxon Rank-Sum Test and Effect Size Analyses of F-measure of FSCS-ART, LimBal-KDFC, and SWFC-ART for Subject Programs}
\label{tbl_programs_effectiveness}
\begin{tabular}{|l||c||ccc||cc||cc|}
\hline
\multirow{3}{*}{Program} &
\multirow{3}{*}{$d$} &
\multicolumn{3}{c||}{F-measure} &
\multicolumn{2}{c||}{\makecell{FSCS-ART vs \\ SWFC-ART}} &
\multicolumn{2}{c|}{\makecell{LimBal-\\KDFC} vs \makecell{SWFC-\\ART}}
\\\cline{3-9}
 &  & \makecell{FSCS-\\ART} & \makecell{LimBal-\\KDFC} & \makecell{SWFC-\\ART} & $p$-value & effect size & $p$-value & effect size \\ \hline
bessj0 & 1 & 444.46 & 448.53 & \textbf{440.48} & 0.5161 & 0.0065 & 0.2342 & 0.0119 \\
airy & 1  & \textbf{789.54} & 806.50 & 807.18 & 0.1045 & 0.0162 & 0.6903 & 0.0040 \\
asinh  & 1 & 5689 & 5612 & \textbf{5583} & 0.0773 & 0.0177 & 0.5439 & 0.0061 \\
erfcc & 1  & 1037.81 & \textbf{1024.31} & 1032.24 & 0.3863 & 0.0087 & 0.8743 & 0.0016 \\
probks & 1  & \textbf{1453.62} & 1460.71 & 1456.30 & 0.9354 & 0.0008 & 0.7251 & 0.0035 \\
tanh & 1  & 313.05 & 311.31 & \textbf{310.70} & 0.5812 & 0.0055 & 0.7955 & 0.0026 \\
bessj & 2  & 454.54 & \textbf{440.02} & 442.08 & 0.1974 & 0.0129 & 0.3242 & 0.0099 \\
gammq  & 2 & 1086.95 & \textbf{1045.17} & 1097.29 & 0.9064 & 0.0012 & 0.1866 & 0.0132 \\
sncndn  & 2 & 631.45 & 631.47 &\textbf{ 631.23} & 0.4489 & 0.0076 & 0.8246 & 0.0022 \\
binomial  & 2 & \textbf{5089} & 5243 & 5166 & 0.6713  & 0.0042 & 0.5927 & 0.0053 \\
plgndr & 3  & 1618.00 & \textbf{1606.05} & 1608.47 & 0.7092 & 0.0037 & 0.8469 & 0.0019 \\
golden & 3  & \textbf{1802.60} & 1808.43 & 1804.03 & 0.8994 & 0.0013 & 0.8369 & 0.0021 \\
cel & 4  & 1547.32 & 1571.71 & \textbf{1542.93} & 0.6888 & 0.0040 & 0.1664 & 0.0138 \\
el2 & 4  & \textbf{721.55 }& 724.94 & 728.06 & 0.8781 & 0.0015 & 0.3901 & 0.0086 \\
period   & 4 & 29794 & 30307 & \textbf{29765} & 0.5266 & 0.0091 & 0.0321 & 0.0309 \\
calDay  & 5 & \textbf{1259.38} & 1314.12 & 1280.12 & 0.8884 & 0.0014 & 0.0768 & 0.0177 \\
complex  & 6 & 1223.95 & 1214.50 & \textbf{1195.77} & 0.2619 & 0.0112 & 0.2391 & 0.0118 \\
pntLinePos  & 6 & 1503.62 & \textbf{1462.88} & 1477.83 & 0.9587 & 0.0005 & 0.3696 & 0.0090 \\
triangle  & 6 & \textbf{1350.33} & 1389.29 & 1379.52 & 0.0783 & 0.0176 & 0.6655 & 0.0043 \\
line  & 8 & 3370.27 & \textbf{3326.32} & 3385.39 & 0.3786 & 0.0088 & 0.3627 & 0.0091 \\
pntTrianglePos & 8 & 4713.73 & \textbf{4238.11} & 4657.62 & 0.4050 & 0.0107 & 0.0593 & 0.0242 \\
twoLinesPos  & 8 & 8177.49 & \textbf{7009.96} & 7613.29 & 0.0199 & 0.0301 & 0.0298 & 0.0280 \\
nearestDistance & 10  &\textbf{ 1934.32} & 2015.68 & 2065.56 & 0.0716 & 0.0233 & 0.3113 & 0.0131 \\
calGCD & 10  & 1035.47 & 1023.50 & \textbf{1003.32} & 0.4538 & 0.0097 & 0.6892 & 0.0052 \\
select  & 11 & 5583.22 & \textbf{5174.07} & 5432.77 & 0.7461 & 0.0042 & 0.0506 & 0.0252 \\
tcas  & 12 & 1681.02 & \textbf{1642.95} & 1736.33 & 0.3609 & 0.0118 & 0.1526 & 0.0185 \\
matrixProcessor  & 12 & 5003 & \textbf{4978} & 5152 & 0.2663 & 0.0111 & 0.4101 & 0.0082 \\
java.util.Arrays  & 15 & 10108 & \textbf{9866} & 10313 & 0.0994 & 0.0164 & 0.0198 & 0.0233 \\
\hline
\end{tabular}
\end{table*}


Table \ref{tbl_programs_effectiveness} presents the F-measure effectiveness results, showing that, for the 28 programs, FSCS-ART, LimBal-KDFC, and SWFC-ART have the best results in 8, 12, and 8 programs, respectively.

Comparing FSCS-ART and SWFC-ART:
The Wilcoxon rank-sum tests for the FSCS-ART and SWFC-ART F-measure data have $p$-values greater than $0.05$ for all subject programs, except \textit{twoLinesPos}, with extremely low effect sizes
---
although the $p$-value for \textit{twoLinesPos} is $0.0199$, the effect size is much less than $0.1$.
Moreover, for \textit{twoLinesPos}, the mean F-measure of SWFC-ART is $7613.96$ while that for FSCS-ART is $8177.49$.  The F-measure results for \textit{sncndn} are very similar for both approaches;
for 13 programs
(\textit{airy, probks, gammq, binomial, golden, el2, calDay, triangle, line, nearestDistance, tcas, matrixProcessor}, and \textit{java.util.Arrays}),
FSCS-ART has better results than SWFC-ART; and
for the remaining 15 programs
(\textit{bessj0, asinh, erfcc, tanh, bessj, plgndr, cel, period, complex, pntLinePos, pntTrianglePos, twoLinesPos, calGCD} and \textit{select}),
SWFC-ART has the better performance.
However, the $p$-value and effect size analyses are such that it is not statistically clear that either method does actually perform better than the  other.

Comparing LimBal-KDFC and SWFC-ART: The $p$-values for the comparisons of the LimBal-KDFC and SWFC-ART F-measure results are all greater than $0.05$ (except for  \textit{twoLinesPos} and \textit{java.util.Arrays}), and all effect sizes are much less than $0.1$.
This means that the failure-detection effectiveness of both methods is similar.
For \textit{twoLinesPos} and \textit{java.util.Arrays}, although the $p$-value is significant, the effect size is not large enough for any conclusion to be drawn.


\nrs{In summary, the results from the studies with real-life programs again align with those of the simulations:
There was insufficient evidence to reject the null hypothesis, and thus we conclude that there are no significant differences among the observed F-measures of all the methods
---
the failure-finding effectiveness of all methods is similar.}

\subsection{Discussion}
\label{discussion}

In this section, we summarize our results by giving answers to the research questions.
We also discuss the asymptotic complexities of the methods.

\subsubsection{RQ1}

\textit{Efficiency Tier-1 (Scalability):}
\nrs{Our studies have shown that, as the the number of executed tests increases, FSCS-ART incurs considerable time overheads, but LimBal-KDFC and SWFC-ART perform significantly better.}
This can also be understood through a theoretical analysis of the complexities of the algorithms:
The FSCS-ART time complexity is in quadratic relation with $n$, but both LimBal-KDFC and SWFC-ART are in a log-linear relation.
Because real-life software can typically have very low failure rates~\cite{Arcuri2011}, it may be necessary to generate and execute many test cases  before finding a first failure
---
LimBal-KDFC and SWFC-ART, therefore, would perform much faster than FSCS-ART in such situations.


\textit{Efficiency Tier-2 (Consistency):}
Our studies have also shown that all three methods take increasing amounts of time to generate a fixed number of test cases in increasing dimensional input spaces.
\nrs{When there were more than five SUT input parameters, the \textit{curse of dimensionality}~\cite{Domingos2012, Bellman1957, Bellman1954} started to impact on the performance of LimBal-KDFC and FSCS-ART, but SWFC-ART retained consistency (LimBal-KDFC was the least consistent).}
\nrs{Again, this can be understood from the theoretical complexity analysis, where LimBal-KDFC has quadratic, while FSCS-ART and SWFC-ART have a linear dimensional dependence on their time complexities.}
As real-life programs  typically have high-dimensional input domains (many program input parameters)~\cite{Lin2009}, SWFC-ART would perform much faster than the other two methods in such situations.

\begin{center}
\fbox{%
    \nrs{\parbox{0.95\linewidth }{%
        \textbf{\em Answer to RQ1}:
        \textit{FSCS-ART has a quadratic time complexity relation with the number of executed tests ($n$), and LimBal-KDFC has a quadratic time complexity relation with the dimensionality ($d$) of the SUT.
        Because real-life programs may often have low failure rates and high dimensional input spaces, neither FSCS-ART nor LimBal-KDFC solve the double-tier efficiency problem.
        In contrast, SWFC-ART's consistency and scalability, regardless of $n$ and $d$, make it the preferred method.}
    }}%
}
\end{center}

\subsubsection{RQ2}
The results show that the failure-detection effectiveness of SWFC-ART is similar to that of FSCS-ART and LimBal-KDFC.
No significant deterioration in the failure-detection effectiveness was observed
---
\nrs{the null hypotensis ($H_0$) could not be rejected.}

\nrs{We also found that the introduction of ANNS strategies delivered similar failure-detection effectiveness to NNS, while significantly reducing the computational overhead.}
Although LimBal-KDFC implements the ANNS through a limited back-tracking strategy and SWFC-ART maintains a Delaunay graph at its bottom layer, both ANNS strategies have similar failure-detection effectiveness.


\begin{center}
\fbox{%
    \nrs{\parbox{0.95\linewidth }{%
        \textbf{\em Answer to RQ2}:
        \textit{SWFC-ART employs a unique ANNS strategy and its failure-detection effectiveness is comparable to that of the state-of-the-art method,  KDFC-ART. }
    }}%
}
\end{center}

\subsubsection{RQ3}
The discrepancy values for SWFC-ART in the experiments were usually better than those for FSCS-ART and LimBal-KDFC, with no observed significant deterioration in even-spreading.
\nrs{We can therefore conclude that, although SWFC-ART incorporates an ANNS, its test case distribution remains comparable to that of the state-of-the-art method, KDFC-ART.}

\begin{center}
\fbox{%
    \nrs{\parbox{0.95\linewidth }{%
        \textbf{\em Answer to RQ3}:
        \textit{SWFC-ART distributes test cases as evenly as the state-of-the-art method, KDFC-ART, even though it employs an ANNS strategy. }
    }%
    }
}
\end{center}

\section{Threats to Validity}
\label{threats-to-validity}

This section discusses some potential limitations and threats to the validity of our work.

\subsection{Construct Validity}
\label{construct-validity}

Construct validity refers to how well the evaluation measures support the investigation of the research questions~\cite{experiments_in_SE}.
We used $T_G$ and $T_E$ to evaluate the efficiency of the methods (RQ1).
\nrs{Although {\em F-time} (the sum of the generation, execution, and evaluation times for a test case)~\cite{chen2006efficient} has also been used in some ART studies, reporting the component times separately appears to be the preferred option when evaluating FSCS-ART computational overhead reduction strategies~\cite{Huang2015, Mao2017, Mao2019} }
---
this removes ambiguity in scenarios where a method may have less $T_G$ but more $T_E$.

We measured the failure-detection effectiveness (RQ2) using the F-measure (or F-ratio).
Alternative metrics do exist, including the {\em P-measure} (the probability of detecting at least one failure) and {\em E-measure} (the expected number of failures detected)~\cite{Chen2004art, chen2006statistical, Ackah-Arthur2019}.
However, both P-measure and E-measure assume that the size of the test set is known in advance.
\nrs{The research in this paper involves the incremental generation of test cases by ART, and so the F-measure (or F-ratio)  is the preferred measure for such comparisons. }
Nevertheless, we look forward to exploring alternative evaluation metrics in our future work.

\nrs{In addition to discrepancy, the dispersion and edge-to-center ratio~\cite{Chen2007a, 932820} have also been used to measure test case distribution (RQ3). }
However, the edge-to-center ratio metric is rarely used, and low discrepancy corresponds to lower dispersion (and vice versa).
Given its simplicity and ease of implementation, we are confident in our choice of discrepancy as the test case distribution metric in our studies.

\subsection{External Validity}
\label{external-validity}


External validity refers to how generalizable the experimental results are for other scenarios.
We used FSCS-ART and LimBal-KDFC as the baseline comparison methods
---
FSCS-ART was selected due to its failure-detection effectiveness, and LimBal-KDFC was selected as the state-of-the-art enhancement of FSCS-ART for both low and high dimensional programs.
Although other overhead reduction strategies exist for FSCS-ART that claim to have ``linear-order'' time complexity, most of them have dimension-related exponential time complexity (as discussed in Section \ref{related-work}).
Our efficiency comparisons were performed in up to 15-dimensional input spaces, with up to 20,000 test cases generated in each.
We also used 28 real-life programs of various sizes and dimensions, more than most other comparable studies.

\nrs{The study reported on in this paper only examined numeric programs, and only used the Euclidean distance to measure the similarity between test cases.}
SWFC-ART supports a wide variety of distance metrics, and so we look forward to exploring its performance with non-numeric programs and other distance metrics in our future work.
Furthermore, because the scope of this study was to increase the efficiency of the FSCS-ART method by employing a graph-based approach, the effectiveness could be increased by integrating SWFC-ART with existing methods that aim to enhance the effectiveness of FSCS-ART~\cite{9040713, Schneckenburger2008, Kuo2008, Kuo2007, Chen2007b, Kuo2007a, Mayer2006}.
\nrs{In particular, given their demonstrated potential for increasing the failure-detection effectiveness without impacting the efficiency,
we look forward to experimenting using the Manhattan distance~\cite{euclidean_art_2020} metric and Inverted FSCS-ART~\cite{Kuo2007} in our future work.}

\subsection{Internal Validity}
\label{Internal-validity}

Internal validity refers to confidence in the findings of the study.
We have double-checked and cross-validated the implemented algorithms to ensure that there is no mistake in the experimental setup.
The methods were rigorously evaluated under different settings
---
dimensions, failure patterns, number of generated test cases, real-life programs
---
and statistical tests were performed on the results.
The number of trials (sample sizes) in the simulations and experiments were significant enough to achieve the desired confidence level.
\nrs{Overall, we have confidence in the correctness of our evaluation setup}

\section{Related Work}
\label{related-work}

This section highlights some related work in ART computational overhead reduction.
Each method discussed is described according to its approach, time complexity, failure-detection effectiveness, and the effect of dimensionality on the method's efficiency.

\subsection{FSCS-ART overhead reduction methods}

\subsubsection{C.G. FSCS-ART}
One of the earliest attempts to reduce FSCS-ART computational overheads involved a \textit{Center-of-Gravity} (C.G.) constraint~\cite{ChanCG2004}.
This method reduces FSCS-ART distance computations by selecting a test case from the candidate set such that the resulting C.G. is as close as possible to the input domain's C.G.
Although this method was reported to reduce time overheads by 74\% in a small experimental study with a real-life program, it still faces quadratic time complexity.
Moreover, the method is unable to maintain the failure-finding effectiveness of the original FSCS-ART, with drops of up to 11\% reported in its F-measure values.

\subsubsection{D-FSCS-ART}
Descending distance FSCS-ART (D-FSCS-ART)~\cite{JixinGeng2010} aimed to reduce the number of distance computations of FSCS-ART by sorting the elements of $E$ according to their x-coordinates, and recording $Lastmin$ (the last maximum value of the shortest distance).
For each new candidate test case, only the x-distance (the difference in x-coordinates) is calculated.
To calculate the maximum value of the shortest distances, $E$ is split in half and the direction of distance calculation is determined by the x-distance.
In this way, the number of distance calculations is reduced
---
in 4-$d$ when $\theta = 0.001$, for example,  D-FSCS-ART achieved an 18\% reduction in distance calculations.
However, the efficiency improvement was not discussed in terms of F-time or $T_G$ (Section \ref{evaluation-metrics}), and it was not clear whether or not the reduction in distance calculations would impact the overall efficiency.
\nrs{Furthermore, it was also not clear whether or not the reduction in distance calculations preserved the failure-detection effectiveness of the original FSCS-ART.}

\subsubsection{FSCS-ART by Temporal Forgetting}
FSCS-ART by temporal forgetting has three types:
complete restart, random forgetting, and consecutive retention~\cite{chan2006forgetting}.
Complete restart simply resets the algorithm after a certain number of test case executions, determined by $k$, the ``memory parameter''.
\nrs{Random forgetting maintains a fixed-sized ($k$) executed test case set by randomly filtering out $n$-$k$ executed tests.}
Consecutive retention limits distance computations to only the $k$ most-recently-executed test cases.
While temporal forgetting may have ``linear-order'' complexity regardless of the SUT dimensionality, it can suffer from severe degradation in failure-detection effectiveness.
As reported, larger values for the memory parameter provide better effectiveness, but lower efficiency.
\nrs{Selection of an appropriate value for the memory parameter, therefore, remains challenging.}

\subsubsection{FSCS-ART by Distance-aware Forgetting}
\label{DF-FSCS-ART}
The original forgetting strategy may not make much use of information about forgotten test cases, which led to the development of
Distance-aware Forgetting FSCS-ART (DF-FSCS-ART)~\cite{Mao2017}, which uses executed test case spatial information.
\nrs{DF-FSCS-ART uses grid partitioning, with test cases located within the neighboring partition of a cell considered to be ``in-sight'', and those outside the partition considered ``out-of-sight'', and thus forgotten.}

\nrs{DF-FSCS-ART claims to preserve the FSCS-ART failure-detection ability, but it has a complexity of O($\tau \cdot 3^d \cdot k \cdot n $)
---
where $\tau$, $d$, $k$, $n$ are pre-set constants for the dynamic partitioning, the input domain dimensionality, the candidate set size, and the executed set size, respectively.}
\nrs{Although DF-FSCS-ART can generate test cases in linear-order, it still has exponential complexity in terms of dimensionality.}
As we saw in our experimental studies (Section \ref{experimental-studies}), even though LimBal-KDFC had a quadratic complexity relation with the SUT dimension (in the worst-case), the performance was severely impacted by the dimensionality.
As most real-life software can have high dimensionality, methods that have exponential complexity with respect to the SUT dimensionality may not be practical.

\subsubsection{KDFC-ART}
\label{KDFC-ART}
There are three variants of the recently proposed KD-tree approach to FSCS-ART~\cite{Mao2019}:
Naive-KDFC; SemiBal-KDFC; and LimBal-KDFC.
LimBal-KDFC is an efficient ART algorithm that attempts to resolve the dimensionality-related FSCS-ART overhead problem while maintaining comparable failure-detection effectiveness.
Although LimBal-KDFC has a log-linear time complexity with respect to the number of generated test cases
---
which is a little worse than linear-order algorithms
---
it only has an $O\,(d^2)$ relation with the SUT dimensionality in its worst-case, which is far better than the $O\,(3^d)$ for DF-FSCS-ART (Section \ref{DF-FSCS-ART}).
LimBal-KDFC also supports incremental generation of test cases, an advantage over RBCVT-Fast (Section \ref{rbcvt-fast}).
Both LimBal-KDFC and SWFC-ART use sophisticated data structures to store the previously executed, non-failure-causing test cases, but the actual structures are very different:
While LimBal-KDFC uses a KD-tree, which is efficient and straightforward, but delivers inconsistent performance in high dimensional input spaces~\cite{Lee1977}, SWFC-ART uses a graph-based data structure that is explicitly designed for high dimensional input domains.
\nrs{As discussed in Section \ref{experimental-results}, LimBal-KDFC appeared to become inconsistent when the input domain dimensionality went above five, but SWFC-ART continued to perform consistently regardless of the number of dimensions.}

\subsection{Other efficient ART methods}

\subsubsection{RBCVT-Fast}
\label{rbcvt-fast}
RBCVT-Fast~\cite{Shahbazi2013} combines a search-based algorithm with Voronoi region centroids to reduce ART overheads.
Although the method has linear-order time complexity and comparable failure-detection effectiveness, the size of the executed test set ($|E|$) must be specified in advance, which is usually impossible (because $|E|$ depends on the SUT failure rate, which is unknown before testing).
Furthermore, RBCVT-Fast faces a potential risk for high dimensional programs:
Because it makes use of a parameter ($\alpha$) to control the probability of test cases generated within the ``random border''
---
an imaginary border outside the real input domain borders
---
if an appropriate value for the parameter is not chosen, the effectiveness may suffer~\cite{Mao2019, Mao2017}.

\subsubsection{ART by Partitioning}
Although ART by Bisection and Random Partition~\cite{10.5555/1018442.1022058} report complexities of $O\,(n)$ and $O\,(n \log n)$, respectively,
they fail to preserve the failure-detection effectiveness of the original FSCS-ART~\cite{Mayer2006a}.
Chow~et~al.~\cite{Chow2013} proposed a \textit{divide-and-conquer} approach by partitioning the input domain into sub-domains and using FSCS-ART to generate a \textit{fixed} number of test cases in each sub-domain.
\nrs{When the number of test cases in a partition reaches a certain \textit{threshold}, then the algorithm further partitions the sub-domains.
However, determining the threshold value is a potential limitation of the method.}

ARTsum~\cite{Barus2016} is another linear-order algorithm, based on category-choice partitioning~\cite{10.1145/62959.62964}.
However, ARTsum is mainly used for non-numeric programs, whereas our work is primarily focused on numeric programs.

\subsubsection{MART}

Mirror Adaptive Random Testing (MART)~\cite{Chen2004mart} is an ART overhead reduction strategy that leverages the relative computational cheapness of mapping functions (compared to test case generation).
One SUT input domain partition is selected as the {\em source} domain, where the ART method is implemented.
Test cases generated in the source domain are then systematically mapped into the remaining partitions ({\em mirror} domains).
Dynamic MART (D-MART)~\cite{Huang2015} incorporated a ``divide-and-conquer'' approach, but suffered from mirroring deficiency because only one mirror test case was generated at a time.
E-MART~\cite{Omari2019a} added flexibility to the mirror generation option of D-MART using a smart mirror allocation scheme.

Although mirror-based ART techniques are efficient, their complexities depend on the adopted ART algorithm (in the source domain), which may be quadratic
---
MART combined with FSCS-ART, for example, has a quadratic complexity of $O\,(n^2/m^2)$.
Possibly because the ``edge-effect''~\cite{Chen2008} in the source domain will be inherited in all mirror domains, MART failure-detection has been found to be inferior to FSCS-ART.
Furthermore, preparing the initial MART testing setup, including selection of the partitioning scheme, remains an open issue.
Finally, even the most recent MART enhancements (E-MART and D-MART, for example) still have an exponential time complexity relationship with the dimensions of the SUT.
SWFC-ART differs from MART and other partition-based approaches:
Instead of partitioning the SUT input domain,  it partitions the executed test cases into a unique graph-based data structure for efficient NNS queries.

\section{Conclusion and Future Work}
\label{conclusion}

FSCS-ART is a well-known ART method that has been widely researched and applied to test real-life programs of varying input types.
Unfortunately, it faces severe computational issues relating to the growth in number of executed test cases and the dimensionality of the SUT's input domain:
\nrs{These issues have been referred to as the double-tier efficiency problem. }
Some existing FSCS-ART efficiency enhancement methods may only solve one-tier of the problem, by reducing time complexity related to the number of test cases generated.
\nrs{However, these methods can have quadratic (Section \ref{KDFC-ART}) or exponential (Section \ref{DF-FSCS-ART}) time complexity with respect to the dimensions of the SUT.}
Although a method may perform well in low dimensional input domains, it may experience severe performance degradation in higher dimensions.
\nrs{Because real-life programs often have both many dimensions, and very low failure rates, addressing only the complexity related to the number of test cases can only solve one aspect of the efficiency problem.}

Our method, SWFC-ART, comprehensively solves the FSCS-ART efficiency problems by using a hierarchical navigable small world graph (HNSWG) to store the executed, non-failure-causing test cases, thus improving the nearest neighbor search (NNS) efficiency.
SWFC-ART reduces the FSCS-ART complexity from $O\,(n^2)$ to $O\,(n \cdot \log n)$, with a negligible computational impact when increasing the dimensionality.
\nrs{SWFC-ART preserves the original FSCS-ART failure-detection effectiveness, and delivers efficient incremental test case generation.}

As this work addresses the FSCS-ART double-tier efficiency problem, there is great potential for many other interesting and promising research directions.


\nrs{The primary requirement for the application of any ART method to non-numeric programs is the capability to support a suitable similarity measure~\cite{Huang2019}.
The underlying HNSWG structure of SWFC-ART fully supports a wide variety of numeric and non-numeric distance (similarity) metrics\footnote{https://github.com/jelmerk/hnswlib}, including:
BrayCurtis dissimilarity;
Canberra distance;
correlation distance;
cosine distance;
Euclidean distance;
inner product; and
Manhattan distance~\cite{Deza2009}.
Non-numeric programs, including object-oriented (OO) software, can also be tested:
Our proposed method can be combined with ART methods designed explicitly for OO programs, such as ARTOO \cite{ciupa2008}.
However, further research will be necessary to enable correct formulating of the framework.
Empirical studies will also be needed to analyze the proposed method's performance against other non-numeric ART methods.
Even for numeric programs, we will examine employing alternatives to the Euclidean distance in SWFC-ART:
Such alternatives may result in better effectiveness while maintaining similar efficiency~\cite{euclidean_art_2020}.}

\nrs{Although SWFC-ART shows similar failure-detection effectiveness to FSCS-ART, it will be interesting to examine combining it with other failure-detection enhancement methods for FSCS-ART~\cite{9040713, Schneckenburger2008, Kuo2008, Kuo2007, Chen2007b, Kuo2007a, Mayer2006}. }
Similarly, using SWFC-ART in source domains of MART~\cite{Chen2004mart} may further boost its efficiency.
The use of HNSWGs in ART strategies other than FSCS-ART is also worth exploring, and will also form part of our future work.


\nrs{
Finally, other ANNS approaches that have proven their efficiency in similarity search applications, including vector quantization \cite{720541, 5432202} and hashing techniques \cite{Indyk1998}, will also be examined for applicability to ART.
Furthermore, boosting ANNS by exploiting the capabilities of modern hardware, such as using SIMD~\cite{9251093} and GPUs~\cite{JDH17}, will also be explored in our future work.}

\nolinenumbers   

 \section*{Acknowledgements}
This work is in part supported by the National Natural Science Foundation of China, under Grant Nos.~61872167, and 61502205.

\bibliographystyle{elsarticle-num-names}
\bibliography{main.bib}

\begin{thebibliography}{134}
\expandafter\ifx\csname natexlab\endcsname\relax\def\natexlab#1{#1}\fi
\providecommand{\url}[1]{\texttt{#1}}
\providecommand{\href}[2]{#2}
\providecommand{\path}[1]{#1}
\providecommand{\DOIprefix}{doi:}
\providecommand{\ArXivprefix}{arXiv:}
\providecommand{\URLprefix}{URL: }
\providecommand{\Pubmedprefix}{pmid:}
\providecommand{\doi}[1]{\href{http://dx.doi.org/#1}{\path{#1}}}
\providecommand{\Pubmed}[1]{\href{pmid:#1}{\path{#1}}}
\providecommand{\bibinfo}[2]{#2}
\ifx\xfnm\relax \def\xfnm[#1]{\unskip,\space#1}\fi
\bibitem[{Myers(1979)}]{Myers:2004:AST:983238}
\bibinfo{author}{G.~J. Myers}, \bibinfo{title}{The art of software testing},
  \bibinfo{publisher}{Wiley New York}, \bibinfo{year}{1979}.
\bibitem[{Duran and Ntafos(1981)}]{Duran1981}
\bibinfo{author}{J.~W. Duran}, \bibinfo{author}{S.~Ntafos},
\newblock \bibinfo{title}{A report on random testing},
\newblock in: \bibinfo{booktitle}{5th International Conference on Software
  Engineering}, volume~\bibinfo{volume}{81}, \bibinfo{publisher}{IEEE},
  \bibinfo{year}{1981}, pp. \bibinfo{pages}{179--183}.
\bibitem[{Agrawal(1978)}]{whenToUseRT}
\bibinfo{author}{V.~D. Agrawal},
\newblock \bibinfo{title}{When to use random testing},
\newblock \bibinfo{journal}{IEEE Transactions on Computers}
  \bibinfo{volume}{27} (\bibinfo{year}{1978}) \bibinfo{pages}{1054--1055}.
\bibitem[{Hamlet and Maciniak(1994)}]{hamlet1994random}
\bibinfo{author}{R.~Hamlet}, \bibinfo{author}{J.~Maciniak},
\newblock \bibinfo{title}{Random testing. encyclopedia of software
  engineering},
\newblock \bibinfo{journal}{Wiley: New York}  (\bibinfo{year}{1994})
  \bibinfo{pages}{970--978}.
\bibitem[{Forrester and Miller(2000)}]{10.5555/1267102.1267108}
\bibinfo{author}{J.~E. Forrester}, \bibinfo{author}{B.~P. Miller},
\newblock \bibinfo{title}{An empirical study of the robustness of {Windows NT}
  applications using random testing},
\newblock in: \bibinfo{booktitle}{4th Conference on USENIX Windows Systems
  Symposium}, volume~\bibinfo{volume}{4}, \bibinfo{publisher}{USENIX
  Association}, \bibinfo{year}{2000}, pp. \bibinfo{pages}{59--68}.
\bibitem[{Regehr(2005)}]{Regehr2005}
\bibinfo{author}{J.~Regehr},
\newblock \bibinfo{title}{Random testing of interrupt-driven software},
\newblock in: \bibinfo{booktitle}{5th ACM International Conference on Embedded
  Software}, \bibinfo{publisher}{ACM}, \bibinfo{year}{2005}, pp.
  \bibinfo{pages}{290--298}.
\bibitem[{Bati et~al.(2007)Bati, Giakoumakis, Herbert, and
  Surna}]{bati2007genetic}
\bibinfo{author}{H.~Bati}, \bibinfo{author}{L.~Giakoumakis},
  \bibinfo{author}{S.~Herbert}, \bibinfo{author}{A.~Surna},
\newblock \bibinfo{title}{A genetic approach for random testing of database
  systems},
\newblock in: \bibinfo{booktitle}{33rd international conference on Very Large
  Data Bases}, \bibinfo{publisher}{VLDB Endowment}, \bibinfo{year}{2007}, pp.
  \bibinfo{pages}{1243--1251}.
\bibitem[{Slutz(1998)}]{slutz1998massive}
\bibinfo{author}{D.~R. Slutz},
\newblock \bibinfo{title}{Massive stochastic testing of {SQL}},
\newblock in: \bibinfo{booktitle}{24th International Conference on Very Large
  Data Bases}, \bibinfo{publisher}{Morgan Kaufmann Publishers},
  \bibinfo{year}{1998}, pp. \bibinfo{pages}{618--622}.
\bibitem[{Muangsiri and Takada(2017)}]{Muangsiri2017}
\bibinfo{author}{W.~Muangsiri}, \bibinfo{author}{S.~Takada},
\newblock \bibinfo{title}{Random {GUI} testing of {Android} application using
  behavioral model},
\newblock \bibinfo{journal}{International Journal of Software Engineering and
  Knowledge Engineering} \bibinfo{volume}{27} (\bibinfo{year}{2017})
  \bibinfo{pages}{1603--1612}.
\bibitem[{{Yoshikawa} et~al.(2003){Yoshikawa}, {Shimura}, and
  {Ozawa}}]{yoshikawa2003random}
\bibinfo{author}{T.~{Yoshikawa}}, \bibinfo{author}{K.~{Shimura}},
  \bibinfo{author}{T.~{Ozawa}},
\newblock \bibinfo{title}{Random program generator for {Java JIT} compiler test
  system},
\newblock in: \bibinfo{booktitle}{3rd International Conference on Quality
  Software}, \bibinfo{organization}{IEEE}, \bibinfo{year}{2003}, pp.
  \bibinfo{pages}{20--23}.
\bibitem[{Pacheco et~al.(2008)Pacheco, Lahiri, and Ball}]{pacheco2008finding}
\bibinfo{author}{C.~Pacheco}, \bibinfo{author}{S.~K. Lahiri},
  \bibinfo{author}{T.~Ball},
\newblock \bibinfo{title}{Finding errors in {.NET} with feedback-directed
  random testing},
\newblock in: \bibinfo{booktitle}{2008 international symposium on Software
  testing and analysis}, \bibinfo{publisher}{ACM}, \bibinfo{year}{2008}, pp.
  \bibinfo{pages}{87--96}.
\bibitem[{Godefroid et~al.(2008)Godefroid, Kiezun, and
  Levin}]{10.1145/1379022.1375607}
\bibinfo{author}{P.~Godefroid}, \bibinfo{author}{A.~Kiezun},
  \bibinfo{author}{M.~Y. Levin},
\newblock \bibinfo{title}{Grammar-based whitebox fuzzing},
\newblock in: \bibinfo{booktitle}{29th ACM SIGPLAN Conference on Programming
  Language Design and Implementation}, \bibinfo{publisher}{ACM},
  \bibinfo{year}{2008}, pp. \bibinfo{pages}{206--215}.
\bibitem[{Miller et~al.(2006)Miller, Cooksey, and
  Moore}]{10.1145/1145735.1145743}
\bibinfo{author}{B.~P. Miller}, \bibinfo{author}{G.~Cooksey},
  \bibinfo{author}{F.~Moore},
\newblock \bibinfo{title}{An empirical study of the robustness of {MacOS}
  applications using random testing},
\newblock in: \bibinfo{booktitle}{1st international workshop on Random
  testing}, \bibinfo{publisher}{ACM}, \bibinfo{year}{2006}, pp.
  \bibinfo{pages}{46--54}.
\bibitem[{{Daboczi} et~al.(2003){Daboczi}, {Kollar}, {Simon}, and
  {Megyeri}}]{Daboczi2003}
\bibinfo{author}{T.~{Daboczi}}, \bibinfo{author}{I.~{Kollar}},
  \bibinfo{author}{G.~{Simon}}, \bibinfo{author}{T.~{Megyeri}},
\newblock \bibinfo{title}{Automatic testing of graphical user interfaces},
\newblock in: \bibinfo{booktitle}{20th IEEE Instrumentation Technology
  Conference}, volume~\bibinfo{volume}{1}, \bibinfo{organization}{IEEE},
  \bibinfo{year}{2003}, pp. \bibinfo{pages}{441--445}.
\bibitem[{Miller et~al.(1990)Miller, Fredriksen, and So}]{10.1145/96267.96279}
\bibinfo{author}{B.~P. Miller}, \bibinfo{author}{L.~Fredriksen},
  \bibinfo{author}{B.~So},
\newblock \bibinfo{title}{An empirical study of the reliability of {UNIX}
  utilities},
\newblock \bibinfo{journal}{Commun. ACM} \bibinfo{volume}{33}
  (\bibinfo{year}{1990}) \bibinfo{pages}{32--44}.
\bibitem[{Miller et~al.(1995)Miller, Koski, Lee, Maganty, Murthy, Natarajan,
  and Steidl}]{millerfuzz1998}
\bibinfo{author}{B.~P. Miller}, \bibinfo{author}{D.~Koski},
  \bibinfo{author}{C.~P. Lee}, \bibinfo{author}{V.~Maganty},
  \bibinfo{author}{R.~Murthy}, \bibinfo{author}{A.~Natarajan},
  \bibinfo{author}{J.~Steidl}, \bibinfo{title}{Fuzz revisited: {A}
  re-examination of the reliability of {UNIX} utilities and services},
  \bibinfo{type}{Technical Report}, University of Wisconsin--Madison Department
  of Computer Sciences, \bibinfo{year}{1995}.
\bibitem[{{Duran} and {Ntafos}(1984)}]{5010257}
\bibinfo{author}{J.~W. {Duran}}, \bibinfo{author}{S.~C. {Ntafos}},
\newblock \bibinfo{title}{An evaluation of random testing},
\newblock \bibinfo{journal}{IEEE Transactions on Software Engineering}
  \bibinfo{volume}{10} (\bibinfo{year}{1984}) \bibinfo{pages}{438--444}.
\bibitem[{Chen et~al.(2013)Chen, Kuo, Liu, and Wong}]{chen2013code}
\bibinfo{author}{T.~Y. Chen}, \bibinfo{author}{F.-C. Kuo},
  \bibinfo{author}{H.~Liu}, \bibinfo{author}{W.~E. Wong},
\newblock \bibinfo{title}{Code coverage of adaptive random testing},
\newblock \bibinfo{journal}{IEEE Transactions on Reliability}
  \bibinfo{volume}{62} (\bibinfo{year}{2013}) \bibinfo{pages}{226--237}.
\bibitem[{Schneckenburger and Mayer(2007)}]{10.1145/1295074.1295091}
\bibinfo{author}{C.~Schneckenburger}, \bibinfo{author}{J.~Mayer},
\newblock \bibinfo{title}{Towards the determination of typical failure
  patterns},
\newblock in: \bibinfo{booktitle}{4th international workshop on Software
  quality assurance: in conjunction with the 6th ESEC/FSE joint meeting},
  \bibinfo{publisher}{ACM}, \bibinfo{year}{2007}, pp. \bibinfo{pages}{90--93}.
\bibitem[{Arcuri and Briand(2011)}]{Arcuri2011}
\bibinfo{author}{A.~Arcuri}, \bibinfo{author}{L.~Briand},
\newblock \bibinfo{title}{Adaptive random testing: An illusion of
  effectiveness?},
\newblock in: \bibinfo{booktitle}{International Symposium on Software Testing
  and Analysis}, \bibinfo{publisher}{ACM}, \bibinfo{year}{2011}, pp.
  \bibinfo{pages}{265--275}.
\bibitem[{Hamlet(2002)}]{10.1145/566171.566203}
\bibinfo{author}{D.~Hamlet},
\newblock \bibinfo{title}{Continuity in software systems},
\newblock \bibinfo{journal}{SIGSOFT Softw. Eng. Notes} \bibinfo{volume}{27}
  (\bibinfo{year}{2002}) \bibinfo{pages}{196--200}.
\bibitem[{{Bishop}(1993)}]{Bishop}
\bibinfo{author}{P.~G. {Bishop}},
\newblock \bibinfo{title}{The variation of software survival time for different
  operational input profiles},
\newblock in: \bibinfo{booktitle}{23rd International Symposium on
  Fault-Tolerant Computing}, \bibinfo{publisher}{{IEEE} Computer Society},
  \bibinfo{year}{1993}, pp. \bibinfo{pages}{98--107}.
\bibitem[{Chan et~al.(1996)Chan, Chen, Mak, and Yu}]{Chan1996}
\bibinfo{author}{F.~T. Chan}, \bibinfo{author}{T.~Y. Chen},
  \bibinfo{author}{I.~K. Mak}, \bibinfo{author}{Y.~T. Yu},
\newblock \bibinfo{title}{Proportional sampling strategy: Guidelines for
  software testing practitioners},
\newblock \bibinfo{journal}{Information and Software Technology}
  \bibinfo{volume}{38} (\bibinfo{year}{1996}) \bibinfo{pages}{775--782}.
\bibitem[{Finelli(1991)}]{Finelli1991}
\bibinfo{author}{G.~B. Finelli},
\newblock \bibinfo{title}{{NASA} software failure characterization
  experiments},
\newblock \bibinfo{journal}{Reliability Engineering {\&} System Safety}
  \bibinfo{volume}{32} (\bibinfo{year}{1991}) \bibinfo{pages}{155--169}.
\bibitem[{Ammann and Knight(1988)}]{Ammann1988}
\bibinfo{author}{P.~Ammann}, \bibinfo{author}{J.~Knight},
\newblock \bibinfo{title}{{Data diversity: an approach to software fault
  tolerance}},
\newblock \bibinfo{journal}{IEEE Transactions on Computers}
  \bibinfo{volume}{37} (\bibinfo{year}{1988}) \bibinfo{pages}{418--425}.
\bibitem[{White and Cohen(1980)}]{White1980}
\bibinfo{author}{L.~White}, \bibinfo{author}{E.~Cohen},
\newblock \bibinfo{title}{A domain strategy for computer program testing},
\newblock \bibinfo{journal}{IEEE Transactions on Software Engineering}
  \bibinfo{volume}{6} (\bibinfo{year}{1980}) \bibinfo{pages}{247--257}.
\bibitem[{Chen et~al.(2005)Chen, Leung, and Mak}]{Chen2004art}
\bibinfo{author}{T.~Y. Chen}, \bibinfo{author}{H.~Leung},
  \bibinfo{author}{I.~K. Mak},
\newblock \bibinfo{title}{Adaptive random testing},
\newblock in: \bibinfo{booktitle}{Advances in Computer Science - ASIAN 2004.
  Higher-Level Decision Making}, \bibinfo{publisher}{Springer},
  \bibinfo{year}{2005}, pp. \bibinfo{pages}{320--329}.
\bibitem[{{Huang} et~al.(2019){Huang}, {Sun}, {Xu}, {Chen}, {Towey}, and
  {Xia}}]{Huang2019}
\bibinfo{author}{R.~{Huang}}, \bibinfo{author}{W.~{Sun}},
  \bibinfo{author}{Y.~{Xu}}, \bibinfo{author}{H.~{Chen}},
  \bibinfo{author}{D.~{Towey}}, \bibinfo{author}{X.~{Xia}},
\newblock \bibinfo{title}{A survey on adaptive random testing},
\newblock \bibinfo{journal}{IEEE Transactions on Software Engineering}
  (\bibinfo{year}{2019}) \bibinfo{pages}{1--35}.
\bibitem[{Chen et~al.(2019)Chen, Ackah-Arthur, Kudjo, and Mao}]{Chen2019}
\bibinfo{author}{J.~Chen}, \bibinfo{author}{H.~Ackah-Arthur},
  \bibinfo{author}{P.~K. Kudjo}, \bibinfo{author}{C.~Mao},
\newblock \bibinfo{title}{A taxonomic review of adaptive random testing for
  numeric programs},
\newblock \bibinfo{journal}{arXiv preprint arXiv:1909.10879}
  (\bibinfo{year}{2019}).
\bibitem[{Chen et~al.(2010)Chen, Kuo, Merkel, and
  Tse}]{10.1016/j.jss.2009.02.022}
\bibinfo{author}{T.~Y. Chen}, \bibinfo{author}{F.-C. Kuo},
  \bibinfo{author}{R.~G. Merkel}, \bibinfo{author}{T.~H. Tse},
\newblock \bibinfo{title}{Adaptive random testing: The {ART} of test case
  diversity},
\newblock \bibinfo{journal}{Journal of Systems and Software}
  \bibinfo{volume}{83} (\bibinfo{year}{2010}) \bibinfo{pages}{60--66}.
\bibitem[{Mayer and Schneckenburger(2006)}]{DBLP:conf/serp/MayerS06}
\bibinfo{author}{J.~Mayer}, \bibinfo{author}{C.~Schneckenburger},
\newblock \bibinfo{title}{Statistical analysis and enhancement of random
  testing methods also under constrained resources},
\newblock in: \bibinfo{booktitle}{International Conference on Software
  Engineering Research and Practice {\&} Conference on Programming Languages
  and Compilers}, volume~\bibinfo{volume}{1}, \bibinfo{publisher}{{CSREA}
  Press}, \bibinfo{year}{2006}, pp. \bibinfo{pages}{16--23}.
\bibitem[{Chen et~al.(2007)Chen, Kuo, and Zhou}]{Chen2007c}
\bibinfo{author}{T.~Y. Chen}, \bibinfo{author}{F.-C. Kuo},
  \bibinfo{author}{Z.~Q. Zhou},
\newblock \bibinfo{title}{On favourable conditions for adaptive random
  testing},
\newblock \bibinfo{journal}{International Journal of Software Engineering and
  Knowledge Engineering} \bibinfo{volume}{17} (\bibinfo{year}{2007})
  \bibinfo{pages}{805--825}.
\bibitem[{{Chen} et~al.(2013){Chen}, {Kuo}, {Liu}, and {Wong}}]{6449335}
\bibinfo{author}{T.~Y. {Chen}}, \bibinfo{author}{F.-C. {Kuo}},
  \bibinfo{author}{H.~{Liu}}, \bibinfo{author}{W.~E. {Wong}},
\newblock \bibinfo{title}{Code coverage of adaptive random testing},
\newblock \bibinfo{journal}{IEEE Transactions on Reliability}
  \bibinfo{volume}{62} (\bibinfo{year}{2013}) \bibinfo{pages}{226--237}.
\bibitem[{{Chen} et~al.(2008){Chen}, {Kuo}, {Liu}, and {Wong}}]{4601538}
\bibinfo{author}{T.~Y. {Chen}}, \bibinfo{author}{F.-C. {Kuo}},
  \bibinfo{author}{H.~{Liu}}, \bibinfo{author}{W.~E. {Wong}},
\newblock \bibinfo{title}{Does adaptive random testing deliver a higher
  confidence than random testing?},
\newblock in: \bibinfo{booktitle}{The 8th International Conference on Quality
  Software}, \bibinfo{organization}{IEEE}, \bibinfo{year}{2008}, pp.
  \bibinfo{pages}{145--154}.
\bibitem[{Bueno et~al.(2014)Bueno, Jino, and Wong}]{10.1016/j.ins.2011.01.025}
\bibinfo{author}{P.~M.~S. Bueno}, \bibinfo{author}{M.~Jino},
  \bibinfo{author}{W.~E. Wong},
\newblock \bibinfo{title}{Diversity oriented test data generation using
  metaheuristic search techniques},
\newblock \bibinfo{journal}{Inf. Sci.} \bibinfo{volume}{259}
  (\bibinfo{year}{2014}) \bibinfo{pages}{490--509}.
\bibitem[{Chen and Merkel(2008)}]{10.1145/1363102.1363107}
\bibinfo{author}{T.~Y. Chen}, \bibinfo{author}{R.~Merkel},
\newblock \bibinfo{title}{An upper bound on software testing effectiveness},
\newblock \bibinfo{journal}{ACM Trans. Softw. Eng. Methodol.}
  \bibinfo{volume}{17} (\bibinfo{year}{2008}) \bibinfo{pages}{16:1--16:27}.
\bibitem[{{Yan} et~al.(2020){Yan}, {Wang}, and {Fei}}]{8944083}
\bibinfo{author}{M.~{Yan}}, \bibinfo{author}{L.~{Wang}},
  \bibinfo{author}{A.~{Fei}},
\newblock \bibinfo{title}{{ARTDL}: Adaptive random testing for deep learning
  systems},
\newblock \bibinfo{journal}{IEEE Access} \bibinfo{volume}{8}
  (\bibinfo{year}{2020}) \bibinfo{pages}{3055--064}.
\bibitem[{Lv et~al.(2019)Lv, Zhang, Zeng, and Zhang}]{Lv2019}
\bibinfo{author}{C.~Lv}, \bibinfo{author}{L.~Zhang}, \bibinfo{author}{F.~Zeng},
  \bibinfo{author}{J.~Zhang},
\newblock \bibinfo{title}{Adaptive random testing for {XSS} vulnerability},
\newblock in: \bibinfo{booktitle}{The 26th Asia-Pacific Software Engineering
  Conference}, \bibinfo{organization}{IEEE}, \bibinfo{year}{2019}, pp.
  \bibinfo{pages}{63--69}.
\bibitem[{Zhang et~al.(2019)Zhang, Zhang, Wang, Zhao, and Zhang}]{Zhang2019}
\bibinfo{author}{L.~Zhang}, \bibinfo{author}{D.~Zhang},
  \bibinfo{author}{C.~Wang}, \bibinfo{author}{J.~Zhao},
  \bibinfo{author}{Z.~Zhang},
\newblock \bibinfo{title}{{ART4SQLi}: The {ART} of {SQL} injection
  vulnerability discovery},
\newblock \bibinfo{journal}{IEEE Transactions on Reliability}
  \bibinfo{volume}{68} (\bibinfo{year}{2019}) \bibinfo{pages}{1470--1489}.
\bibitem[{Ciupa et~al.(2008)Ciupa, Leitner, Oriol, and Meyer}]{ciupa2008}
\bibinfo{author}{I.~Ciupa}, \bibinfo{author}{A.~Leitner},
  \bibinfo{author}{M.~Oriol}, \bibinfo{author}{B.~Meyer},
\newblock \bibinfo{title}{{ARTOO}: adaptive random testing for object-oriented
  software},
\newblock in: \bibinfo{booktitle}{The 30th international conference on Software
  engineering}, \bibinfo{publisher}{ACM}, \bibinfo{year}{2008}, pp.
  \bibinfo{pages}{71--80}.
\bibitem[{Lin et~al.(2009)Lin, Tang, Chen, and Zhao}]{Lin2009}
\bibinfo{author}{Y.~Lin}, \bibinfo{author}{X.~Tang}, \bibinfo{author}{Y.~Chen},
  \bibinfo{author}{J.~Zhao},
\newblock \bibinfo{title}{A divergence-oriented approach to adaptive random
  testing of {Java} programs},
\newblock in: \bibinfo{booktitle}{2009 IEEE/ACM International Conference on
  Automated Software Engineering}, \bibinfo{publisher}{IEEE Computer Society},
  \bibinfo{year}{2009}, pp. \bibinfo{pages}{221--232}.
\bibitem[{Indyk and Motwani(1998)}]{Indyk1998}
\bibinfo{author}{P.~Indyk}, \bibinfo{author}{R.~Motwani},
\newblock \bibinfo{title}{Approximate nearest neighbors: {T}owards removing the
  curse of dimensionality},
\newblock in: \bibinfo{booktitle}{The 30th Annual ACM Symposium on Theory of
  Computing}, \bibinfo{publisher}{ACM}, \bibinfo{year}{1998}, pp.
  \bibinfo{pages}{604--613}.
\bibitem[{Bin~Ali et~al.(2019)Bin~Ali, Engstr{\"o}m, Taromirad, Mousavi,
  Minhas, Helgesson, Kunze, and Varshosaz}]{Ali2019}
\bibinfo{author}{N.~Bin~Ali}, \bibinfo{author}{E.~Engstr{\"o}m},
  \bibinfo{author}{M.~Taromirad}, \bibinfo{author}{M.~R. Mousavi},
  \bibinfo{author}{N.~M. Minhas}, \bibinfo{author}{D.~Helgesson},
  \bibinfo{author}{S.~Kunze}, \bibinfo{author}{M.~Varshosaz},
\newblock \bibinfo{title}{On the search for industry-relevant regression
  testing research},
\newblock \bibinfo{journal}{Empirical Software Engineering}
  \bibinfo{volume}{24} (\bibinfo{year}{2019}) \bibinfo{pages}{2020--2055}.
\bibitem[{{Miranda} et~al.(2018){Miranda}, {Cruciani}, {Verdecchia}, and
  {Bertolino}}]{8453081}
\bibinfo{author}{B.~{Miranda}}, \bibinfo{author}{E.~{Cruciani}},
  \bibinfo{author}{R.~{Verdecchia}}, \bibinfo{author}{A.~{Bertolino}},
\newblock \bibinfo{title}{{FAST} approaches to scalable similarity-based test
  case prioritization},
\newblock in: \bibinfo{booktitle}{The 2018 IEEE/ACM 40th International
  Conference on Software Engineering}, \bibinfo{publisher}{{ACM}},
  \bibinfo{year}{2018}, pp. \bibinfo{pages}{222--232}.
\bibitem[{Cartaxo et~al.(2011)Cartaxo, Machado, and Neto}]{cartaxo2011use}
\bibinfo{author}{E.~G. Cartaxo}, \bibinfo{author}{P.~D. Machado},
  \bibinfo{author}{F.~O. Neto},
\newblock \bibinfo{title}{On the use of a similarity function for test case
  selection in the context of model-based testing},
\newblock \bibinfo{journal}{Software Testing, Verification and Reliability}
  \bibinfo{volume}{21} (\bibinfo{year}{2011}) \bibinfo{pages}{75--100}.
\bibitem[{Mao et~al.(2019)Mao, Zhan, Tse, and Chen}]{Mao2019}
\bibinfo{author}{C.~Mao}, \bibinfo{author}{X.~Zhan}, \bibinfo{author}{T.~H.
  Tse}, \bibinfo{author}{T.~Y. Chen},
\newblock \bibinfo{title}{{KDFC-ART}: a {KD-tree} approach to enhancing
  fixed-size-candidate-set adaptive random testing},
\newblock \bibinfo{journal}{IEEE Transactions on Reliability}
  \bibinfo{volume}{68} (\bibinfo{year}{2019}) \bibinfo{pages}{1444--1469}.
\bibitem[{Devroey et~al.(2016)Devroey, Perrouin, Legay, Schobbens, and
  Heymans}]{10.1145/2866614.2866627}
\bibinfo{author}{X.~Devroey}, \bibinfo{author}{G.~Perrouin},
  \bibinfo{author}{A.~Legay}, \bibinfo{author}{P.-Y. Schobbens},
  \bibinfo{author}{P.~Heymans},
\newblock \bibinfo{title}{Search-based similarity-driven behavioural {SPL}
  testing},
\newblock in: \bibinfo{booktitle}{The 10th International Workshop on
  Variability Modelling of Software-Intensive Systems},
  \bibinfo{publisher}{ACM}, \bibinfo{year}{2016}, pp. \bibinfo{pages}{89--96}.
\bibitem[{{Cruciani} et~al.(2019){Cruciani}, {Miranda}, {Verdecchia}, and
  {Bertolino}}]{8812048}
\bibinfo{author}{E.~{Cruciani}}, \bibinfo{author}{B.~{Miranda}},
  \bibinfo{author}{R.~{Verdecchia}}, \bibinfo{author}{A.~{Bertolino}},
\newblock \bibinfo{title}{Scalable approaches for test suite reduction},
\newblock in: \bibinfo{booktitle}{The IEEE/ACM 41st International Conference on
  Software Engineering}, \bibinfo{organization}{IEEE}, \bibinfo{year}{2019},
  pp. \bibinfo{pages}{419--429}.
\bibitem[{Bertolino et~al.(2020)Bertolino, Cruciani, Miranda, and
  Verdecchia}]{bertolino2020know}
\bibinfo{author}{A.~Bertolino}, \bibinfo{author}{E.~Cruciani},
  \bibinfo{author}{B.~Miranda}, \bibinfo{author}{R.~Verdecchia},
  \bibinfo{title}{Know Your Neighbor: Fast Static Prediction of Test
  Flakiness}, \bibinfo{type}{Technical Report}, ISTI Technical Reports
  2020/001, \bibinfo{year}{2020}.
\bibitem[{Li et~al.(2019)Li, Zhang, Sun, Wang, Li, Zhang, and Lin}]{Li2019}
\bibinfo{author}{W.~Li}, \bibinfo{author}{Y.~Zhang}, \bibinfo{author}{Y.~Sun},
  \bibinfo{author}{W.~Wang}, \bibinfo{author}{M.~Li},
  \bibinfo{author}{W.~Zhang}, \bibinfo{author}{X.~Lin},
\newblock \bibinfo{title}{Approximate nearest neighbor search on high
  dimensional data-experiments, analyses, and improvement},
\newblock \bibinfo{journal}{IEEE Transactions on Knowledge and Data
  Engineering} \bibinfo{volume}{32} (\bibinfo{year}{2019})
  \bibinfo{pages}{1475--1488}.
\bibitem[{Aum{\"{u}}ller et~al.(2020)Aum{\"{u}}ller, Bernhardsson, and
  Faithfull}]{Aumuller2017}
\bibinfo{author}{M.~Aum{\"{u}}ller}, \bibinfo{author}{E.~Bernhardsson},
  \bibinfo{author}{A.~J. Faithfull},
\newblock \bibinfo{title}{Ann-benchmarks: {A} benchmarking tool for approximate
  nearest neighbor algorithms},
\newblock \bibinfo{journal}{Inf. Syst.} \bibinfo{volume}{87}
  (\bibinfo{year}{2020}) \bibinfo{pages}{34--49}.
\bibitem[{Ponomarenko et~al.(2014)Ponomarenko, Avrelin, Naidan, and
  Boytsov}]{Ponomarenko2014}
\bibinfo{author}{A.~Ponomarenko}, \bibinfo{author}{N.~Avrelin},
  \bibinfo{author}{B.~Naidan}, \bibinfo{author}{L.~Boytsov},
\newblock \bibinfo{title}{Comparative analysis of data structures for
  approximate nearest neighbor search},
\newblock \bibinfo{journal}{Data Analytics}  (\bibinfo{year}{2014})
  \bibinfo{pages}{125--130}.
\bibitem[{{Malkov} and {Yashunin}(2020)}]{Malkov2018}
\bibinfo{author}{Y.~A. {Malkov}}, \bibinfo{author}{D.~A. {Yashunin}},
\newblock \bibinfo{title}{Efficient and robust approximate nearest neighbor
  search using hierarchical navigable small world graphs},
\newblock \bibinfo{journal}{IEEE Transactions on Pattern Analysis and Machine
  Intelligence} \bibinfo{volume}{42} (\bibinfo{year}{2020})
  \bibinfo{pages}{824--836}.
\bibitem[{O'Neil(2017)}]{ONeil2017}
\bibinfo{author}{D.~J. O'Neil},
\newblock \bibinfo{title}{Nearest neighbor problem},
\newblock in: \bibinfo{booktitle}{Encyclopedia of GIS},
  \bibinfo{publisher}{Springer International Publishing},
  \bibinfo{address}{Cham}, \bibinfo{year}{2017}, pp.
  \bibinfo{pages}{1421--1426}.
\bibitem[{Arya and Mount(1993)}]{Arya1993}
\bibinfo{author}{S.~Arya}, \bibinfo{author}{D.~M. Mount},
\newblock \bibinfo{title}{Approximate nearest neighbor queries in fixed
  dimensions},
\newblock in: \bibinfo{booktitle}{The 4th Annual ACM-SIAM Symposium on Discrete
  Algorithms}, \bibinfo{publisher}{Society for Industrial and Applied
  Mathematics}, \bibinfo{year}{1993}, pp. \bibinfo{pages}{271--280}.
\bibitem[{Lee and Wong(1977)}]{Lee1977}
\bibinfo{author}{D.-T. Lee}, \bibinfo{author}{C.~Wong},
\newblock \bibinfo{title}{Worst-case analysis for region and partial region
  searches in multidimensional binary search trees and balanced quad trees},
\newblock \bibinfo{journal}{Acta Informatica} \bibinfo{volume}{9}
  (\bibinfo{year}{1977}) \bibinfo{pages}{23--29}.
\bibitem[{Ch{\'a}vez and Tellez(2010)}]{Chavez2010}
\bibinfo{author}{E.~Ch{\'a}vez}, \bibinfo{author}{E.~S. Tellez},
\newblock \bibinfo{title}{Navigating k-nearest neighbor graphs to solve nearest
  neighbor searches},
\newblock in: \bibinfo{booktitle}{Mexican Conference on Pattern Recognition},
  \bibinfo{organization}{Springer}, \bibinfo{year}{2010}, pp.
  \bibinfo{pages}{270--280}.
\bibitem[{Wang et~al.(2015)Wang, Wang, Zeng, Gan, Li, and Guo}]{Wang2015}
\bibinfo{author}{J.~Wang}, \bibinfo{author}{J.~Wang},
  \bibinfo{author}{G.~Zeng}, \bibinfo{author}{R.~Gan}, \bibinfo{author}{S.~Li},
  \bibinfo{author}{B.~Guo},
\newblock \bibinfo{title}{Fast neighborhood graph search using cartesian
  concatenation},
\newblock in: \bibinfo{booktitle}{Multimedia Data Mining and Analytics -
  Disruptive Innovation}, \bibinfo{publisher}{Springer}, \bibinfo{year}{2015},
  pp. \bibinfo{pages}{397--417}.
\bibitem[{Aoyama et~al.(2011)Aoyama, Saito, Sawada, and Ueda}]{Aoyama2011}
\bibinfo{author}{K.~Aoyama}, \bibinfo{author}{K.~Saito},
  \bibinfo{author}{H.~Sawada}, \bibinfo{author}{N.~Ueda},
\newblock \bibinfo{title}{{Fast approximate similarity search based on
  degree-reduced neighborhood graphs}},
\newblock in: \bibinfo{booktitle}{17th ACM International Conference on
  Knowledge Discovery and Data Mining}, \bibinfo{publisher}{ACM Press},
  \bibinfo{year}{2011}, pp. \bibinfo{pages}{1055--1063}.
\bibitem[{Paredes(2008)}]{Paredes2008}
\bibinfo{author}{R.~Paredes}, \bibinfo{title}{Graphs for metric space
  searching}, Ph.D. thesis, University of Chile, Chile, \bibinfo{address}{Dept.
  of Computer Science Tech Report TR/DCC-2008-10}, \bibinfo{year}{2008}.
\bibitem[{Hajebi et~al.(2011)Hajebi, Abbasi-Yadkori, Shahbazi, and
  Zhang}]{Hajebi2011}
\bibinfo{author}{K.~Hajebi}, \bibinfo{author}{Y.~Abbasi-Yadkori},
  \bibinfo{author}{H.~Shahbazi}, \bibinfo{author}{H.~Zhang},
\newblock \bibinfo{title}{Fast approximate nearest-neighbor search with
  k-nearest neighbor graph},
\newblock in: \bibinfo{booktitle}{22nd International Joint Conference on
  Artificial Intelligence}, volume~\bibinfo{volume}{2},
  \bibinfo{publisher}{AAAI Press}, \bibinfo{year}{2011}, pp.
  \bibinfo{pages}{1312--1317}.
\bibitem[{Wang and Li(2012)}]{Wang2012}
\bibinfo{author}{J.~Wang}, \bibinfo{author}{S.~Li},
\newblock \bibinfo{title}{{Query-driven iterated neighborhood graph search for
  large scale indexing}},
\newblock in: \bibinfo{booktitle}{20th ACM International Conference on
  Multimedia}, \bibinfo{publisher}{ACM Press}, \bibinfo{year}{2012}, pp.
  \bibinfo{pages}{179--188}.
\bibitem[{Jiang et~al.(2016)Jiang, Xie, Deng, Xu, and Wang}]{Jiang2016}
\bibinfo{author}{Z.~Jiang}, \bibinfo{author}{L.~Xie},
  \bibinfo{author}{X.~Deng}, \bibinfo{author}{W.~Xu},
  \bibinfo{author}{J.~Wang},
\newblock \bibinfo{title}{Fast nearest neighbor search in the {Hamming} space},
\newblock in: \bibinfo{booktitle}{International Conference on Multimedia
  Modeling}, \bibinfo{organization}{Springer}, \bibinfo{year}{2016}, pp.
  \bibinfo{pages}{325--336}.
\bibitem[{Caretta~Cartozo and De~Los~Rios(2009)}]{CarettaCartozo2009}
\bibinfo{author}{C.~Caretta~Cartozo}, \bibinfo{author}{P.~De~Los~Rios},
\newblock \bibinfo{title}{Extended navigability of small world networks: exact
  results and new insights},
\newblock \bibinfo{journal}{Physical Review Letters} \bibinfo{volume}{102}
  (\bibinfo{year}{2009}) \bibinfo{pages}{238703:1--238703:4}.
\bibitem[{Dong et~al.(2011)Dong, Moses, and Li}]{Dong2011}
\bibinfo{author}{W.~Dong}, \bibinfo{author}{C.~Moses}, \bibinfo{author}{K.~Li},
\newblock \bibinfo{title}{{Efficient k-nearest neighbor graph construction for
  generic similarity measures}},
\newblock in: \bibinfo{booktitle}{20th International Conference on World Wide
  Web}, \bibinfo{publisher}{ACM Press}, \bibinfo{year}{2011}, pp.
  \bibinfo{pages}{577--586}.
\bibitem[{Milgram(1967)}]{Milgram1967}
\bibinfo{author}{S.~Milgram},
\newblock \bibinfo{title}{The small world problem},
\newblock \bibinfo{journal}{Psychology today} \bibinfo{volume}{2}
  (\bibinfo{year}{1967}) \bibinfo{pages}{60--67}.
\bibitem[{Watts and Strogatz(1998)}]{Watts1998}
\bibinfo{author}{D.~Watts}, \bibinfo{author}{S.~Strogatz},
\newblock \bibinfo{title}{{Collective dynamics of `small-world' networks}},
\newblock \bibinfo{journal}{Nature} \bibinfo{volume}{393}
  (\bibinfo{year}{1998}) \bibinfo{pages}{440--442}.
\bibitem[{Mehlhorn and Schreiber(2013)}]{Mehlhorn2013}
\bibinfo{author}{H.~Mehlhorn}, \bibinfo{author}{F.~Schreiber},
  \bibinfo{title}{Small-world property}, \bibinfo{publisher}{Springer},
  \bibinfo{year}{2013}, pp. \bibinfo{pages}{1957--1959}.
\bibitem[{Kleinberg(2000{\natexlab{a}})}]{Kleinberg2000}
\bibinfo{author}{J.~M. Kleinberg},
\newblock \bibinfo{title}{{Navigation in a small world}},
\newblock \bibinfo{journal}{Nature} \bibinfo{volume}{406}
  (\bibinfo{year}{2000}{\natexlab{a}}) \bibinfo{pages}{845}.
\bibitem[{Kleinberg(2000{\natexlab{b}})}]{Kleinberg}
\bibinfo{author}{J.~Kleinberg},
\newblock \bibinfo{title}{The small-world phenomenon: An algorithmic
  perspective},
\newblock in: \bibinfo{booktitle}{32nd Annual ACM Symposium on Theory of
  Computing}, \bibinfo{publisher}{ACM}, \bibinfo{year}{2000}{\natexlab{b}}, pp.
  \bibinfo{pages}{163--170}.
\bibitem[{Lifshits and Zhang(2009)}]{Lifshits2009}
\bibinfo{author}{Y.~Lifshits}, \bibinfo{author}{S.~Zhang},
\newblock \bibinfo{title}{{Combinatorial algorithms for nearest neighbors,
  near-duplicates and small-world design}},
\newblock in: \bibinfo{booktitle}{20th Annual ACM-SIAM Symposium on Discrete
  Algorithms}, \bibinfo{publisher}{Society for Industrial and Applied
  Mathematics}, \bibinfo{year}{2009}, pp. \bibinfo{pages}{318--326}.
\bibitem[{Karbasi et~al.(2015)Karbasi, Ioannidis, and Massoulie}]{Karbasi2015}
\bibinfo{author}{A.~Karbasi}, \bibinfo{author}{S.~Ioannidis},
  \bibinfo{author}{L.~Massoulie},
\newblock \bibinfo{title}{{From small-world networks to comparison-based
  search}},
\newblock \bibinfo{journal}{IEEE Transactions on Information Theory}
  \bibinfo{volume}{61} (\bibinfo{year}{2015}) \bibinfo{pages}{3056--3074}.
\bibitem[{Beaumont et~al.(2007{\natexlab{a}})Beaumont, Kermarrec, Marchal, and
  Riviere}]{Beaumont2007}
\bibinfo{author}{O.~Beaumont}, \bibinfo{author}{A.-M. Kermarrec},
  \bibinfo{author}{L.~Marchal}, \bibinfo{author}{E.~Riviere},
\newblock \bibinfo{title}{{VoroNet: A scalable object network based on Voronoi
  tessellations}},
\newblock in: \bibinfo{booktitle}{IEEE International Parallel and Distributed
  Processing Symposium}, \bibinfo{publisher}{IEEE},
  \bibinfo{year}{2007}{\natexlab{a}}, pp. \bibinfo{pages}{1--10}.
\bibitem[{Beaumont et~al.(2007{\natexlab{b}})Beaumont, Kermarrec, and
  Rivi{\`e}re}]{Beaumont}
\bibinfo{author}{O.~Beaumont}, \bibinfo{author}{A.-M. Kermarrec},
  \bibinfo{author}{{\'E}.~Rivi{\`e}re},
\newblock \bibinfo{title}{Peer to peer multidimensional overlays:
  {Approximating} complex structures},
\newblock in: \bibinfo{booktitle}{Principles of Distributed Systems},
  \bibinfo{publisher}{Springer}, \bibinfo{year}{2007}{\natexlab{b}}, pp.
  \bibinfo{pages}{315--328}.
\bibitem[{Malkov and Ponomarenko(2016)}]{Malkov2015}
\bibinfo{author}{Y.~Malkov}, \bibinfo{author}{A.~Ponomarenko},
\newblock \bibinfo{title}{Growing homophilic networks are natural navigable
  small worlds},
\newblock \bibinfo{journal}{PloS One} \bibinfo{volume}{11(6)}
  (\bibinfo{year}{2016}) \bibinfo{pages}{e0158162:1--e0158162:14)}.
\bibitem[{Malkov et~al.(2012)Malkov, Ponomarenko, Logvinov, and
  Krylov}]{Malkov2012}
\bibinfo{author}{Y.~Malkov}, \bibinfo{author}{A.~Ponomarenko},
  \bibinfo{author}{A.~Logvinov}, \bibinfo{author}{V.~Krylov},
\newblock \bibinfo{title}{Scalable distributed algorithm for approximate
  nearest neighbor search problem in high dimensional general metric spaces},
\newblock in: \bibinfo{booktitle}{International Conference on Similarity Search
  and Applications}, \bibinfo{organization}{Springer}, \bibinfo{year}{2012},
  pp. \bibinfo{pages}{132--147}.
\bibitem[{Malkov et~al.(2014)Malkov, Ponomarenko, Logvinov, and
  Krylov}]{Malkov2014}
\bibinfo{author}{Y.~Malkov}, \bibinfo{author}{A.~Ponomarenko},
  \bibinfo{author}{A.~Logvinov}, \bibinfo{author}{V.~Krylov},
\newblock \bibinfo{title}{{Approximate nearest neighbor algorithm based on
  navigable small world graphs}},
\newblock \bibinfo{journal}{Information Systems} \bibinfo{volume}{45}
  (\bibinfo{year}{2014}) \bibinfo{pages}{61--68}.
\bibitem[{Pugh(1990)}]{Vitter1990}
\bibinfo{author}{W.~Pugh},
\newblock \bibinfo{title}{Skip lists: {A} probabilistic alternative to balanced
  trees},
\newblock \bibinfo{journal}{Commun. ACM} \bibinfo{volume}{33}
  (\bibinfo{year}{1990}) \bibinfo{pages}{668--676}.
\bibitem[{Liu et~al.(2011)Liu, Xie, Yang, Lu, and Chen}]{liu2011adaptive}
\bibinfo{author}{H.~Liu}, \bibinfo{author}{X.~Xie}, \bibinfo{author}{J.~Yang},
  \bibinfo{author}{Y.~Lu}, \bibinfo{author}{T.~Y. Chen},
\newblock \bibinfo{title}{Adaptive random testing through test profiles},
\newblock \bibinfo{journal}{Software: Practice and Experience}
  \bibinfo{volume}{41} (\bibinfo{year}{2011}) \bibinfo{pages}{1131--1154}.
\bibitem[{Chen et~al.(2007)Chen, Kuo, and Liu}]{Chen2007a}
\bibinfo{author}{T.~Y. Chen}, \bibinfo{author}{F.-C. Kuo},
  \bibinfo{author}{H.~Liu},
\newblock \bibinfo{title}{On test case distributions of adaptive random
  testing},
\newblock in: \bibinfo{booktitle}{19th International Conference on Software
  Engineering and Knowledge Engineering}, \bibinfo{organization}{Knowledge
  Systems Institute--Graduate School}, \bibinfo{year}{2007}, pp.
  \bibinfo{pages}{141--144}.
\bibitem[{{Chow} et~al.(2013){Chow}, {Chen}, and {Tse}}]{Chow2013}
\bibinfo{author}{C.~{Chow}}, \bibinfo{author}{T.~Y. {Chen}},
  \bibinfo{author}{T.~H. {Tse}},
\newblock \bibinfo{title}{The {ART} of divide and conquer: An innovative
  approach to improving the efficiency of adaptive random testing},
\newblock in: \bibinfo{booktitle}{13th International Conference on Quality
  Software}, \bibinfo{publisher}{IEEE}, \bibinfo{year}{2013}, pp.
  \bibinfo{pages}{268--275}.
\bibitem[{Ackah-Arthur et~al.(2019)Ackah-Arthur, Chen, Towey, Omari, Xi, and
  Huang}]{Ackah-Arthur2019}
\bibinfo{author}{H.~Ackah-Arthur}, \bibinfo{author}{J.~Chen},
  \bibinfo{author}{D.~Towey}, \bibinfo{author}{M.~Omari},
  \bibinfo{author}{J.~Xi}, \bibinfo{author}{R.~Huang},
\newblock \bibinfo{title}{One-domain-one-input: {Adaptive} random testing by
  orthogonal recursive bisection with restriction},
\newblock \bibinfo{journal}{IEEE Transactions on Reliability}
  \bibinfo{volume}{68} (\bibinfo{year}{2019}) \bibinfo{pages}{1404--1428}.
\bibitem[{Liu et~al.(2010)Liu, Xie, Yang, Lu, and Chen}]{Liu2010}
\bibinfo{author}{H.~Liu}, \bibinfo{author}{X.~Xie}, \bibinfo{author}{J.~Yang},
  \bibinfo{author}{Y.~Lu}, \bibinfo{author}{T.~Y. Chen},
\newblock \bibinfo{title}{Adaptive random testing by exclusion through test
  profile},
\newblock in: \bibinfo{booktitle}{10th International Conference on Quality
  Software}, \bibinfo{organization}{IEEE}, \bibinfo{year}{2010}, pp.
  \bibinfo{pages}{92--101}.
\bibitem[{Jia and Harman(2010)}]{Jia2011}
\bibinfo{author}{Y.~Jia}, \bibinfo{author}{M.~Harman},
\newblock \bibinfo{title}{An analysis and survey of the development of mutation
  testing},
\newblock \bibinfo{journal}{IEEE Transactions on Software Engineering}
  \bibinfo{volume}{37} (\bibinfo{year}{2010}) \bibinfo{pages}{649--678}.
\bibitem[{573(2010)}]{5733835}
\bibinfo{title}{{ISO/IEC/IEEE} international standard - systems and software
  engineering--vocabulary},
\newblock \bibinfo{journal}{ISO/IEC/IEEE 24765:2010(E)}  (\bibinfo{year}{2010})
  \bibinfo{pages}{1--418}.
\bibitem[{Press et~al.(2007)Press, Teukolsky, Vetterling, and
  Flannery}]{10.5555/1403886}
\bibinfo{author}{W.~H. Press}, \bibinfo{author}{S.~A. Teukolsky},
  \bibinfo{author}{W.~T. Vetterling}, \bibinfo{author}{B.~P. Flannery},
  \bibinfo{title}{Numerical Recipes: {The} Art of Scientific Computing},
  \bibinfo{edition}{3} ed., \bibinfo{publisher}{Cambridge University Press},
  \bibinfo{address}{USA}, \bibinfo{year}{2007}.
\bibitem[{col(2020)}]{collected_algorithms_acm}
\bibinfo{title}{{Collected Algorithms of the ACM}}, \bibinfo{year}{Accessed on:
  Jan 14, 2020}. \URLprefix \url{http://calgo.acm.org/}.
\bibitem[{Ferrer et~al.(2012)Ferrer, Chicano, and Alba}]{Ferrer2012}
\bibinfo{author}{J.~Ferrer}, \bibinfo{author}{F.~Chicano},
  \bibinfo{author}{E.~Alba},
\newblock \bibinfo{title}{{Evolutionary algorithms for the multi-objective test
  data generation problem}},
\newblock \bibinfo{journal}{Software: Practice and Experience}
  \bibinfo{volume}{42} (\bibinfo{year}{2012}) \bibinfo{pages}{1331--1362}.
\bibitem[{{Liang, Yong Daniel}(2017)}]{Y.DanielLiang2017}
\bibinfo{author}{{Liang, Yong Daniel}}, \bibinfo{title}{Introduction to {Java}
  programming and data structures, comprehensive version},
  \bibinfo{publisher}{Pearson Education}, \bibinfo{year}{2017}.
\bibitem[{May(2007)}]{May2007}
\bibinfo{author}{P.~S. May}, \bibinfo{title}{Test data generation: {Two}
  evolutionary approaches to mutation testing}, Ph.D. thesis, Computing
  Laboratory, \bibinfo{address}{The University of Kent}, \bibinfo{year}{2007}.
\bibitem[{Do et~al.(2005)Do, Elbaum, and Rothermel}]{Do2005}
\bibinfo{author}{H.~Do}, \bibinfo{author}{S.~Elbaum},
  \bibinfo{author}{G.~Rothermel},
\newblock \bibinfo{title}{Supporting controlled experimentation with testing
  techniques: An infrastructure and its potential impact},
\newblock \bibinfo{journal}{Empirical Software Engineering}
  \bibinfo{volume}{10} (\bibinfo{year}{2005}) \bibinfo{pages}{405--435}.
\bibitem[{Walkinshaw and Fraser(2017)}]{7927980}
\bibinfo{author}{N.~Walkinshaw}, \bibinfo{author}{G.~Fraser},
\newblock \bibinfo{title}{Uncertainty-driven black-box test data generation},
\newblock in: \bibinfo{booktitle}{International Conference on Software Testing,
  Verification and Validation}, \bibinfo{publisher}{{IEEE} Computer Society},
  \bibinfo{year}{2017}, pp. \bibinfo{pages}{253--263}.
\bibitem[{mat(2020)}]{matrix_processor}
\bibinfo{title}{{JetBrains Academy}}, \bibinfo{year}{Accessed on: Dec 12,
  2020}. \URLprefix \url{https://hyperskill.org/projects/60?track=1}.
\bibitem[{jav(2020)}]{java_util_arrays}
\bibinfo{title}{{Java API Documentation}}, \bibinfo{year}{Accessed on: Dec 12,
  2020}. \URLprefix
  \url{https://docs.oracle.com/javase/8/docs/api/java/util/Arrays.html}.
\bibitem[{Giorgetti et~al.(2010)Giorgetti, March\'{e}, Tushkanova, and
  Kouchnarenko}]{10.1145/1868281.1868289}
\bibinfo{author}{A.~Giorgetti}, \bibinfo{author}{C.~March\'{e}},
  \bibinfo{author}{E.~Tushkanova}, \bibinfo{author}{O.~Kouchnarenko},
\newblock \bibinfo{title}{Specifying generic {Java} programs: {Two} case
  studies},
\newblock in: \bibinfo{booktitle}{10th Workshop on Language Descriptions, Tools
  and Applications}, \bibinfo{publisher}{ACM}, \bibinfo{year}{2010}, pp.
  \bibinfo{pages}{8:1--8:8}.
\bibitem[{Freund(1988)}]{freund1988modern}
\bibinfo{author}{J.~E. Freund}, \bibinfo{title}{Modern elementary statistics},
  \bibinfo{publisher}{Prentice-Hall, Inc.}, \bibinfo{year}{1988}.
\bibitem[{Wilcoxon(1945)}]{Wilcoxon1945}
\bibinfo{author}{F.~Wilcoxon},
\newblock \bibinfo{title}{Individual comparisons by ranking methods},
\newblock \bibinfo{journal}{Biometrics Bulletin} \bibinfo{volume}{1}
  (\bibinfo{year}{1945}) \bibinfo{pages}{80--83}.
\bibitem[{Mann and Whitney(1947)}]{mann1947}
\bibinfo{author}{H.~B. Mann}, \bibinfo{author}{D.~R. Whitney},
\newblock \bibinfo{title}{On a test of whether one of two random variables is
  stochastically larger than the other},
\newblock \bibinfo{journal}{The Annals of Mathematical Statistics}
  (\bibinfo{year}{1947}) \bibinfo{pages}{50--60}.
\bibitem[{Sullivan and Feinn(2012)}]{Sullivan2012}
\bibinfo{author}{G.~M. Sullivan}, \bibinfo{author}{R.~Feinn},
\newblock \bibinfo{title}{Using effect size---or why the $p$ value is not
  enough},
\newblock \bibinfo{journal}{Journal of Graduate Medical Education}
  \bibinfo{volume}{4} (\bibinfo{year}{2012}) \bibinfo{pages}{279--282}.
\bibitem[{Pallant(2016)}]{pallant_2016}
\bibinfo{author}{J.~Pallant}, \bibinfo{title}{SPSS survival manual: A step by
  step guide to data analysis using SPSS}, \bibinfo{publisher}{Open University
  Press}, \bibinfo{year}{2016}.
\bibitem[{Cohen(1988)}]{jacobcohen1988}
\bibinfo{author}{J.~Cohen}, \bibinfo{title}{Statistical Power Analysis for the
  Behavioral Sciences}, \bibinfo{edition}{2} ed.,
  \bibinfo{publisher}{Hillsdale, N.J. : L. Erlbaum Associates},
  \bibinfo{year}{1988}.
\bibitem[{Domingos(2012)}]{Domingos2012}
\bibinfo{author}{P.~Domingos},
\newblock \bibinfo{title}{A few useful things to know about machine learning},
\newblock \bibinfo{journal}{Commun. ACM} \bibinfo{volume}{55}
  (\bibinfo{year}{2012}) \bibinfo{pages}{78--87}.
\bibitem[{Bellman(1957)}]{Bellman1957}
\bibinfo{author}{R.~Bellman}, \bibinfo{title}{Dynamic Programming},
  \bibinfo{edition}{1} ed., \bibinfo{publisher}{Princeton University Press},
  \bibinfo{address}{Princeton, NJ, USA}, \bibinfo{year}{1957}.
\bibitem[{Bellman(1954)}]{Bellman1954}
\bibinfo{author}{R.~Bellman},
\newblock \bibinfo{title}{{The theory of dynamic programming}},
\newblock \bibinfo{journal}{Bulletin of the American Mathematical Society}
  \bibinfo{volume}{60} (\bibinfo{year}{1954}) \bibinfo{pages}{503--515}.
\bibitem[{Wohlin et~al.(2012)Wohlin, Runeson, H{\"{o}}st, Ohlsson, and
  Regnell}]{experiments_in_SE}
\bibinfo{author}{C.~Wohlin}, \bibinfo{author}{P.~Runeson},
  \bibinfo{author}{M.~H{\"{o}}st}, \bibinfo{author}{M.~C. Ohlsson},
  \bibinfo{author}{B.~Regnell}, \bibinfo{title}{Experimentation in Software
  Engineering}, \bibinfo{publisher}{Springer}, \bibinfo{year}{2012}.
\bibitem[{Chen and Merkel(2006)}]{chen2006efficient}
\bibinfo{author}{T.~Y. Chen}, \bibinfo{author}{R.~Merkel},
\newblock \bibinfo{title}{Efficient and effective random testing using the
  {Voronoi} diagram},
\newblock in: \bibinfo{booktitle}{17th Australian Software Engineering
  Conference}, \bibinfo{publisher}{{IEEE} Computer Society},
  \bibinfo{year}{2006}, pp. \bibinfo{pages}{300--305}.
\bibitem[{Huang et~al.(2015)Huang, Liu, Xie, and Chen}]{Huang2015}
\bibinfo{author}{R.~Huang}, \bibinfo{author}{H.~Liu}, \bibinfo{author}{X.~Xie},
  \bibinfo{author}{J.~Chen},
\newblock \bibinfo{title}{{Enhancing mirror adaptive random testing through
  dynamic partitioning}},
\newblock \bibinfo{journal}{Information and Software Technology}
  \bibinfo{volume}{67} (\bibinfo{year}{2015}) \bibinfo{pages}{13--29}.
\bibitem[{Mao et~al.(2017)Mao, Chen, and Kuo}]{Mao2017}
\bibinfo{author}{C.~Mao}, \bibinfo{author}{T.~Y. Chen}, \bibinfo{author}{F.-C.
  Kuo},
\newblock \bibinfo{title}{{Out of sight, out of mind: {A} distance-aware
  forgetting strategy for adaptive random testing}},
\newblock \bibinfo{journal}{Science China Information Sciences}
  \bibinfo{volume}{60} (\bibinfo{year}{2017}) \bibinfo{pages}{1--21}.
\bibitem[{Chen et~al.(2006)Chen, Kuo, and Merkel}]{chen2006statistical}
\bibinfo{author}{T.~Y. Chen}, \bibinfo{author}{F.-C. Kuo},
  \bibinfo{author}{R.~Merkel},
\newblock \bibinfo{title}{On the statistical properties of testing
  effectiveness measures},
\newblock \bibinfo{journal}{Journal of Systems and Software}
  \bibinfo{volume}{79} (\bibinfo{year}{2006}) \bibinfo{pages}{591--601}.
\bibitem[{{Branicky} et~al.(2001){Branicky}, {LaValle}, {Olson}, and {Libo
  Yang}}]{932820}
\bibinfo{author}{M.~S. {Branicky}}, \bibinfo{author}{S.~M. {LaValle}},
  \bibinfo{author}{K.~{Olson}}, \bibinfo{author}{{Libo Yang}},
\newblock \bibinfo{title}{Quasi-randomized path planning},
\newblock in: \bibinfo{booktitle}{IEEE International Conference on Robotics and
  Automation (Cat. No.01CH37164)}, volume~\bibinfo{volume}{2},
  \bibinfo{publisher}{{IEEE}}, \bibinfo{year}{2001}, pp.
  \bibinfo{pages}{1481--1487}.
\bibitem[{{Li} et~al.(2019){Li}, {Li}, {Li}, and {Wang}}]{9040713}
\bibinfo{author}{Z.~{Li}}, \bibinfo{author}{Q.~{Li}},
  \bibinfo{author}{R.~{Li}}, \bibinfo{author}{L.~{Wang}},
\newblock \bibinfo{title}{An enhanced {ART} in high dimensional input domain},
\newblock in: \bibinfo{booktitle}{10th International Conference on Software
  Engineering and Service Science}, \bibinfo{publisher}{IEEE},
  \bibinfo{year}{2019}, pp. \bibinfo{pages}{495--497}.
\bibitem[{Schneckenburger and Schweiggert(2008)}]{Schneckenburger2008}
\bibinfo{author}{C.~Schneckenburger}, \bibinfo{author}{F.~Schweiggert},
\newblock \bibinfo{title}{Investigating the dimensionality problem of adaptive
  random testing incorporating a local search technique},
\newblock in: \bibinfo{booktitle}{1st International Conference on Software
  Testing Verification and Validation Workshop}, \bibinfo{publisher}{{IEEE}
  Computer Society}, \bibinfo{year}{2008}, pp. \bibinfo{pages}{241--250}.
\bibitem[{Kuo et~al.(2008)Kuo, Chen, Liu, and Chan}]{Kuo2008}
\bibinfo{author}{F.-C. Kuo}, \bibinfo{author}{T.~Y. Chen},
  \bibinfo{author}{H.~Liu}, \bibinfo{author}{W.~K. Chan},
\newblock \bibinfo{title}{Enhancing adaptive random testing for programs with
  high dimensional input domains or failure-unrelated parameters},
\newblock \bibinfo{journal}{Software Quality Journal} \bibinfo{volume}{16}
  (\bibinfo{year}{2008}) \bibinfo{pages}{303--327}.
\bibitem[{Kuo et~al.(2007)Kuo, Chen, Liu, and Chan}]{Kuo2007}
\bibinfo{author}{F.-C. Kuo}, \bibinfo{author}{T.~Y. Chen},
  \bibinfo{author}{H.~Liu}, \bibinfo{author}{W.~K. Chan},
\newblock \bibinfo{title}{Enhancing adaptive random testing in high dimensional
  input domains},
\newblock in: \bibinfo{booktitle}{ACM Symposium on Applied Computing},
  \bibinfo{publisher}{ACM}, \bibinfo{year}{2007}, pp.
  \bibinfo{pages}{1467--1472}.
\bibitem[{Chen et~al.(2007)Chen, Huang, and Kuo}]{Chen2007b}
\bibinfo{author}{T.~Y. Chen}, \bibinfo{author}{D.~H. Huang},
  \bibinfo{author}{F.-C. Kuo},
\newblock \bibinfo{title}{{Adaptive random testing by balancing}},
\newblock in: \bibinfo{booktitle}{2nd international workshop on Random testing
  co-located with the 22nd IEEE/ACM International Conference on Automated
  Software Engineering}, \bibinfo{publisher}{ACM Press}, \bibinfo{year}{2007},
  pp. \bibinfo{pages}{2--9}.
\bibitem[{Kuo et~al.(2007)Kuo, Sim, Sun, Tang, and Zhou}]{Kuo2007a}
\bibinfo{author}{F.-C. Kuo}, \bibinfo{author}{K.~Y. Sim},
  \bibinfo{author}{C.~Sun}, \bibinfo{author}{S.~Tang},
  \bibinfo{author}{Z.~Zhou},
\newblock \bibinfo{title}{Enhanced random testing for programs with high
  dimensional input domains},
\newblock in: \bibinfo{booktitle}{International Conference on Software
  Engineering and Knowledge Engineering}, \bibinfo{publisher}{Knowledge Systems
  Institute}, \bibinfo{year}{2007}, pp. \bibinfo{pages}{135--140}.
\bibitem[{Mayer(2006)}]{Mayer2006}
\bibinfo{author}{J.~Mayer},
\newblock \bibinfo{title}{Towards effective adaptive random testing for
  higher-dimensional input domains},
\newblock in: \bibinfo{booktitle}{8th Annual Conference on Genetic and
  Evolutionary Computation}, \bibinfo{publisher}{ACM}, \bibinfo{year}{2006},
  pp. \bibinfo{pages}{1955--1956}.
\bibitem[{Huang et~al.(2020)Huang, Cui, Sun, and Towey}]{euclidean_art_2020}
\bibinfo{author}{R.~Huang}, \bibinfo{author}{C.~Cui}, \bibinfo{author}{W.~Sun},
  \bibinfo{author}{D.~Towey},
\newblock \bibinfo{title}{Poster: Is {Euclidean} distance the best distance
  measurement for adaptive random testing?},
\newblock in: \bibinfo{booktitle}{13th International Conference on Software
  Testing, Validation and Verification}, \bibinfo{publisher}{{IEEE}},
  \bibinfo{year}{2020}, pp. \bibinfo{pages}{406--409}.
\bibitem[{{Chan} et~al.(2004){Chan}, {Chan}, {Chen}, and {Yiu}}]{ChanCG2004}
\bibinfo{author}{F.~T. {Chan}}, \bibinfo{author}{K.~P. {Chan}},
  \bibinfo{author}{T.~Y. {Chen}}, \bibinfo{author}{S.~M. {Yiu}},
\newblock \bibinfo{title}{Adaptive random testing with {CG} constraint},
\newblock in: \bibinfo{booktitle}{28th Annual International Computer Software
  and Applications Conference}, volume~\bibinfo{volume}{2},
  \bibinfo{publisher}{{IEEE} Computer Society}, \bibinfo{year}{2004}, pp.
  \bibinfo{pages}{96--99}.
\bibitem[{{Jixin Geng} and {Jiongmin Zhang}(2010)}]{JixinGeng2010}
\bibinfo{author}{{Jixin Geng}}, \bibinfo{author}{{Jiongmin Zhang}},
\newblock \bibinfo{title}{A new method to solve the {``Boundary Effect''} of
  adaptive random testing},
\newblock in: \bibinfo{booktitle}{2010 International Conference on Educational
  and Information Technology}, volume~\bibinfo{volume}{1},
  \bibinfo{organization}{IEEE}, \bibinfo{year}{2010}, pp.
  \bibinfo{pages}{298--302}.
\bibitem[{Chan et~al.(2006)Chan, Chen, and Towey}]{chan2006forgetting}
\bibinfo{author}{K.~P. Chan}, \bibinfo{author}{T.~Y. Chen},
  \bibinfo{author}{D.~Towey},
\newblock \bibinfo{title}{Forgetting test cases},
\newblock in: \bibinfo{booktitle}{30th Annual International Computer Software
  and Applications Conference}, volume~\bibinfo{volume}{1},
  \bibinfo{organization}{IEEE}, \bibinfo{year}{2006}, pp.
  \bibinfo{pages}{485--494}.
\bibitem[{Shahbazi et~al.(2013)Shahbazi, Tappenden, and Miller}]{Shahbazi2013}
\bibinfo{author}{A.~Shahbazi}, \bibinfo{author}{A.~F. Tappenden},
  \bibinfo{author}{J.~Miller},
\newblock \bibinfo{title}{{Centroidal Voronoi tessellations--a new approach to
  random testing}},
\newblock \bibinfo{journal}{IEEE Transactions on Software Engineering}
  \bibinfo{volume}{39} (\bibinfo{year}{2013}) \bibinfo{pages}{163--183}.
\bibitem[{Chen et~al.(2004)Chen, Merkel, Wong, and
  Eddy}]{10.5555/1018442.1022058}
\bibinfo{author}{T.~Y. Chen}, \bibinfo{author}{R.~Merkel},
  \bibinfo{author}{P.~Wong}, \bibinfo{author}{G.~Eddy},
\newblock \bibinfo{title}{Adaptive random testing through dynamic
  partitioning},
\newblock in: \bibinfo{booktitle}{4th International Conference on Quality
  Software}, \bibinfo{publisher}{{IEEE} Computer Society},
  \bibinfo{year}{2004}, pp. \bibinfo{pages}{79--86}.
\bibitem[{Mayer and Schneckenburger(2006)}]{Mayer2006a}
\bibinfo{author}{J.~Mayer}, \bibinfo{author}{C.~Schneckenburger},
\newblock \bibinfo{title}{An empirical analysis and comparison of random
  testing techniques},
\newblock in: \bibinfo{booktitle}{2006 ACM/IEEE International Symposium on
  Empirical Software Engineering}, \bibinfo{publisher}{ACM},
  \bibinfo{year}{2006}, pp. \bibinfo{pages}{105--114}.
\bibitem[{Barus et~al.(2016)Barus, Chen, Kuo, Liu, Merkel, and
  Rothermel}]{Barus2016}
\bibinfo{author}{A.~C. Barus}, \bibinfo{author}{T.~Y. Chen},
  \bibinfo{author}{F.-C. Kuo}, \bibinfo{author}{H.~Liu},
  \bibinfo{author}{R.~Merkel}, \bibinfo{author}{G.~Rothermel},
\newblock \bibinfo{title}{A cost-effective random testing method for programs
  with non-numeric inputs},
\newblock \bibinfo{journal}{IEEE Transactions on Computers}
  \bibinfo{volume}{65} (\bibinfo{year}{2016}) \bibinfo{pages}{3609--3623}.
\bibitem[{Ostrand and Balcer(1988)}]{10.1145/62959.62964}
\bibinfo{author}{T.~J. Ostrand}, \bibinfo{author}{M.~J. Balcer},
\newblock \bibinfo{title}{The category-partition method for specifying and
  generating functional tests},
\newblock \bibinfo{journal}{Commun. ACM} \bibinfo{volume}{31}
  (\bibinfo{year}{1988}) \bibinfo{pages}{676--686}.
\bibitem[{Chen et~al.(2004)Chen, Kuo, Merkel, and Ng}]{Chen2004mart}
\bibinfo{author}{T.~Y. Chen}, \bibinfo{author}{F.-C. Kuo},
  \bibinfo{author}{R.~G. Merkel}, \bibinfo{author}{S.~P. Ng},
\newblock \bibinfo{title}{Mirror adaptive random testing},
\newblock \bibinfo{journal}{Information and Software Technology}
  \bibinfo{volume}{46} (\bibinfo{year}{2004}) \bibinfo{pages}{1001--1010}.
\bibitem[{Omari et~al.(2019)Omari, Chen, Ackah-Arthur, and Kudjo}]{Omari2019a}
\bibinfo{author}{M.~Omari}, \bibinfo{author}{J.~Chen},
  \bibinfo{author}{H.~Ackah-Arthur}, \bibinfo{author}{P.~K. Kudjo},
\newblock \bibinfo{title}{Elimination by linear association: An effective and
  efficient static mirror adaptive random testing},
\newblock \bibinfo{journal}{IEEE Access} \bibinfo{volume}{7}
  (\bibinfo{year}{2019}) \bibinfo{pages}{71038--71060}.
\bibitem[{Chen et~al.(2008)Chen, Kuo, and Liu}]{Chen2008}
\bibinfo{author}{T.~Y. Chen}, \bibinfo{author}{F.-C. Kuo},
  \bibinfo{author}{H.~Liu},
\newblock \bibinfo{title}{Distributing test cases more evenly in adaptive
  random testing},
\newblock \bibinfo{journal}{Journal of Systems and Software}
  \bibinfo{volume}{81} (\bibinfo{year}{2008}) \bibinfo{pages}{2146--2162}.
\bibitem[{Deza and Deza(2009)}]{Deza2009}
\bibinfo{author}{M.~M. Deza}, \bibinfo{author}{E.~Deza},
  \bibinfo{title}{{Encyclopedia of distances}}, \bibinfo{publisher}{Springer
  Berlin Heidelberg}, \bibinfo{year}{2009}.
\bibitem[{{Gray} and {Neuhoff}(1998)}]{720541}
\bibinfo{author}{R.~M. {Gray}}, \bibinfo{author}{D.~L. {Neuhoff}},
\newblock \bibinfo{title}{Quantization},
\newblock \bibinfo{journal}{IEEE Transactions on Information Theory}
  \bibinfo{volume}{44} (\bibinfo{year}{1998}) \bibinfo{pages}{2325--2383}.
\bibitem[{{Jégou} et~al.(2011){Jégou}, {Douze}, and {Schmid}}]{5432202}
\bibinfo{author}{H.~{Jégou}}, \bibinfo{author}{M.~{Douze}},
  \bibinfo{author}{C.~{Schmid}},
\newblock \bibinfo{title}{Product quantization for nearest neighbor search},
\newblock \bibinfo{journal}{IEEE Transactions on Pattern Analysis and Machine
  Intelligence} \bibinfo{volume}{33} (\bibinfo{year}{2011})
  \bibinfo{pages}{117--128}.
\bibitem[{{Ashfaq} et~al.(2020){Ashfaq}, {Huang}, and {Omari}}]{9251093}
\bibinfo{author}{M.~{Ashfaq}}, \bibinfo{author}{R.~{Huang}},
  \bibinfo{author}{M.~{Omari}},
\newblock \bibinfo{title}{{FSCS-SIMD: An efficient implementation of
  Fixed-Size-Candidate-Set adaptive random testing using SIMD instructions}},
\newblock in: \bibinfo{booktitle}{31st International Symposium on Software
  Reliability Engineering}, \bibinfo{publisher}{{IEEE}}, \bibinfo{year}{2020},
  pp. \bibinfo{pages}{277--288}.
\bibitem[{{Johnson} et~al.(2019){Johnson}, {Douze}, and {Jégou}}]{JDH17}
\bibinfo{author}{J.~{Johnson}}, \bibinfo{author}{M.~{Douze}},
  \bibinfo{author}{H.~{Jégou}},
\newblock \bibinfo{title}{Billion-scale similarity search with {GPUs}},
\newblock \bibinfo{journal}{IEEE Transactions on Big Data}
  (\bibinfo{year}{2019}) \bibinfo{pages}{1--14}.

\end{thebibliography}

\vbox{}
\vbox{}
\scriptsize
\noindent
\textbf{Muhammad Ashfaq}
received the Bachelor's degree in Information Technology in 2017 from University of Gujrat, Pakistan. Currently he is pursuing Master's degree in Computer Science and Technology from the School of Computer Science and Communication Engineering, Jiangsu University, China. His current research interests include software testing and software debugging. His work has been published in International Symposium on Software Reliability Engineering (ISSRE 2020). One of his works has been accepted to be presented in the proceedings of 12th Asia-Pacific Symposium on Internetware (Internetware 2020).

\vbox{}
\noindent
\textbf{Rubing Huang}
received the Ph.D. degree in computer science and technology from the Huazhong University of Science and Technology, Wuhan, China, in 2013. From 2016 to 2018, he was a visiting scholar at Swinburne University of Technology and at Monash University, Australia. He is an associate professor at the Faculty of Information and Technology, Macau University of Science and Technology (MUST). Before joining MUST, he worked as an associate professor at Jiangsu University, China. His current research interests include software testing (including adaptive random testing, random testing, failure-based testing, combinatorial testing, and regression testing), debugging, and maintenance. He has more than 50 publications in journals and proceedings, including in IEEE Transactions on Software Engineering, IEEE Transactions on Reliability, IEEE Transactions on Emerging Topics in Computational Intelligence, Journal of Systems and Software, Information and Software Technology, IET Software, The Computer Journal, International Journal of Software Engineering and Knowledge Engineering, Information Sciences, ICSE, ISSRE, ICST, COMPSAC, QRS, SEKE, and SAC. He is a senior member of the IEEE and the China Computer Federation, and a member of the ACM. More information about him and his work is available online at https://huangrubing.github.io/.

\vbox{}
\noindent
\textbf{Dave Towey}
received the B.A. and M.A. degrees in computer science, linguistics, and languages from the University of Dublin, Trinity College, Ireland; the M.Ed. degree in education leadership from the University of Bristol, U.K.; and the Ph.D. degree in computer science from The University of Hong Kong, China. He is an associate professor at University of Nottingham Ningbo China (UNNC), in Zhejiang, China, where he serves as the deputy head of the School of Computer Science. He is also the deputy director of the International Doctoral Innovation Centre. He is a member of the UNNC Artificial Intelligence and Optimization research group. His current research interests include software testing (especially adaptive random testing, for which he was amongst the earliest researchers who established the field, and metamorphic testing), computer security, and technology-enhanced education. He co-founded the ICSE International Workshop on Metamorphic Testing in 2016. He is a fellow of the HEA, a senior member of the IEEE, and a member of the ACM.

\vbox{}
\noindent
\textbf{Michael Omari}
received the B.Sc. degree in computer science from the University of Ghana, Legon, in 2007, and the Master's degree in information technology from Coventry University, U.K., in 2014. He holds a Ph.D. degree in computer applied technology from the School of Computer Science and Telecommunication Engineering, Jiangsu University, China. He is a lecturer at the Department of Computer Science, School of Applied Science, Takoradi Technical University specializing in assembly language programming and software engineering. His research interests include software testing and embedded systems.

\vbox{}
\noindent
\textbf{Dmitry Yashunin}
received a Master's degree in physics from Nizhny Novgorod State University in 2009, and a PhD degree in laser physics from the Institute of Applied Physics RAS in 2015. From 2008 to 2012 he was working at Mera Networks (currently Orion Innovation) as a software engineer. From 2016 to 2019 he was working at Intelli-Vision in the position of a leading research engineer. Dmitry currently works at Harman (a Samsung company) as an associate director. He is author of more than 10 papers on physics and computer science. His current research interests include scalable similarity search, computer vision and deep learning.

\vbox{}
\noindent
\textbf{Patrick Kwaku Kudjo}
is a Lecturer at the Department of Information Technology, University of Professional Studies, Ghana. He has a Ph.D. in Computer Application Technology awarded by Jiangsu University, China. He holds a Master's degree in Information Technology from Sikkim Manipal University, India, and a Bachelor's degree in Computer Science and Management from Wisconsin International University College, Accra Ghana. Patrick is an avid researcher and practitioner with more than twenty high-quality research papers in reputable journals and conferences such as Journal of Systems and Software, Software Quality Journal, Software: Practice and Experience, Applied Intelligence, Service-Oriented Computing, and Applications ISSTA, QRS, ASE, SEKE, etc. His current research interest includes Information Security, Machine Learning, and Blockchain Analysis. He is a member of the Association for Computing Machinery (ACM), and the Institute of Electrical and Electronics Engineers (IEEE).

\vbox{}
\noindent
\textbf{Tao Zhang}
received the BS degree in automation, the MEng degree in software engineering from Northeastern University, China, and the PhD degree in computer science from the University of Seoul, South Korea. After that, he spent one year with the Hong Kong Polytechnic University as a postdoctoral research fellow. Currently, he is an associate professor with the Faculty of Information Technology, Macau University of Science and Technology (MUST). Before joining MUST, he was the faculty member of Harbin Engineering University and Nanjing University of Posts and Telecommunications, China. He published more than 50 high-quality papers at renowned software engineering and security journals and conferences such as the IEEE Transactions on Software Engineering, IEEE Transactions on Information Forensics and Security, IEEE Transactions on Dependable and Secure Computing, IEEE Software, ICSE, etc. His current research interests include mining software repositories and mobile software security. He is a senior member of ACM, IEEE, and CCF.

\end{document}